\documentclass[11pt]{article}
\pdfoutput=1

\usepackage{epsfig}
\usepackage{psfrag}
\usepackage{latexsym}
\usepackage{indentfirst}
\usepackage{fancyhdr}
\usepackage{dsfont}
\usepackage{amsmath}
\usepackage{amssymb}
\usepackage{amsfonts}
\usepackage{mathrsfs}
\usepackage{amsthm}
\usepackage{pifont}
\usepackage{dsfont}
\usepackage{multirow}
\usepackage{array}
\usepackage{chngpage}
\usepackage{longtable}
\usepackage{cite}
\usepackage{bbold}
\usepackage{color}
\usepackage{braket}
\usepackage{colordvi}
\usepackage{fancybox}
\usepackage[footnotesize]{caption2}
\usepackage{graphicx}
\usepackage[center,footnotesize,hang]{subfigure}
\usepackage{bbm}
\usepackage[colorlinks, linkcolor=red, anchorcolor=black, citecolor=green]{hyperref}
\usepackage{multirow}
\usepackage{array}
\usepackage{textcomp}  
\usepackage{enumitem}
\usepackage{mathrsfs}  
\newcommand{\PreserveBackslash}[1]{\let\temp=\\#1\let\\=\temp}
\newcolumntype{C}[1]{>{\PreserveBackslash\centering}p{#1}}
\newcolumntype{R}[1]{>{\PreserveBackslash\raggedleft}p{#1}}
\newcolumntype{L}[1]{>{\PreserveBackslash\raggedright}p{#1}}
\allowdisplaybreaks  

\newcommand{\bq}{\begin{eqnarray}}
\newcommand{\nq}{\end{eqnarray}}

\makeatletter
\@addtoreset{equation}{section}
\makeatother

\begin{document}

\title{\hfill ~\\[-30mm] \hfill\mbox{\small USTC-ICTS-18-06}\\[10mm]
        \textbf{{\Large Quark and Lepton Mixing Patterns from a Common Discrete Flavor Symmetry with Generalized CP     }  }}

\date{}

\author{\\[1mm]Jun-Nan Lu\footnote{Email: {\tt hitman@mail.ustc.edu.cn}}~,~Gui-Jun Ding\footnote{Email: {\tt dinggj@ustc.edu.cn}}\\ \\
\it{\small Interdisciplinary Center for Theoretical Study and  Department of Modern Physics, }\\
\it{\small University of Science and
    Technology of China, Hefei, Anhui 230026, China}\\[4mm] }
\maketitle

\begin{abstract}

We have studied two approaches to predict quark and lepton mixing patterns from the same flavor symmetry group in combination with CP symmetry. The first approach is based on the residual symmetry $Z_2$ in the charged lepton sector and $Z_2\times CP$ in the neutrino sector. All lepton mixing angles and CP violation phases depend on three real parameters $\theta_{l}$, $\delta_{l}$ and $\theta_{\nu}$. This approach is extended to the quark sector, the up and down quark mass matrices are assumed to be invariant under a $Z_2$ subgroup and $Z_2\times CP$. The necessary and sufficient conditions for the equivalence of two mixing patterns are derived. The second approach has an abelian subgroup and a single CP transformation as residual symmetries of the charged lepton and neutrino sectors respectively. The lepton mixing would be determined up to a real orthogonal matrix multiplied from the right hand side. Analogously we assume that a single CP transformation is preserved by the down (or up) quark mass matrix, and the residual symmetry of the up (or down) quark sector is be an abelian subgroup. As an example, we analyze the possible mixing patterns which can be obtained from the breaking of $\Delta(6n^2)$ and CP symmetries. We find $\Delta(294)$ combined with CP is the smallest flavor group which can give a good fit to the experimental data of quark and lepton mixing in both approaches.

\end{abstract}
\thispagestyle{empty}
\vfill

\newpage
\setcounter{page}{1}

\section{\label{sec:Introduction}Introduction}

Over the past few decades, the quark CKM mixing matrix has been measured quite precisely in meson decays~\cite{Patrignani:2016xqp}. It is established that quark mixing angles are small and the largest one is the Cabibbo angle $\theta_{C}\simeq13.02^{\circ}$ between the first and the second generations. Regarding the CP violation in the quark sector, the unitarity triangle approximately is a right triangle, and the angle $\alpha$ of the unitary triangle is constrained as $\alpha\simeq
(87.6^{+3.5}_{-3.3})^{\circ}$~\cite{Patrignani:2016xqp}. The discovery of neutrino oscillation is a great progress in particle physics and it leads to the 2015 physics Nobel prizes. The neutrino mixing parameters have been rather well measured by a lot of neutrino oscillation experiments. The latest global fitting of neutrino oscillation data gives $31.42^{\circ}\leq\theta_{12}\leq36.05^{\circ}$, $40.3^{\circ}\leq\theta_{23}\leq51.5^{\circ}$ and $8.09^{\circ}\leq\theta_{13}\leq8.98^{\circ}$ at $3\sigma$ confidence level~\cite{Esteban:2016qun,nufit}. Similar results are obtained by other oscillations global fit~\cite{deSalas:2017kay,Capozzi:2018ubv}. Concerning the CP violation in neutrino oscillation, the hit for a Dirac CP $\delta_{CP}$ around $3\pi/2$ is reported by T2K~\cite{Abe:2017uxa} and NO$\nu$A~\cite{Adamson:2017gxd} although the significance of this signal is still low. Establishing the existence of leptonic CP violation is an important goal of future long baseline neutrino experiments. We can easily see that the observed pattern of neutrino mixing differs drastically from the quark mixing pattern.

The standard model can only accommodate but not explain these data. Understanding the origin of the quark and lepton mixing and mass hierarchy is a longstanding fundamental open question in particle physics. There have been many attempts in the literature to try and explain these structures. It turns out that a broken flavour symmetry based on the non-abelian discrete groups is particularly suitable to reproduce certain neutrino mixing pattern, for review see~\cite{Altarelli:2010gt,Ishimori:2010au,King:2013eh,King:2014nza,King:2015aea}.
The non-abelian discrete flavour symmetry is also exploited to
explain the mixing pattern among quarks~\cite{Lam:2007qc,Blum:2007jz,deAdelhartToorop:2011re,Holthausen:2013vba,Araki:2013rkf,Yao:2015dwa,Varzielas:2016zuo}.
It is notable that the experimentally favored Cabibbo angle (for example $\theta_{C}\simeq\pi/14$) can be generated in terms of group-theoretical quantities, if the three generations of left-handed quarks are assigned to an irreducible triplet or the direct sum of a two-dimensional and a one-dimensional representation of the flavor symmetry group~\cite{Lam:2007qc,Blum:2007jz,Yao:2015dwa}. However, the three small hierarchical mixing angles together with quark CP violation can not generated simultaneously if the quark mixing matrix is fully determined by the residual symmetries in the down quark and up quark mass matrices.~\cite{Yao:2015dwa}.

In order to constrain the CP violating phases, a powerful extension is to amend the flavor symmetry with a CP symmetry~\cite{Harrison:2002kp,Harrison:2002et,Harrison:2004he,Grimus:2003yn,Grimus:2012hu,Feruglio:2012cw,Chen:2014wxa,Chen:2015nha,Everett:2015oka,Everett:2016jsk}.
The CP transformation matrix is generally represented by a non-diagonal matrix in the flavor space and consequently it is usually called generalized CP. In order to consistently combine flavor symmetry with CP symmetry, certain consistency condition has to be fulfilled such the generalized CP symmetry is dictated by the flavor symmetry group~\cite{Grimus:1995zi,Feruglio:2012cw,Holthausen:2012dk,Chen:2014tpa}. In the most extensively studied scenario, the flavor and generalized CP symmetries are broken to an abelian subgroup and $Z_2\times CP$ in the charged lepton and neutrino sectors respectively, the lepton mixing angles and CP phases would be expressed in terms of a single real parameter $\theta$ which can take values in the range $0\leq\theta<\pi$~\cite{Feruglio:2012cw,Ding:2013hpa,Ding:2013bpa,Feruglio:2013hia,Li:2013jya,Ding:2013nsa,Ding:2014hva,Ding:2014ssa,Li:2014eia,
Hagedorn:2014wha,Ding:2014ora,Branco:2015hea,Branco:2015gna,Li:2015jxa,DiIura:2015kfa,Ballett:2015wia,Ding:2015rwa,Li:2016ppt,Li:2016nap}.
An exhaustive scan over discrete groups of order less than 2000 is performed in~\cite{Yao:2016zev}, and a classification of all phenomenologically viable mixing patterns obtained in the semi-direct approach is presented~\cite{Yao:2016zev}. Moreover, the generalized CP symmetry can not only constrain the Dirac and Majorana phases but also possibly the CP violation in leptogenesis~\cite{Chen:2016ptr,Hagedorn:2016lva,Li:2017zmk}.

Other possible schemes of predicting lepton mixing parameters from flavor symmetry and generalized CP have also been considered in the literature. The scenario with the residual symmetry $Z_2\times CP$ in both neutrino and charged lepton mass terms is suggested in~\cite{Lu:2016jit,Rong:2016cpk}. The resulting PMNS mixing matrix would depend on two rotation angles $\theta_{\nu}$ and $\theta_{l}$ which freely vary between $0$ and $\pi$. Most importantly, a remarkable advantage of this scheme is that
the experimentally measured quark mixing angles and CP violation phase can be reproduced if the flavor group and generalized CP are broken to two distinct $Z_2\times CP$ subgroups as well in the up and down quark sectors~\cite{Lu:2016jit,Li:2017abz}. The CKM mixing matrix would be predicted in terms of two real parameters $\theta_{u}$ and $\theta_{d}$ which can be chosen to lie in the interval $0\leq\theta_{u,d}<\pi$. What's more, the observed patterns of quark and lepton flavor mixings can be simultaneously understood from the same flavor
symmetry in combination with generalized CP symmetry, and the smallest flavor group is $\Delta(294)$~\cite{Li:2017abz}.

A three-parameter model based on the remnant symmetry $Z_2$ in the charged lepton sector and $Z_2\times CP$ in the neutrino sector is proposed in~\cite{Turner:2015uta,Penedo:2017vtf}. Then the lepton mixing angles and CP phases are determined in terms of three parameters $\theta_{\nu}$, $\theta_{l}$ and $\delta_{l}$ in this scheme. The possible lepton mixing patterns which can be obtained from the popular flavor symmetries $S_4$~\cite{Penedo:2017vtf} and $A_5$~\cite{Turner:2015uta} have been studied, and all possible residual symmetries of this type are considered~\cite{Turner:2015uta,Penedo:2017vtf}. In the present work, we shall investigate whether the experimentally preferred CKM mixing matrix can be derived in a similar way, assuming the parental flavor symmetry and CP symmetry are broken to $Z_2$ and $Z_2\times CP$ in the up and down quark sectors. Furthermore, we shall discuss whether realistic lepton and quark mixing patterns can be achieved from a single flavor symmetry group combined with CP. In order to show concrete examples and find new interesting mixing patterns, we have performed a comprehensive study for the infinite group series $\Delta(6n^2)$ which are broken to all possible residual symmetries of the structure indicated above.

Another three-parameter model is the residual symmetry pattern
for which the charged lepton and neutrino mass matrices are invariant under an abelian subgroup and a single CP transformation respectively. The lepton mixing matrix would be fixed up to a real orthogonal matrix which contains three free rotation angles $\theta_{1,2,3}$~\cite{Li:2017zmk,Chen:2015siy,Chen:2016ica,Chen:2018lsv,Joshipura:2018rit}. A benchmark example of this type model is the $\mu-\tau$ reflection symmetry~\cite{Harrison:2002kp,Harrison:2002et,Harrison:2004he,Grimus:2003yn,Grimus:2012hu,Xing:2015fdg}, which exchanges the muon (tau) neutrino with the tau (muon) antineutrino in the charged lepton mass basis. It is well-known that the $\mu-\tau$ reflection predicts maximal atmospheric mixing $\theta_{23}=\pi/4$ and maximal Dirac phase $\delta_{CP}=\pm\pi/2$. In light of the experimental indications that $\theta_{23}$ and $\delta_{CP}$ deviate from maximal values~\cite{Esteban:2016qun,nufit,deSalas:2017kay,Capozzi:2018ubv}, one could break the $\mu-\tau$ reflection symmetry~\cite{Zhao:2017yvw,Liu:2017frs,Xing:2017mkx,Nath:2018hjx} or consider its possible variations such as the generalized $\mu-\tau$, $e-\mu$ or $e-\tau$ reflections~\cite{Chen:2015siy,Chen:2016ica,Chen:2018lsv}. We shall extend this approach to the quark sector in this work.
As an example, we shall analyze all possible lepton and quark mixing patterns which can be obtained from $\Delta(6n^2)$ flavor group and CP in this scheme.

The motivation of this paper is to investigate whether the observed
pattern of neutrino and quark flavor mixing, which drastically differs from the each other, can be naturally understood in the three-parameter models mentioned above. The paper is organized as follows. In section~\ref{sec:Lepton_flavor_mixing_Z2Z2CP} we present the general formula of lepton mixing matrix when the residual symmetries of the neutrino and charged lepton mass terms are $Z_2\times CP$ and $Z_2$ respectively. Subsequently we derive the criterion to determine whether two distinct residual symmetries give rise to the same mixing pattern. As an example, we perform a detailed study of lepton mixing patterns arising from the breaking of the flavor group $\Delta(6n^2)$ and CP to $Z_2$ in the charged lepton and to $Z_2\times CP$ in the neutrino sector. The predictions for the lepton mixing parameters and the effective Majorana mass of the neutrinoless double beta decay are analyzed numerically. In section~\ref{sec:quark_flavor_mixing_Z2Z2CP} the extension of the analysis of section~\ref{sec:Lepton_flavor_mixing_Z2Z2CP} to quark flavor mixing is discussed, and we investigate the possible quark mixing patterns if the $\Delta(6n^2)$ and CP are broken to $Z_2$ and $Z_2\times CP$ in the up and down quark sectors respectively. In section~\ref{sec:Lepton_flavor_mixing_OneCP} we study another three-parameter model in which the charged lepton and neutrino mass matrices are invariant under the action of a residual abelian subgroup and a single CP transformation respectively. Then we apply the approach of section~\ref{sec:Lepton_flavor_mixing_OneCP} to quark flavor mixing in section~\ref{sec:onecp_quark}. Finally we summarize our main results and make some concluding remarks in section~\ref{sec:conclusion}.

\section{\label{sec:Lepton_flavor_mixing_Z2Z2CP}Lepton flavor mixing from flavor and CP symmetries breaking to residual symmetries $Z_{2}$ and $Z_{2}\times CP$}

In order to understand lepton flavor mixing and CP violation, we shall impose a discrete flavor symmetry $G_{f}$ which can be consistently combined with the generalized CP symmetry.
The three generations of left-handed leptons are assigned to transform as a faithful irreducible triplet $\mathbf{3}$ of $G_f$. We assume that the flavor and CP symmetries are broken into the subgroups $Z_{2}$ and $Z_{2} \times CP$ in the charged lepton and neutrino sectors respectively. Such approach has been studied for $S_4$ and $A_5$ flavor symmetry in Refs.~\cite{Turner:2015uta,Penedo:2017vtf }. In the following, we shall recapitulate how lepton mixing can be predicted from this symmetry breaking scheme.

As regards the charged leptons, the remnant symmetry $Z_{2}$ is denoted as $Z_{2}^{g_{l}}$, where $g_{l}$ refers to the generator and it is of order two with $g_{l}^{2}=1$. The requirement that $Z_{2}^{g_{l}}$ is preserved entails that the hermitian combination $m^{\dagger}_{l}m_{l}$ should be invariant under the action of $g_{l}$\footnote{The charged lepton mass matrix $m_l$ is given in the right-left basis.},
\begin{equation}
\label{eq:mlsq-invariant}
\rho_{\mathbf{3}}^{\dagger}(g_{l})m_{l}^{\dagger}m_{l}\rho_{\mathbf{3}}(g_{l})=m_{l}^{\dagger}m_{l}\,,
\end{equation}
which implies
\begin{equation}
\label{eq:commute}[m_{l}^{\dagger}m_{l}, \rho_{\mathbf{3}}(g_{l})]=0\,,
\end{equation}
where $\rho_{\mathbf{3}}(g_{l})$ denotes the representation matrix of $g_{l}$ in the triplet representation $\mathbf{3}$. The unitary matrix $U_{l}$ realizes the transformation to the physical basis where the
product $m^{\dagger}_{l}m_{l}$ is diagonal,
\begin{equation}
\label{eq:charged-lepton_diagonalization}
U_{l}^{\dagger}m_{l}^{\dagger}m_{l}U_{l}=\text{diag}(m_{e}^{2},m_{\mu}^{2},m_{\tau}^{2})\,,
\end{equation}
where $m_{e},m_{\mu}$ and $m_{\tau}$ are the masses of $e,\mu$ and $\tau$ respectively.
From Eq.~\eqref{eq:commute}, we can see that $\rho_{\mathbf{3}}(g_{l})$ can be diagonalized by $U_{l}$ as well, in addition, $g_{l}$ is of order two and consequently the eignevalue of $\rho_{\mathbf{3}}(g_{l})$ is either $+1$ or $-1$. Therefore we have
\begin{equation}
\label{eq:diag-rho_l-1}
U_{l}^{\dagger}\rho_{\mathbf{3}}(g_{l})U_{l}=P_{l}\text{diag}(1,-1,-1)P_{l}^{T}\,,
\end{equation}
where $P_{l}$ is a generic three dimensional permutation matrix. Since the $Z^{g_l}_2$ charges of the three generations of lepton doublets are partially degenerate, the residual symmetry $Z^{g_l}_2$ can not fully distinguish the three generations. Solving the constraint equation of Eq.~\eqref{eq:diag-rho_l-1}, we find that the remnant flavor symmetry $Z^{g_l}_2$ fixes the unitary transformation $U_{l}$ to be of the following form
\begin{equation}
\label{eq:Ul_expression}
U_{l}=\Sigma_{l}U_{23}^{\dagger}(\theta_{l},\delta_{l})Q_{l}^{\dagger}P_{l}^{T}\,,
\end{equation}
where $\Sigma_{l}$ is a constant diagonalization matrix of $\rho_{\mathbf{3}}(g_{l})$ and it satisfy
\begin{equation}
\Sigma_{l}^{\dagger}\rho_{\mathbf{3}}(g_{l})\Sigma_{l}=\text{diag}(1,-1,-1)\,,
\end{equation}
and $U_{23}(\theta_{l},\delta_{l})$ is a block diagonal unitary rotation, and $Q_l$ is a phase matrix with
\begin{equation}
U_{23}(\theta_{l},\delta_{l})=\left( \begin{array}{ccc}
1 & 0 & 0 \\
0 & \cos \theta_{l} & \sin \theta_{l} \\
0 & -\sin \theta_{l} & \cos \theta_{l} \end{array} \right)
\left( \begin{array}{ccc}
1 & 0 & 0 \\
0 & e^{i\delta_{l}} & 0 \\
0 & 0 & e^{-i\delta_{l}} \end{array} \right),~~
Q_{l}=\left( \begin{array}{ccc}
e^{-i\gamma_{1}} & 0 & 0 \\
0 & e^{-i\gamma_{2}} & 0 \\
0 & 0 & e^{-i\gamma_{3}}  \end{array} \right)\,,
\end{equation}
where $\theta_{l}$, $\delta_{l}$ and $\gamma_{1,2,3}$ are free real parameters.

In this work we assume neutrinos are Majorana particles. The predictions for $\theta_{12}$, $\theta_{13}$, $\theta_{23}$ and $\delta_{CP}$ are identical for Majorana or Dirac neutrinos, while the Majorana phases are unphysical if neutrinos are Dirac particles. The neutrino remnant symmetries $Z_{2}\times CP$ are denoted as $Z_{2}^{g_{\nu}} \times X_{\nu}$ with $g_{\nu}^{2}=1$. The remnant CP transformation $X_{\nu}$ should be a $3\times 3$ unitary and symmetric matrix otherwise the light neutrino masses would be completely or partially degenerate. The residual symmetry $Z_{2}^{g_{\nu}}\times X_{\nu}$ is well defined if and only if the restricted consistency condition is satisfied~\cite{Feruglio:2012cw,Chen:2014wxa,Everett:2015oka,Chen:2015nha},
\begin{equation}
\label{eq:nu_consistency_condition}
X_{\nu}\rho_{\mathbf{3}}^{*}(g_{\nu})X_{\nu}^{-1}=\rho_{\mathbf{3}}(g_{\nu})\,.
\end{equation}
Requiring that $Z_{2}^{g_{\nu}} \times X_{\nu}$ is a symmetry of the neutrino mass matrix $m_{\nu}$ entails that $m_{\nu}$ should be invariant under $Z_{2}^{g_{\nu}} \times X_{\nu}$,
\begin{equation}
\label{eq:nu-mass-sym}
\rho_{\mathbf{3}}^{T}(g_{\nu})m_{\nu}\rho_{\mathbf{3}}(g_{\nu})=m_{\nu},\qquad
X_{\nu}^{T}m_{\nu}X_{\nu}=m_{\nu}^{*}\,.
\end{equation}
The neutrino mass matrix $m_{\nu}$ can be expressed in terms of the neutrino masses $m_{1,2,3}$ and the unitary rotation $U_{\nu}$ as $m_{\nu}=U_{\nu}^{*}\text{diag}(m_{1},m_{2},m_{3})U^{\dagger}_{\nu}$. Inserting this identity into Eq.~\eqref{eq:nu-mass-sym}, we find the imposed residual symmetry $Z_{2}^{g_{\nu}} \times X_{\nu}$ leads to the following constraints on $U_{\nu}$,
\begin{eqnarray}
\label{eq:Unu_Gnu}&&U_{\nu}^{\dagger}\rho_{\mathbf{3}}(g_\nu)U_\nu=\text{diag}(\pm 1,\pm 1,\pm 1)\,,\\
\label{eq:Unu_Xnu}&&U_\nu^{\dagger}X_{\nu}U_{\nu}^*=\text{diag}(\pm 1,\pm 1,\pm 1)\equiv Q^2_{\nu}\,,
\end{eqnarray}
where $Q_{\nu}$ is a diagonal and unitary matrix with non-vanishing entries equal to $\pm1$ and $\pm i$. The (11) entry of $Q_{\nu}$ can be set to be one by choosing the overall phase of $U_{\nu}$. Therefore $Q_{\nu}$ can be parameterized as
\begin{equation}
\label{eq:Q_nu}
 Q_{\nu}=\left( \begin{array}{ccc}
1 & 0 & 0 \\
0 & i^{k_{1}} & 0 \\
0 & 0 & i^{k_{2}}
\end{array} \right)\,,
\end{equation}
with $k_{1,2}=0,1,2,3$. As shown in Ref.~\cite{Yao:2016zev}, the constraint equations in Eqs.~(\ref{eq:Unu_Gnu}, \ref{eq:Unu_Xnu}) can be conveniently solved by performing Takagi factorization of the residual CP transformation $X_{\nu}$,
\begin{equation}
\label{eq:z2lep38}
X_{\nu}=\Sigma_{\nu}\Sigma_{\nu}^{T}\,,
\end{equation}
where $\Sigma_{\nu}$ is a unitary matrix and it diagonalizes the residual flavor transformation $\rho_{\mathbf{3}}(g_{\nu})$ as well,
\begin{equation}
\label{eq:z2lep39}
\Sigma_{\nu}^{\dagger}\rho_{\mathbf{3}}(g_{\nu})\Sigma_{\nu}=\pm\text{diag}(1,-1,-1)\,.
\end{equation}
Then the unitary transformation $U_{\nu}$ would be fixed to take the form~\cite{Yao:2016zev}
\begin{equation}
\label{eq:Unu_expression}
U_{\nu}=\Sigma_{\nu}S_{23}(\theta_{\nu})P_{\nu}Q_{\nu}\,.
\end{equation}
where $P_{\nu}$ is a permutation matrix, $S_{23}(\theta_{\nu})$ is a rotation matrix in the (23)-plane,
\begin{equation}
S_{23}(\theta_{\nu})=
\left( \begin{array}{ccc}
1 & 0 & 0 \\
0 & \cos \theta_{\nu} & \sin \theta_{\nu} \\
0 & -\sin \theta_{\nu} & \cos \theta_{\nu} \\ \end{array} \right)\,,
\end{equation}
with $\theta_{\nu}$ real. The lepton flavor mixing arises from the mismatch between $U_{l}$ and $U_{\nu}$. Hence the residual symmetry $\left\{Z^{g_{l}}_2, Z^{g_{\nu}}_2\times X_{\nu}\right\}$ enforces that the lepton mixing matrix $U_{PMNS}$ is given by
\begin{equation}
\label{eq:PMNS_z2z2cp}
U_{PMNS}=U_{l}^{\dagger}U_{\nu}=Q_{l}P_{l}U_{23}(\theta_{l},\delta_{l})\Sigma_{l}^{\dagger}
\Sigma_{\nu}S_{23}(\theta_{\nu})P_{\nu}Q_{\nu}\,,
\end{equation}
where we have redefined $P_{l}Q_{l}P_{l}^{T}$ as $Q_{l}$ which can be absorbed into the charged lepton fields. Since the lepton masses can not be predicted in this approach, the lepton mixing matrix is determined only up to exchanges of rows and columns. The permutation matrices $P_{l}$ and $P_{\nu}$ can take six possible forms and it can be generated from
\begin{equation}
P_{12}=\left(\begin{array}{ccc}
0 & 1 & 0 \\
1 & 0 & 0 \\
0 & 0 & 1 \end{array}\right),~~~~
P_{13}=\left(\begin{array}{ccc}
0 & 0 & 1 \\
0 & 1 & 0 \\
1 & 0 & 0  \end{array} \right),~~~~
P_{23}=\left( \begin{array}{ccc}
1 & 0 & 0 \\
0 & 0 & 1 \\
0 & 1 & 0 \end{array} \right)\,.
\end{equation}
We see that the lepton mixing matrix depends on three free continuous parameters $\theta_{l},\delta_{l}$ and $\theta_{\nu}$, and one entry (e.g. the (11) element of $\Sigma_{l}^{\dagger}\Sigma_{\nu}$) is fixed to be certain constant value by the postulated residual symmetry. Moreover,
the PMNS mixing matrix in Eq.~\eqref{eq:PMNS_z2z2cp} has the following properties
\begin{eqnarray}
&&U_{PMNS}(\theta_{l}+\pi,\delta_{l},\theta_{\nu})=P_{l}\text{diag}(1,-1,-1)P_{l}^{T}U_{PMNS}(\theta_{l},\delta_{l},\theta_{\nu})\,\\
&&U_{PMNS}(\pi-\theta_{l},\delta_{l},\theta_{\nu})=P_{l}\text{diag}(1,-i,i)P_{l}^{T}U_{PMNS}(\theta_{l},\delta_{l}-\pi/2,\theta_{\nu})\,\\
&&U_{PMNS}(\theta_{l},\delta_{l}+\pi,\theta_{\nu})=P_{l}\text{diag}(1,-1,-1)P_{l}^{T}U_{PMNS}(\theta_{l},\delta_{l},\theta_{\nu})\,\\
&&U_{PMNS}(\theta_{l},\delta_{l},\theta_{\nu}+\pi)=U_{PMNS}(\theta_{l},\delta_{l},\theta_{\nu})P^{T}_{\nu}\text{diag}(1,-1,-1)P_{\nu}\,
\end{eqnarray}
where the diagonal matrices $P_{l}\text{diag}(1,-1,-1)P_{l}^{T}$, $P_{l}\text{diag}(1,-i, i)P_{l}^{T}$ can be absorbed into $Q_{l}$ and $P^{T}_{\nu}\text{diag}(1,-1,-1)P_{\nu}$ can be absorbed into $Q_{\nu}$. Thus the fundamental intervals of $\theta_{l},\delta_{l}$ and $\theta_{\nu}$ are $[0,\pi/2],[0,\pi)$ and $[0,\pi)$ respectively. Notice that two pair of subgroups $\{Z^{g'_{l}}_2, Z^{g'_{\nu}}_2\times X'_{\nu}\}$ and $\{Z^{g_{l}}_2, Z^{g_{\nu}}_2\times X_{\nu}\}$ would lead to the same $U_{PMNS}$ if they are related by a basis transformation, i.e. if these pair of groups are conjugate under an element belonging to $G_{f}$.

\subsection{\label{subsec:criterion_Z2xZ2xCP_lepton}The criterion for the equivalence of two lepton mixing patterns}

In some cases, two distinct residual symmetries lead to the same mixing pattern, if a possible shift in the continuous free parameters $\theta_{l}$, $\delta_{l}$ and $\theta_{\nu}$ is taken into account. Then we shall call these two mixing patterns are equivalent. In this section, we shall derive the criterion to determine whether two resulting mixing patterns are equivalent or not. In our approach, the lepton mixing matrices derived from two distinct residual symmetries of the structure $\{Z^{g_{l}}_2, Z^{g_{\nu}}_2\times X_{\nu}\}$  take the form
\begin{align}
\nonumber U_{PMNS}&=Q_{l}P_{l}U_{23}(\theta_{l},\delta_{l})\Sigma_{l}^{\dagger}
\Sigma_{\nu}S_{23}(\theta_{\nu})P_{\nu}Q_{\nu},\\
\label{eq:UPMNS_equv}U'_{PMNS}&=Q'_{l}P'_{l}U_{23}(\theta'_{l},\delta'_{l})\Sigma'^{\dagger}_{l}\Sigma'_{\nu}S_{23}(\theta'_{\nu})P'_{\nu}Q'_{\nu}\,.
\end{align}
If the two mixing patterns are equivalent, the fixed element has to be equal, and without loss of generality we assume it is the (11) entry of the PMNS matrix. As a result, the permutation matrices $P_{l}$, $P_{\nu}$, $P'_{l}$ and $P'_{\nu}$ can only be $1$ and $P_{23}$. In addition, the following identities are satisfied,
\begin{equation}
\label{eq:z2lep61}
P_{23}U_{23}(\theta_{l},\delta_{l})=\text{diag}(1, -1, 1)U_{23}(\theta_{l}-\pi/2,\delta_{l}),\quad S_{23}(\theta_{\nu})P_{23}=S_{23}(\theta_{\nu}+\pi/2)\text{diag}(1, -1, 1)\,,
\end{equation}
where the diagonal matrices can be absorbed into $Q_{l}$ and $Q_{\nu}$. Hence we could choose $P_{l}=P_{\nu}=P'_{l}=P'_{\nu}=1$, and then we have
\begin{equation}
U_{PMNS}=Q_{l}U_{23}(\theta_{l},\delta_{l})US_{23}(\theta_{\nu})Q_{\nu},~~U'_{PMNS}=Q'_{l}U_{23}(\theta'_{l},\delta'_{l})U'S_{23}(\theta'_{\nu})Q'_{\nu}\,.
\end{equation}
with $U\equiv\Sigma_{l}^{\dagger}\Sigma_{\nu}$ and $U'\equiv\Sigma_{l}^{'\dagger}\Sigma_{\nu}$. Generally $U$ and $U'$ can be denoted as
\begin{equation}
\label{eq:z2lep82}
U=\left( \begin{array}{ccc}
a_{1} &a_{2} &a_{3}\\
a_{4} &a_{5} &a_{6}\\
a_{7} &a_{8} &a_{9} \end{array}
\right),~~
U'=\left( \begin{array}{ccc}
b_{1} &b_{2} &b_{3}\\
b_{4} &b_{5} &b_{6}\\
b_{7} &b_{8} &b_{9} \end{array}
\right)\,,
\end{equation}
where $a_1$ and $b_1$ are elements fixed by residual symmetries, and they can be chosen to be positive by multiplying an overall phase. A necessary condition for the equivalence of $U_{PMNS}$ and $U'_{PMNS}$ is
\begin{equation}
a_1=b_1,\qquad a_1, b_2\in\mathbb{R}\,.
\end{equation}
It is quite convenient to make an equivalent transformation of $U_{PMNS}$ by extracting constant matrices $U_{23}(\theta_{l}^{c},\delta_{l}^{c})$ and $S_{23}(\theta_{\nu}^{c})$ from $U_{23}(\theta_{l},\delta_{l})$ and $S_{23}(\theta_{\nu})$ respectively. Then the mixing matrix $U_{PMNS}$ can be written as
\begin{equation}
\label{eq:widetilde_U_PMNS}
U_{PMNS}=\widetilde{Q}_{l}U_{23}(\widetilde{\theta}_{l},\widetilde{\delta}_{l})\widetilde{U}S_{23}(\widetilde{\theta}_{\nu})\widetilde{Q}_{\nu}\,,
\end{equation}
where $\widetilde{Q}_{\nu}=Q_{\nu}$ and
\begin{equation}
\label{eq:widetilde_U}
\widetilde{U}=\text{diag}(1,e^{-i\delta_{2}},e^{-i\delta_{2}})U_{23}(\theta_{l}^{c},\delta_{l}^{c})US_{23}(\theta_{\nu}^{c})\,.
\end{equation}
If we take the values of $\delta_{2}, \theta_{l}^{c}, \delta_{l}^{c}$ and $\theta_{\nu}^{c}$ to be
\begin{eqnarray}
\nonumber&&\qquad\qquad\qquad\delta_{l}^{c}=\frac{\text{arg}(a_{7})-\text{arg}(a_{4})}{2},~~
\delta_{2}=\frac{\text{arg}(a_{4})+\text{arg}(a_{7})}{2},~~\\
&&
~~\sin\theta_{l}^{c}=\frac{|a_{7}|-|a_{4}|}{\sqrt{2(|a_4|^2+|a_7|^2)}},
~~\cos\theta_{l}^{c}=\frac{|a_{4}|+|a_{7}|}{\sqrt{2(|a_4|^2+|a_7|^2)}},~~
\cot2\theta_{\nu}^{c}=\frac{2\Re(a_{2}a^{*}_{3})}{|a_{2}|^{2}-|a_{3}|^{2}}\,,
\end{eqnarray}
the unitary matrix $\widetilde{U}$ can be transformed into the ``standard form'',
\begin{equation}
\label{eq:U_tilde}\widetilde{U}=\left( \begin{array}{ccc}
a_1   ~&~  \sqrt{\frac{1}{2}(1-a^2_1)}\,e^{i\rho_2}  ~&~  \sqrt{\frac{1}{2}(1-a^2_1)}\,e^{i\rho_3} \\
\sqrt{\frac{1}{2}(1-a^2_1)} ~&~ \widehat{a}_5 \,e^{i\rho_2}  ~&~ \widehat{a}_6 \,e^{i\rho_3}  \\
\sqrt{\frac{1}{2}(1-a^2_1)} ~&~ \widehat{a}_6 \,e^{i\rho_2}  ~&~ \widehat{a}_5 \,e^{i\rho_3}
\end{array}
\right)\,,
\end{equation}
with
\begin{align}
\label{eq:tansform_target_a}\rho_2=\text{arg}(a_2\cos\theta^{c}_{\nu}-a_3\sin\theta^{c}_{\nu}),~~~ \rho_3=\text{arg}(a_3\cos\theta^{c}_{\nu}+a_2\sin\theta^{c}_{\nu}),~~~a_{1}+\widehat{a}_{5}+\widehat{a}_{6}=0\,.
\end{align}
Notice that $\widetilde{Q}_{l}$ in Eq.~\eqref{eq:widetilde_U_PMNS} is an arbitrary diagonal phase matrix, $\widetilde{\theta}_{l}$, $\widetilde{\delta}_{l}$, $\widetilde{\theta}_{\nu}$ are free parameters and they are closely related to $\theta_{l}$, $\delta_{l}$ and $\delta_{\nu}$ as follows,
\begin{eqnarray}
\nonumber&&Q_{l}=\widetilde{Q}_{l}\text{diag}(1,e^{-i(\delta_{2}+\delta_{1})},e^{i(\delta_{1}-\delta_{2})})\,,\\
\nonumber&&U_{23}(\theta_{l},\delta_{l})=\text{diag}(1,e^{i\delta_{1}},e^{-i\delta_{1}})U_{23}(\widetilde{\theta}_{l},\widetilde{\delta}_{l})U_{23}(\theta_{l}^{c},\delta_{l}^{c})\,,\\
&&S_{23}(\theta_{\nu})=S_{23}(\theta_{\nu}^{c})S_{23}(\widetilde{\theta}_{\nu})\,,
\end{eqnarray}
with
\begin{eqnarray}
\nonumber&&\cos\theta_{l}=|e^{i\widetilde{\delta}_{l}}\cos\theta_{l}^{c}\cos\widetilde{\theta}_{l}-e^{-i\widetilde{\delta}_{l}}\sin\theta_{l}^{c}\sin\widetilde{\theta}_{l}|\,,\\
\nonumber&&\sin\theta_{l}=|e^{i\widetilde{\delta}_{l}}\sin\theta_{l}^{c}\cos\widetilde{\theta}_{l}+e^{-i\widetilde{\delta}_{l}}\cos\theta_{l}^{c}\sin\widetilde{\theta}_{l}|\,,\\
\nonumber&&\varphi_{1}=\text{arg}[e^{i\delta_{l}^{c}}(e^{i\widetilde{\delta}_{l}}\cos\theta_{l}^{c}\cos\widetilde{\theta}_{l}-e^{-i\widetilde{\delta}_{l}}\sin\theta_{l}^{c}\sin\widetilde{\theta}_{l})]\,,\\
\nonumber&&\varphi_{2}=\text{arg}[e^{-i\delta_{l}^{c}}(e^{i\widetilde{\delta}_{l}}\sin\theta_{l}^{c}\cos\widetilde{\theta}_{l}+e^{-i\widetilde{\delta}_{l}}\cos\theta_{l}^{c}\sin\widetilde{\theta}_{l})]\,,\\
\nonumber&&\delta_{l}=\frac{\varphi_{1}-\varphi_{2}}{2}\,,\\
\nonumber&&\delta_{1}=-\frac{\varphi_{1}+\varphi_{2}}{2}\,,\\
&&\theta_{\nu}=\theta_{\nu}^{c}+\widetilde{\theta}_{\nu}\,.
\end{eqnarray}
In the same fashion, we can transform another PMNS matrix $U'_{PMNS}$ into the standard form,
\begin{equation}
U'_{PMNS}=\widetilde{Q}'_{l}U_{23}(\widetilde{\theta}'_{l},\widetilde{\delta}'_{l})\widetilde{U}'S_{23}(\widetilde{\theta}'_{\nu})\widetilde{Q}'_{\nu}\,,
\end{equation}
with
\begin{equation}
\label{eq:U_tildep}\widetilde{U}'=\left( \begin{array}{ccc}
b_1   ~&~  \sqrt{\frac{1}{2}(1-b^2_1)}\,e^{i\rho'_2}  ~&~  \sqrt{\frac{1}{2}(1-b^2_1)}\,e^{i\rho'_3} \\
\sqrt{\frac{1}{2}(1-b^2_1)} ~&~ \widehat{b}_5 \,e^{i\rho'_2}  ~&~ \widehat{b}_6 \,e^{i\rho'_3}  \\
\sqrt{\frac{1}{2}(1-b^2_1)} ~&~ \widehat{b}_6 \,e^{i\rho'_2}  ~&~ \widehat{b}_5 \,e^{i\rho'_3}
\end{array}
\right)\,,
\end{equation}
which satisfies
\begin{equation}
b_{1}+\widehat{b}_{5}+\widehat{b}_{6}=0\,.
\end{equation}
The equivalence of $U_{PMNS}$ and $U'_{PMNS}$ means that for any given values of $\widetilde{\theta_l}$, $\widetilde{\delta}_{l}$, $\widetilde{\theta}_{\nu}$ and the matrices $\widetilde{Q}_{l}$, $\widetilde{Q}_{\nu}$, the corresponding solutions of $\widetilde{\theta}'_l$, $\widetilde{\delta}_{l}'$, $\widetilde{\theta}'_{\nu}$  as well as $\widetilde{Q}'_{l}$ and $\widetilde{Q}'_{\nu}$ can be found such that $U_{PMNS}$ and $U'_{PMNS}$ give the same mixing pattern, i.e.
\begin{equation}
\label{eq:widetilde_U_PMNS_equal}
\widetilde{Q}_{l}U_{23}(\widetilde{\theta}_{l},\widetilde{\delta}_{l})\widetilde{U}S_{23}(\widetilde{\theta}_{\nu})\widetilde{Q}_{\nu}=\widetilde{Q}'_{l}U_{23}(\widetilde{\theta}'_{l},\widetilde{\delta}'_{l})\widetilde{U}'S_{23}(\widetilde{\theta}'_{\nu})\widetilde{Q_{\nu}'}\,,
\end{equation}
which yields
\begin{equation}
\label{eq:equiv_2}
U^{\dagger}_{23}(\widetilde{\theta}'_{l},\widetilde{\delta}'_{l})\widetilde{Q}_{L}U_{23}(\widetilde{\theta}_{l},\widetilde{\delta}_{l})\widetilde{U}S_{23}(\widetilde{\theta}_{\nu})\widetilde{Q}_{N}S^{T}_{23}(\widetilde{\theta}'_{\nu})=\widetilde{U'}\,,
\end{equation}
where $\widetilde{Q}_{L}\equiv\widetilde{Q}'^{\dagger}_{l}\widetilde{Q}_{l}=\text{diag}(e^{i\phi_{1}},e^{i\phi_{2}},e^{i\phi_{3}})$ is a generic diagonal phase matrix with $\phi_{1,2,3}$ are arbitrary free parameters, and $\widetilde{Q}_{N}\equiv\widetilde{Q}_{\nu}\widetilde{Q}'^{\dagger}_{\nu}$ is also diagonal with entries $\pm1$ and $\pm i$. The combination $U^{\dagger}_{23}(\widetilde{\theta}'_{l},\widetilde{\delta}'_{l})\widetilde{Q}_{L}U_{23}(\widetilde{\theta}_{l},\widetilde{\delta}_{l})$ in Eq.~\eqref{eq:equiv_2} can be simplified into
\begin{equation}
\label{eq:simplify_U23}
U^{\dagger}_{23}(\widetilde{\theta}'_{l},\widetilde{\delta}'_{l})\widetilde{Q}_{L}U_{23}(\widetilde{\theta}_{l},\widetilde{\delta}_{l})=\text{diag}(e^{i\widehat{\phi}_{1}},e^{i\widehat{\phi}_{2}},e^{i\widehat{\phi}_{3}})U_{23}(\widehat{\theta}_{l},\widehat{\delta}_{l})\,
\end{equation}
with
\begin{eqnarray}
\nonumber&& \widehat{\phi}_{1}= \phi_{1}\,,\\
\nonumber&&\widehat{\phi}_{2}= (\phi_{2}+\phi_{3}+\psi_{1}+\psi_{2}-2\widetilde{\delta}'_{l})/2\,,\\
\nonumber&&\widehat{\phi}_{3}= (\phi_{2}+\phi_{3}-\psi_{1}-\psi_{2}+2\widetilde{\delta}'_{l})/2\,,\\
\nonumber&&\cos\widehat{\theta}_{l}=|e^{i(\phi_{2}-\phi_{3})/2}\cos\widetilde{\theta}'_{l}\cos\widetilde{\theta}_{l}+e^{-i(\phi_{2}-\phi_{3})/2}\sin\widetilde{\theta}'_{l}\sin\widetilde{\theta}_{l}|,\\
\nonumber&&\sin\widehat{\theta}_{l}=|e^{i(\phi_{2}-\phi_{3})/2}\cos\widetilde{\theta}'_{l}\sin\widetilde{\theta}_{l}-e^{-i(\phi_{2}-\phi_{3})/2}\sin\widetilde{\theta}'_{l}\cos\widetilde{\theta}_{l}|,\\
\nonumber&&\psi_{1}=\text{arg}[e^{i\widetilde{\delta}_{l}}(e^{i(\phi_{2}-\phi_{3})/2}\cos\widetilde{\theta}'_{l}\cos\widetilde{\theta}_{l}+e^{-i(\phi_{2}-\phi_{3})/2}\sin\widetilde{\theta}'_{l}\sin\widetilde{\theta}_{l})]\,,\\
\nonumber&&\psi_{2}=\text{arg}[e^{-i\widetilde{\delta}_{l}}(e^{i(\phi_{2}-\phi_{3})/2}\cos\widetilde{\theta}'_{l}\sin\widetilde{\theta}_{l}-e^{-i(\phi_{2}-\phi_{3})/2}\sin\widetilde{\theta}'_{l}\cos\widetilde{\theta}_{l})]\,,\\
&&\widehat{\delta}_{l}=(\psi_{1}-\psi_{2})/2\,.
\end{eqnarray}
The phase matrix $\widetilde{Q}_{N}$ is of the form
\begin{equation}
\widetilde{Q}_{N}=\left( \begin{array}{ccc}
\eta_{1} & 0 & 0 \\
0 & \eta_{2} & 0 \\
0 & 0 & k\eta_{2} \\ \end{array} \right)
\end{equation}
where $\eta_{1}$ and $\eta_{2}$ can be either $\pm 1$ or $\pm i$. After some straightforward algebra, we find that the equivalence condition of Eq.~\eqref{eq:equiv_2} can be satisfied only for $k=\pm1$. In addition, the following equality is fulfilled
\begin{equation}
\label{eq:S_23_simp}
S_{23}(\widetilde{\theta}_{\nu})\widetilde{Q}_{N}S^{T}_{23}(\widetilde{\theta}'_{\nu})=\widetilde{Q}_{N}S_{23}(k\widetilde{\theta}_{\nu}-\widetilde{\theta}'_{\nu}) \equiv \widetilde{Q}_{N}S_{23}(\widehat{\theta}_{\nu}),\quad\text{with}\quad \widehat{\theta}_{\nu}= k\widetilde{\theta}_{\nu}-\widetilde{\theta}'_{\nu} \,.
\end{equation}
Thus the equivalence condition of Eq.~\eqref{eq:equiv_2} is simplified into
\begin{equation}
\label{eq:hat_equal_relation}
\text{diag}(e^{i\widehat{\phi}_{1}},e^{i\widehat{\phi}_{2}},e^{i\widehat{\phi}_{3}})U_{23}(\widehat{\theta}_{l},\widehat{\delta}_{l})\widetilde{U}\widetilde{Q}_{N}S_{23}(\widehat{\theta}_{\nu})=\widetilde{U}'\,.
\end{equation}
If we can find a solution for $\widehat{\phi}_{1}$, $\widehat{\phi}_{2}$, $\widehat{\phi}_{3}$, $\widehat{\theta}_{l}$, $\widehat{\delta}_{l}$, $\widetilde{Q}_{N}$ and $\widehat{\theta}_{\nu}$ such that Eq.~\eqref{eq:hat_equal_relation} is satisfied, $U_{PMNS}$ and $U'_{PMNS}$ would be essentially the same mixing pattern.

In order to determine the parameters $\widehat{\phi}_{2}, \widehat{\phi}_{3}, \widehat{\theta}_{l}$ and $\widehat{\delta}_{l}$, we can compare the $(21)$ and $(31)$ entries of the matrices on both sides of the Eq.~\eqref{eq:hat_equal_relation}, and we obtain
\begin{equation}
\label{eq:a2a3_constriant_lepton}
e^{i(\widehat{\phi}_{2}-\widehat{\delta}_{l})}\eta_{1}(e^{2i\widehat{\delta}_{l}}\cos \widehat{\theta}_{l}+\sin \widehat{\theta_{l}})=1,~~~
e^{i(\widehat{\phi}_{3}-\widehat{\delta}_{l})}\eta_{1}(\cos \widehat{\theta}_{l}-e^{2i\widehat{\delta}_{l}}\sin \widehat{\theta_{l}})=1\,,
\end{equation}
which imply
\begin{equation}
|e^{2i\widehat{\delta}_{l}}\cos \widehat{\theta}_{l}+\sin \widehat{\theta_{l}}|^2=1,~~~|\cos \widehat{\theta}_{l}-e^{2i\widehat{\delta}_{l}}\sin \widehat{\theta_{l}}|^2=1\,.
\end{equation}
Consequently $\widehat{\theta_{l}}$ and $\widehat{\delta}_{l}$ fulfill
\begin{equation}
\label{eq:necessary_consition_2}
\cos 2\widehat{\delta}_{l} \sin 2\widehat{\theta}_{l}=0\,,
\end{equation}
which yields
\begin{equation}
\widehat{\theta}_{l}=0, \pi/2,~~~\text{or}~~~ \widehat{\delta}_{l}=\pm\pi/4, \pm3\pi/4\,.
\end{equation}
\begin{itemize}
\item{$\widehat{\theta}_{l}=0$}

In this case, from Eq.~\eqref{eq:a2a3_constriant_lepton} we can find the values of $\widehat{\phi}_{2}$ and $\widehat{\phi}_{3}$ are
\begin{equation}
e^{i\widehat{\phi}_{2}}=e^{-i\widehat{\delta}_{l}}/\eta_{1},~~~
e^{i\widehat{\phi}_{3}}=e^{i\widehat{\delta}_{l}}/\eta_{1}\,,
\end{equation}
where the parameter $\widehat{\delta}_{l}$ can take any value.

\item{$\widehat{\theta}_{l}=\pi/2$}

The parameters $\widehat{\phi}_{2}$ and $\widehat{\phi}_{3}$ are determined to be
\begin{equation}
e^{i\widehat{\phi}_{2}}=e^{i\widehat{\delta}_{l}}/\eta_{1},~~~
e^{i\widehat{\phi}_{3}}=-e^{-i\widehat{\delta}_{l}}/\eta_{1}\,,
\end{equation}
where $\widehat{\delta}_{l}$ is free.

\item{$\cos2\widehat{\delta}_{l}=0$}

We have $\widehat{\delta}_{l}=\pm\pi/4$ or $\widehat{\delta}_{l}=\pm3\pi/4$ in this case, and the values of $\widehat{\phi}_{2}$ and $\widehat{\phi}_{3}$ are
\begin{equation}
e^{i\widehat{\phi}_{2}}=e^{i(\kappa_{2}\widehat{\theta}_{l}-\widehat{\delta}_{l})}/\eta_{1},~~
e^{i\widehat{\phi}_{3}}=e^{i(\kappa_{2}\widehat{\theta}_{l}+\widehat{\delta}_{l})}/\eta_{1}\,,
\end{equation}
where $\kappa_{2}=-i e^{2i\delta_{l}}$ is $\pm1$.

\end{itemize}

As regards the parameter $\widehat{\theta}_{\nu}$, comparing the $(12)$ and $(13)$ entries of the matrices on both sides of the Eq.~\eqref{eq:hat_equal_relation} we find
\begin{equation}
\label{eq:cons_2}\frac{\eta_{2}e^{-i\rho'_{2}}}{\eta_{1}}(e^{i\rho_{2}}\cos\widehat{\theta}_{\nu}-ke^{i\rho_{3}}\sin\widehat{\theta}_{\nu})=1,~~
\frac{\eta_{2}e^{-i\rho'_{3}}}{\eta_{1}}(ke^{i\rho_{3}}\cos\widehat{\theta}_{\nu}+e^{i\rho_{2}}\sin\widehat{\theta}_{\nu})=1\,,
\end{equation}
which requires
\begin{equation}
\label{eq:necessary_consition_1}
\sin2\widehat{\theta}_{\nu}\cos(\rho_{2}-\rho_{3})=0\,.
\end{equation}
Hence the condition $\widehat{\theta}_{\nu}=0, \pi/2$ or $\cos(\rho_{2}-\rho_{3})=0$ should be fulfilled. In the following, we shall further analyze the equivalence condition of Eq.~\eqref{eq:hat_equal_relation} for the (22), (23), (32) and $(32)$ entries.

\begin{itemize}
\item{$\widehat{\theta}_{\nu}=0$}

Inserting this value of $\widehat{\theta}_{\nu}$ into Eq.~\eqref{eq:cons_2}, we can obtain
\begin{equation}
\label{eq:sol_1}\eta_{2}=\eta_{1}e^{i(\rho'_{2}-\rho_{2})},~~~k=e^{i(\rho'_{3}-\rho_{3}-\rho'_{2}+\rho_{2})}\,.
\end{equation}
Taking into account the constraint in Eq.~\eqref{eq:necessary_consition_2}, we find that the equivalence condition of Eq.~\eqref{eq:hat_equal_relation} entails
\begin{eqnarray}
&&\widehat{a}_{5}=\widehat{b}_{5},\quad \widehat{a}_{6}=\widehat{b}_{6},~~~~\text{for}~~\widehat{\theta}_{l}=0\,,\\
&&\widehat{a}_{5}=\widehat{b}_{6},\quad \widehat{a}_{6}=\widehat{b}_{5},~~~~\text{for}~~\widehat{\theta}_{l}=\pi/2\,.
\end{eqnarray}
For the case of $\cos2\widehat{\delta}_{l}=0$, the parameter $\widehat{\theta}_{l}$ is fixed to be
\begin{eqnarray}
\label{eq:theta_l_equiv1}\tan\widehat{\theta}_{l}=i\kappa_{2}\frac{\widehat{b}_{5}-\widehat{a}_{5}}{\widehat{a}_{6}-\widehat{b}_{5}}\,.
\end{eqnarray}
The solutions of $\tan\widehat{\theta}_{l}$ should be real, i.e.
\begin{equation}
\label{real_constriant_lep}
 i(\widehat{a}_{5}-\widehat{b}_{5})(\widehat{a}_{6}-\widehat{b}_{5})^{*}\in\mathbb{R}\,.
\end{equation}
As both $\widetilde{U}$ and $\widetilde{U}'$ shown in Eqs.~(\ref{eq:U_tilde},\ref{eq:U_tildep}) are unitary matrices, one can show that the above constraint of Eq.~\eqref{real_constriant_lep} is satisfied automatically.
We see the that $U_{PMNS}$ and $U'_{PMNS}$ would give the same lepton mixing if the condition in Eq.~\eqref{eq:sol_1} is fulfilled.

\item{$\widehat{\theta}_{\nu}=\pi/2$}

Solving the equation Eq.~\eqref{eq:cons_2}, we find
\begin{equation}
\label{eq:sol_2}\eta_{2}=\eta_{1}e^{i(\rho'_{3}-\rho_{2})},~~k=-e^{i(\rho'_{2}-\rho_{3}-\rho'_{3}+\rho_{2})}\,.
\end{equation}
Furthermore, the equivalence condition of Eq.~\eqref{eq:hat_equal_relation} requires
\begin{eqnarray}
&&\widehat{a}_{5}=\widehat{b}_{6},\quad \widehat{a}_{6}=\widehat{b}_{5},~~~~\text{for}~~\widehat{\theta}_{l}=0\,,\\
&&\widehat{a}_{5}=\widehat{b}_{5},\quad \widehat{a}_{6}=\widehat{b}_{6},~~~~\text{for}~~\widehat{\theta}_{l}=\pi/2\,,\\
&&\tan\widehat{\theta}_{l}=i\kappa_{2}\frac{\widehat{b}_{5}-\widehat{a}_{6}}{\widehat{a}_{5}-\widehat{b}_{5}},~~~~\text{for}~~\cos2\widehat{\delta}_{l}=0\,.
\end{eqnarray}
This means that the solution for equivalence condition Eq.~\eqref{eq:hat_equal_relation} always exists once the condition of Eq.~\eqref{eq:sol_2} is satisfied.

\item{$\cos(\rho_{2}-\rho_{3})=0$}

In this case, the phases $\rho_{2}$ and $\rho_{3}$ are correlated as follow,
\begin{equation}
\label{eq:sol_3_fir}e^{i\rho_{3}}=i\kappa_{1}e^{i\rho_{2}},~~~~\text{with}~~\kappa_{1}=\pm 1\,.
\end{equation}
Plugging this identity into Eq.~\eqref{eq:cons_2}, we obtain
\begin{equation}
\label{eq:sol_3}
\eta_{2}=\eta_{1}e^{i(\rho'_{2}-\rho_{2}+k\kappa_{1}\widehat{\theta}_{\nu})},~~e^{i\rho'_{3}}=ik\kappa_{1}e^{i\rho'_{2}}\,,
\end{equation}
which implies
\begin{eqnarray}
\nonumber&&\tan\widehat{\theta}_{\nu}=k\kappa_{1}\tan\left(\rho_2-\rho'_2\right), ~~~\text{for}~~~\eta_2/\eta_1=\pm1\,,\\
\label{eq:theta_nu_sol}&&\tan\widehat{\theta}_{\nu}=-k\kappa_{1}\cot\left(\rho_2-\rho'_2\right), ~~~\text{for}~~~\eta_2/\eta_1=\pm i\,.
\end{eqnarray}
In a similar way as previous cases, we find for $\widehat{\theta}_{l}=0, \pi/2$, the equivalence of $U_{PMNS}$ and $U'_{PMNS}$ imposes additional constraint,
\begin{equation}
\tan\left(\rho_2-\rho'_2\right)=i\frac{\widehat{b}_{5}-\widehat{a}_{6}}{\widehat{a}_{5}-\widehat{b}_{5}},~~~\text{or}~~~\tan\left(\rho_2-\rho'_2\right)=i\frac{\widehat{a}_{5}-\widehat{b}_{5}}{\widehat{b}_{5}-\widehat{a}_{6}}\,.
\end{equation}
For the case of $\cos2\widehat{\delta}_{l}=0$, the angle $\widehat{\theta}_{l}$ is determined to be
\begin{equation}
\tan(k\kappa_1\widehat{\theta}_{\nu}+\kappa_2\widehat{\theta}_{l})=i\frac{\widehat{a}_{5}-\widehat{b}_{5}}{\widehat{b}_{5}-\widehat{a}_{6}}\,.
\end{equation}

\end{itemize}

In short summary, $U_{PMNS}$ and $U'_{PMNS}$ in Eq.~\eqref{eq:UPMNS_equv} predicted by two distinct residual symmetry would be the same lepton mixing pattern if one of the conditions in Eq.~\eqref{eq:sol_1}, Eq.~\eqref{eq:sol_2} and Eq.~\eqref{eq:sol_3_fir}, Eq.~\eqref{eq:sol_3} are fulfilled with $a_1=b_1$. These necessary and sufficient conditions for the equivalence of the two mixing patterns in this scenario can be compactly written as
\begin{equation}
e^{4i(\rho'_{2}-\rho_{2})}=1,\quad e^{2i(\rho'_{2}-\rho'_{3}-\rho_{2}+\rho_{3})}=1\,,\label{eq:equv_cond1_Z2Z2CP}
\end{equation}
or
\begin{equation}
e^{4i(\rho'_{2}-\rho_{3})}=1,\quad e^{2i(\rho'_{2}-\rho'_{3}-\rho_{3}+\rho_{2})}=1\,,\label{eq:equv_cond2_Z2Z2CP}
\end{equation}
or
\begin{equation}
e^{2i(\rho_{2}-\rho_{3})}=-1,\quad e^{2i(\rho'_{2}-\rho'_{3})}=-1\,.\label{eq:equv_cond3_Z2Z2CP}
\end{equation}
We would like to remind that the fixed element has to be equal, i.e., $a_1=b_1$.

\subsection{\label{sec:lepton_z2z2cp}Examples of lepton mixing patterns from $\Delta(6n^{2})$ and CP symmetries}

In the following, we shall analyze the lepton mixing patterns that arise from the breaking of $\Delta(6n^2)$ and CP symmetry to the residual symmetries $Z_2$ in the charged lepton sector and to $Z_2 \times CP$ in the neutrino sector. $\Delta(6n^2)$ is a series of $SU(3)$ subgroup, it can be determined in terms of four generators $a$, $b$, $c$ and $d$ which obey the following relations~\cite{Luhn:2007uq,King:2013vna,King:2014rwa,Hagedorn:2014wha,Ding:2014ora}
\begin{align}
\nonumber&\qquad\qquad  a^3=b^2=(ab)^2=c^{n}=d^{n}=1, \quad   cd=dc, \\
\label{eq:relations_Delta}& aca^{-1}=c^{-1}d^{-1},\quad ada^{-1}=c\,, \quad  bcb^{-1}=d^{-1}, \quad bdb^{-1}=c^{-1}\,.
\end{align}
The $\Delta(6n^2)$ group has $6n^2$ elements which is of the form
\begin{equation}
g=a^{\alpha}b^{\beta}c^{\gamma}d^{\delta}\,,
\end{equation}
where $\alpha=0,1,2$, $\beta=0,1$ and $c,d=0,1,\ldots, n-1$. The $\Delta(6n^2)$ group theory including the  conjugate classes, inequivalent irreducible representations and Clebsch-Gordan coefficients has been presented in Ref.~\cite{Ding:2014ora}. We shall assign the three generations of left-handed leptons to a faithful irreducible three-dimensional representation $\mathbf{3}$ in which the four generators are represented by
\begin{equation}
\label{eq:rep_matrix}a=\begin{pmatrix}0 & ~1~ &0 \\ 0&~0~&1 \\
   1&~0~&0\end{pmatrix},~~
   b=-\begin{pmatrix} 0 &~0~ &1 \\ 0&~1~&0 \\
   1&~0~&0\end{pmatrix},~~
   c=\begin{pmatrix} \eta&~0~ &0 \\ 0&~\eta^{-1}~&0 \\
   0&~0~&1\end{pmatrix},~~
   d=\begin{pmatrix}1 &~0~ &0 \\ 0&~\eta~&0 \\
   0&~0~&\eta^{-1}\end{pmatrix}\,,
\end{equation}
with $\eta=e^{2\pi i/n}$. For other faithful three-dimensional representations of $\Delta(6n^2)$ group, the set of all matrices coincide with those of $\mathbf{3}$ up to an overall sign such that all conclusions obtained in a comprehensive study of lepton mixing using $\mathbf{3}$ also hold for other triplets. As shown in Ref.~\cite{Ding:2014ora}, if $n$ is not divisible by three or the doublet representations $\mathbf{2}_2$, $\mathbf{2}_3$ and $\mathbf{2}_4$ are absent for $3|n$, the generalized CP symmetry can be consistently combined with the $\Delta(6n^2)$ flavor symmetry, and it is of the same form as the flavor symmetry transformation in our working basis. For the sake of notational simplicity, we shall not distinguish the abstract elements of $\Delta(6n^2)$ and their representation matrices in the following.

Now we proceed to determine all possible $Z_{2}$ and $Z_2\times CP$ subgroups of the $\Delta(6n^2)$ and CP symmetries. The order two elements of the $\Delta(6n^2)$ group are
\begin{equation}\label{eq:z2_1}
bc^xd^x,~~ abc^x,~~a^2bd^x,\quad x=0,1\ldots n-1\,,
\end{equation}
which are conjugate to each other. It has another three $Z_{2}$ elements
\begin{equation}\label{eq:z2_2}
c^{n/2},\quad d^{n/2},\quad c^{n/2}d^{n/2}\,,
\end{equation}
if the group index $n$ is even. Notice that the three elements in Eq.~\eqref{eq:z2_2} are conjugate to each other as well. The residual CP transformation $X_{\nu}$ should be a symmetric unitary matrix.
Consequently the viable residual CP transformations originating from the generalized CP symmetry compatible with $\Delta(6n^2)$ are
\begin{equation}\label{eq:X_nu}
c^{\gamma}d^{\rho}, ~~ bc^{\gamma}d^{-\gamma},~~ abc^{\gamma}d^{2\gamma}, ~~ a^2bc^{2\gamma}d^{\gamma}, \quad \gamma,\rho=0,1,\ldots,n-1\,.
\end{equation}
As regards the residual symmetry $Z^{g_{\nu}}_2\times X_{\nu}$, the constrained consistency condition of Eq.~\eqref{eq:nu_consistency_condition} has to be satisfied. As a consequence, there are only nine kinds of $Z^{g_{\nu}}_2\times X_{\nu}$ subgroups,
\begin{eqnarray}
\nonumber&&g_{\nu}=bc^{x}d^{x},~~~X_{\nu}=c^{\gamma}d^{-2x-\gamma}, ~ bc^{x+\gamma}d^{-x-\gamma}\,,\\
\nonumber&&g_{\nu}=abc^{x},~~~X_{\nu}=c^{\gamma}d^{2x+2\gamma},~abc^{x+\gamma}d^{2x+2\gamma}\,,\\
\nonumber&&g_{\nu}=a^2bd^{x},~~~X_{\nu}=c^{2x+2\gamma}d^{\gamma},~a^{2}bc^{2x+2\gamma}d^{x+\gamma}\,,\\
\nonumber&&g_{\nu}=c^{n/2},~~~X_{\nu}=c^{\gamma}d^{\rho}\,,\\
\nonumber&&g_{\nu}=c^{n/2},~~~X_{\nu}=abc^{\gamma}d^{2\gamma}\,,\\
\nonumber&&g_{\nu}=d^{n/2},~~~X_{\nu}=c^{\gamma}d^{\rho}\,,\\
\nonumber&&g_{\nu}=d^{n/2},~~~X_{\nu}=a^2bc^{2\gamma}d^{\gamma}\,,\\
\nonumber&&g_{\nu}=c^{n/2}d^{n/2},~~~X_{\nu}=c^{\gamma}d^{\rho}\,,\\
&&g_{\nu}=c^{n/2}d^{n/2},~~~X_{\nu}=bc^{\gamma}d^{-\gamma}\,,
\end{eqnarray}
where $x, \gamma, \rho=0,1,\ldots, n-1$. After considering all possible choices for $Z^{g_{l}}_2$ and $Z^{g_{\nu}}_2\times X_{\nu}$, we find that only the following five combinations of residual symmetries $(Z^{g_{l}}_2, Z^{g_{\nu}}_2, X_{\nu})$ can accommodate the experimental data,
\begin{eqnarray}
\nonumber
&&(Z_{2}^{bc^{x}d^{x}},Z_{2}^{bc^{y}d^{y}},\{c^{\gamma}d^{-2y-\gamma},bc^{y+\gamma}d^{-y-\gamma}\})\,,\\
\nonumber
&&(Z_{2}^{bc^{x}d^{x}},Z_{2}^{abc^{y}},\{c^{\gamma}d^{2y+\gamma},abc^{y+\gamma}d^{2y+2\gamma}\})\,,\\
\nonumber
&&(Z_{2}^{bc^{x}d^{x}},Z_{2}^{c^{n/2}},c^{\gamma}d^{\rho})\,,\\
\nonumber&&(Z_{2}^{bc^{x}d^{x}},Z_{2}^{c^{n/2}},abc^{\gamma}d^{2\gamma})\,,\\
&&(Z_{2}^{c^{n/2}},Z_{2}^{bc^{x}d^{x}},\{c^{\gamma}d^{-2x-\gamma},bc^{x+\gamma}d^{-x-\gamma}\})\,.
\end{eqnarray}
In the following, we shall analyze the predictions for lepton flavor mixing for each possible residual symmetries.

\begin{description}[labelindent=-0.8em, leftmargin=0.3em]

\item[~~(\uppercase\expandafter{\romannumeral1})]

$g_l=bc^xd^x$, $g_{\nu}=bc^yd^y$, $X_{\nu}=\left\{c^{\gamma}d^{-2y-\gamma},
bc^{y+\gamma}d^{-y-\gamma}\right\}$

In this case, the unitary transformation $\Sigma_{l}$ is chosen to be
\begin{equation}
\Sigma_{l}=\frac{1}{\sqrt{2}}\begin{pmatrix}
-e^{-\frac{2i\pi x}{n}}  &0  & e^{-\frac{2i\pi x}{n}}\\
0  &  \sqrt{2}   &  0  \\
1  &  0  &  1
\end{pmatrix}\,.
\end{equation}
The Takagi factorization $\Sigma_{\nu}$ for $X_{\nu}$ is
\begin{equation}
\Sigma_{\nu}=\frac{1}{\sqrt{2}}\left(
\begin{array}{ccc}
 -e^{\frac{i \pi  \gamma }{n}} &~ e^{\frac{i \pi  \gamma }{n}} ~& 0 \\
 0 &~ 0 ~& -\sqrt{2} e^{-\frac{2 i \pi  (y+\gamma )}{n}} \\
 e^{\frac{i \pi  (2 y+\gamma )}{n}} &~ e^{\frac{i \pi  (2 y+\gamma )}{n}} ~& 0
\end{array}
\right)\,.
\end{equation}
Then the matrix $\Sigma=\Sigma_{l}^{\dagger}\Sigma_{\nu}$ is of the form
\begin{equation}
\label{eq:Sigma_I}
\Sigma=\left(
\begin{array}{ccc}
 \cos \varphi_{1} & -i \sin \varphi_{1} & 0 \\
 0 & 0 & -e^{-i \varphi_{2}} \\
 -i \sin \varphi_{1} & \cos \varphi_{1} & 0
\end{array}
\right)\,,
\end{equation}
where an overall phase is omitted and the parameters $\varphi_{1}$ and $\varphi_{2}$ are given by
\begin{equation}\label{eq:varphi_I}
\varphi_1=\frac{x-y}{n}\pi,\qquad
\varphi_2=\frac{x+3(y+\gamma)}{n}\pi\,.
\end{equation}
Using the master formula of Eq.~\eqref{eq:PMNS_z2z2cp}, we find the lepton mixing matrix is given by
\begin{equation}
U_{I}=\left(
\begin{array}{ccc}
 \cos\varphi_{1} &~ s_{\nu}\sin \varphi_{1}  ~& -c_{\nu}\sin \varphi_{1}  \\
 -s_{l}\sin \varphi_{1}  &~ c_{l} c_{\nu}e^{i \delta} +s_{l} s_{\nu}\cos \varphi_{1}  ~& c_{l} s_{\nu}e^{i \delta} -c_{\nu} s_{l}\cos \varphi_{1}  \\
 c_{l}\sin \varphi_{1}  &~ c_{\nu} s_{l}e^{i \delta} -c_{l} s_{\nu}\cos \varphi_{1}  ~& s_{l} s_{\nu}e^{i \delta}+c_{l} c_{\nu}\cos \varphi_{1}   \\
\end{array}
\right)\,,
\end{equation}
up to independent row and column permutations, where both phases matrices $Q_{l}$ and $Q_{\nu}$ are neglected for simplicity, and the parameters $c_{l}$, $s_{l}$, $c_{\nu}$, $s_{\nu}$ and $\delta$ denote
\begin{equation}\label{eq:define_sc}
c_{l}\equiv\cos\theta_{l}, \quad s_{l}\equiv\sin\theta_{l}, \quad c_{\nu}\equiv\cos\theta_{\nu},  \quad s_{\nu}\equiv\sin\theta_{\nu}, \quad \delta\equiv 2\delta_{l}-\varphi_{2}\,.
\end{equation}
Since $\delta_{l}$ is a continuous free parameters, the discrete parameters $\varphi_{2}$ can be absorbed by $\delta_{l}$ and it will not appear in $U_{I}$ explicitly. The parameters $\varphi_{1}$ can take following discrete values
\begin{equation}
\varphi_{1}~(\text{mod}~2\pi)=0,\frac{1}{n}\pi,\frac{2}{n}\pi,...,\frac{2n-1}{n}\pi \,.
\end{equation}
Moreover, we see that the matrix $U_{I}$ satisfies the following symmetry properties,
\begin{eqnarray}
\nonumber
U_{I}(\varphi_{1},\theta_{l}+\pi,\theta_{\nu},\delta)&=&\text{diag}(1,-1,-1)U_{I}(\varphi_{1},\theta_{l},\theta_{\nu},\delta)\,,\\
\nonumber
U_{I}(\varphi_{1},\pi-\theta_{l},\theta_{\nu},\delta)&=&\text{diag}(1,1,-1)U_{I}(\varphi_{1},\theta_{l},\theta_{\nu},\delta-\pi)\,,\\
\nonumber
U_{I}(\varphi_{1},\theta_{l},\pi+\theta_{\nu},\delta)&=&U_{I}(\varphi_{1},\theta_{l},\theta_{\nu},\delta)\text{diag}(1,-1,-1)\,,\\
\nonumber
U_{I}(\varphi_{1}+\pi,\theta_{l},\theta_{\nu},\delta)&=&U_{I}(\varphi_{1},\theta_{l},\pi-\theta_{\nu},\delta)\text{diag}(-1,-1,1)\,,\\
\nonumber
U_{I}(\pi-\varphi_{1},\theta_{l},\theta_{\nu},\delta)&=&\text{diag}(-1,1,1)U_{I}(\varphi_{1},\theta_{l},\pi-\theta_{\nu},\delta)\text{diag}(1,-1,1)\,,\\
\nonumber
U_{I}(\varphi_{1},\theta_{l},\theta_{\nu},\pi+\delta)&=&U_{I}(\varphi_{1},\theta_{l},\pi-\theta_{\nu},\delta)\text{diag}(1,1,-1)\,,\\
U_{I}(\varphi_{1},\theta_{l},\theta_{\nu},\pi-\delta)&=&U_{I}^{*}(\varphi_{1},\theta_{l},\pi-\theta_{\nu},\delta)\text{diag}(1,1,-1)\,.
\end{eqnarray}
Note that the above diagonal matrices can be absorbed into $Q_{l}$ and $Q_{\nu}$. As a result, it is sufficient to focus on the fundamental regions of $0\leq\varphi_1\leq\pi/2$, $0\leq\theta_{l}<\pi/2$, $0\leq\theta_{\nu}<\pi$ and $0\leq\delta<\pi$. We find that the elements $\cos \varphi_{1}$ is independent of the free parameters and it is completely determined by residual symmetry. Depending on the value of the discrete parameter $\varphi_1$, the fixed element $\cos\varphi_{1}$ can be in any position of the PMNS mixing matrix. Hence the row and column permutations lead to nine independent mixing patterns
\begin{equation}
\label{eq:PMNS_caseI}
\begin{array}{lll}
 U_{I,1}=U_{I}, ~~~~&~~~~ U_{I,2}=U_{I}P_{12},  ~~~~&~~~~ U_{I,3}=U_{I}P_{13}\,, \\
U_{I,4}=P_{12}U_{I}, ~~~~&~~~~  U_{I,5}=P_{12}U_{I}P_{12},~~~~&~~~~ U_{I,6}=P_{12}U_{I}P_{13}\,, \\
U_{I,7}=P_{23}P_{12}U_{I}, ~~~~&~~~~ U_{I,8}=P_{23}P_{12}U_{I}P_{12}, ~~~~&~~~~ U_{I,9}=P_{23}P_{12}U_{I}P_{13}\,.
\end{array}
\end{equation}
In the following, we shall perform a numerical analysis of the predictions for the lepton mixing angles and CP violation phases. We shall scan over the free parameters $\theta_{l}$, $\theta_{\nu}$ and $\delta$, and all possible values of the discrete parameter $\varphi_1$ for certain $n$ will be considered. For the group $\Delta(6\cdot2^2)\cong S_4$, the values of $\varphi_1$ can be $0$ and $\pi/2$. Consequently the fixed element $\cos\varphi_1$ is either $0$ or $1$ such that the measured lepton mixing angles can not be accommodated. Thus we consider the next small group with $n=3$, then the possible values of $\varphi_{1}$ are $0$ and $\pi/3$ in the regions of  $\varphi_{1}\in[0,\pi/2]$. Recalling that the fixed elements of the PMNS matrix is $\cos\varphi_{1}$, therefore only the case of $\varphi_{1}=\pi/3$ can generate a phenomenological viable mixing pattern since $\cos0=1$ is excluded by the experimental data.
For $\varphi_{1}=\pi/3$, the fixed elements $\cos\varphi_{1}=1/2$ can be the $(21)$, $(22)$, $(31)$, $(32)$ entries of the PMNS matrix. Hence only $U_{I,4}$, $U_{I,5}$, $U_{I,7}$, $U_{I,8}$
can be in accordance with the experimental data. We require all the three mixing angles $\theta_{12}$, $\theta_{13}$ and $\theta_{23}$ lie in their $3\sigma$ regions of experimental data~\cite{Esteban:2016qun}. The obtained regions of mixing angles and CP violation phases are summarized in table~\ref{Tab:z2z2cp_lepton_v2}.

For the mixing matrix $U_{I,4}$ with $\varphi_{1}=\pi/3$, neearly any value of $\theta_{13}$ in its $3\sigma$ range can be achieved, the solar angle $\theta_{12}$ is found to lie in $[32.989^{\circ}, 36.031^{\circ}]$, and in most cases the atmospheric angle $\theta_{23}$ lies in the first octant since $\theta_{23}\in[40.280^{\circ}, 45.843^{\circ}]$. As regards the CP violation phases, the Dirac CP phase $\delta_{CP}$ is predicted to be around $0/2\pi$, namely, $\delta_{CP}\in[0,0.304\pi]\cup[1.696\pi, 2\pi]$, and the Majorana phases are constrained to lie around $0,\pi$ in the following intervals $\alpha_{21} (\text{mod}~\pi)\in[0,0.138\pi]\cup[0.863\pi,\pi]$ and $\alpha_{31} (\text{mod}~\pi)\in[0,0.085\pi]\cup[0.915\pi,\pi]$. The correlations between different mixing parameters for the mixing pattern $U_{I,4}$ with $\varphi_{1}=\pi/3$ are displayed in figure~\ref{fig:U_I_4_correlations}. We see that the CP violation phases are strongly correlated with each other.

For the mixing pattern $U_{I,5}$ with $\varphi_{1}=\pi/3$, approximately the whole $3\sigma$ ranges of $\theta_{13}$ and $\theta_{12}$ can be reproduced, as can be seen from table~\ref{Tab:z2z2cp_lepton_v2}. The atmospheric angle is in the second octant with $\theta_{23}\in[45.635^{\circ}, 51.531^{\circ}]$. The allowed regions of the three CP violating phases are found to be $\delta_{CP}\in[0,0.464\pi]\cup[1.536\pi,2\pi]$,  $\alpha_{21}(\text{mod}\,\pi)\in[0,0.252\pi]\cup[0.748\pi,\pi]$ and $\alpha_{31}(\text{mod}\,\pi)\in[0,0.162\pi]\cup[0.838\pi,\pi]$ respectively. All the mixing angles and CP phases depend on three free parameters $\theta_{l}$, $\theta_{\nu}$ and $\delta$, in particular we have the sum rule
\begin{equation}
\cos\delta_{CP}=\frac{4\cos^2\theta_{12}\cos^2\theta_{23}+4\sin^2\theta_{12}\sin^2\theta_{13}\sin^2\theta_{23}-1}{2\sin2\theta_{12}\sin2\theta_{23}\sin\theta_{13}\cos\delta_{CP}}
\end{equation}
Hence we expect the mixing parameters should be correlated with each other. Figure~\ref{fig:U_I_5_correlations} shows that there are really peculiar correlations between $\theta_{23}$ and three CP violation phases, and the values of $\delta_{CP}$, $\alpha_{21}$ and $\alpha_{31}$ are highly correlated.

Furthermore we notice that $U_{I,7}$ and $U_{I,8}$ can be obtained from $U_{I,4}$ and $U_{I,5}$ by exchanging the second and third rows of the mixing matrix respectively. As a consequence, $U_{I,7}$ and $U_{I,8}$ lead to same predictions for $\theta_{12}$, $\theta_{13}$, $\alpha_{21}$ and $\alpha_{31}$ as with $U_{I,4}$ and $U_{I,5}$ respectively.
The predicted ranges of $\theta_{23}$ get approximately reflected around $45^{\circ}$, i.e., $\sin^{2}\theta_{23}\rightarrow 1-\sin^{2}\theta_{23}$. The allowed ranges of $\delta_{CP}$ of  $U_{I,7}$ and $U_{I,8}$ with $\varphi_{3}=\pi/3$ can be obtained by shifting $\pi$ of the corresponding ranges of $U_{I,4}$ and $U_{I,5}$ respectively, i.e., $\delta_{CP}\rightarrow \delta_{CP}+\pi$.

Since the three CP violating phases $\delta_{CP}$, $\alpha_{21}$ and $\alpha_{31}$ are predicted to lie in narrow regions
and they are strongly correlated. It is possible to derive specific predictions for the effective mass $|m_{ee}|$ of the
neutrinoless double decay. The effective mass $|m_{ee}|$ is expressed in terms of light neutrino masses and lepton mixing parameters as
\begin{equation}
|m_{ee}|=\left|m_1\cos^2\theta_{12}\cos^2\theta_{13}+m_2\sin^2\theta_{12}\cos^2\theta_{13}e^{i\alpha_{21}}+m_3\sin^2\theta_{13}e^{i(\alpha_{31}-2\delta_{CP})}\right|\,.
\end{equation}

The most general allowed regions of $|m_{ee}|$ versus the lightest neutrino mass $m_{\text{lightest}}$ for $U_{I,4}$ and $U_{I,5}$ are presented in figures~\ref{fig:0nubb_CaseI_456}. Since the first row of  $U_{I,7}$ and $U_{I,8}$ are in common with the corresponding one of $U_{I,4}$ and $U_{I,5}$ respectively, they don't lead to new predictions for the effective mass $|m_{ee}|$. We see that the effective mass $|m_{ee}|$ for inverted ordering (IO) is around the boundaries of the area allowed in the generic case. For the case of normal ordering (NO), strong cancellation in $|m_{ee}|$ can occur for certain values of the lightest neutrino mass. Future experiments searching for $0\nu\beta\beta$ decay is capable of probing all the IO region. Along with the predictions for the neutrino oscillation parameters, the predicted range for $|m_{ee}|$ can also be used to test our approach.

\begin{figure}[hptb!]
\centering
\begin{tabular}{ >{\centering\arraybackslash} m{8.0cm} >{\centering\arraybackslash} m{8.0cm} }
\includegraphics[width=0.48\textwidth]{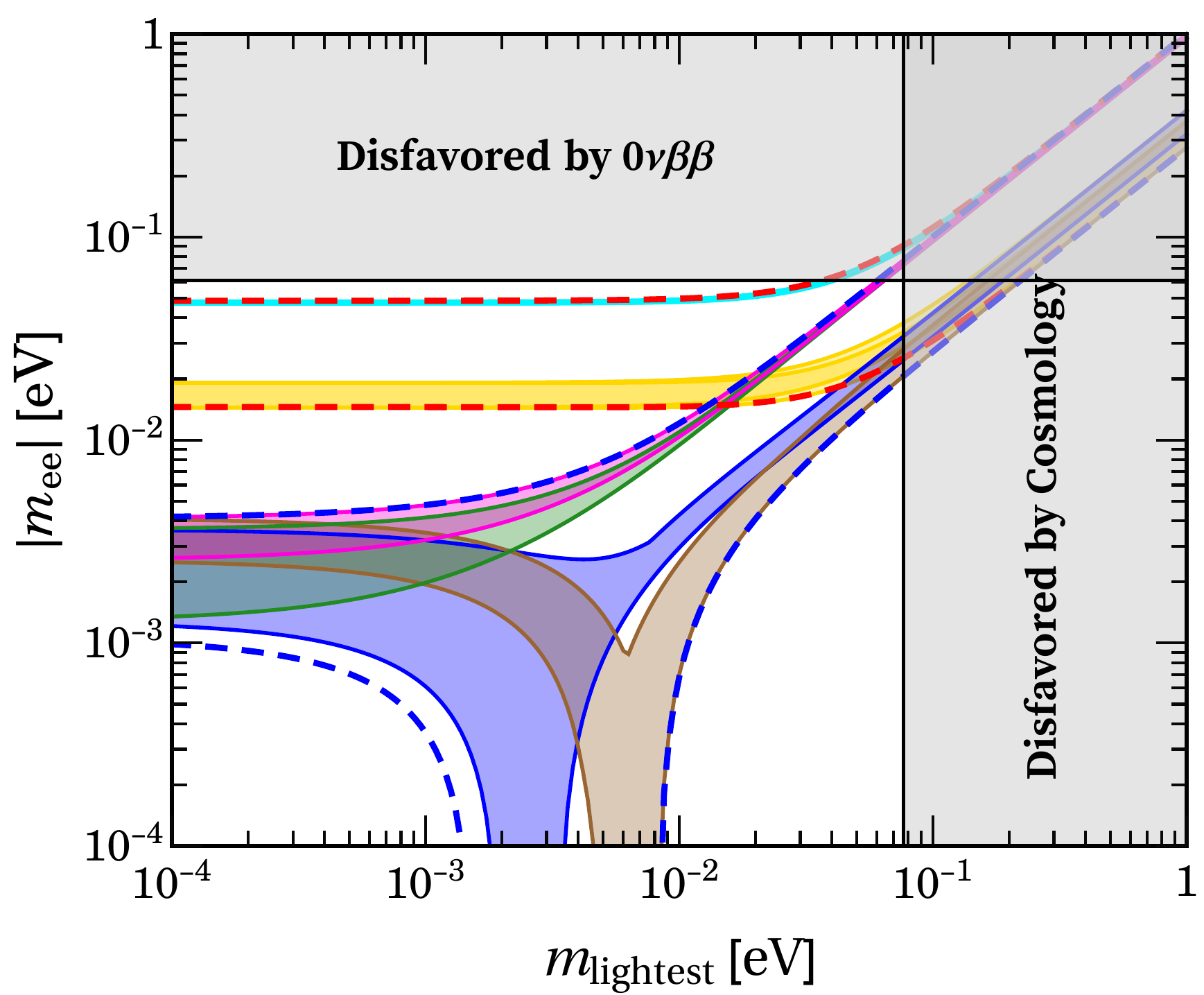}~~&
\includegraphics[width=0.48\textwidth]{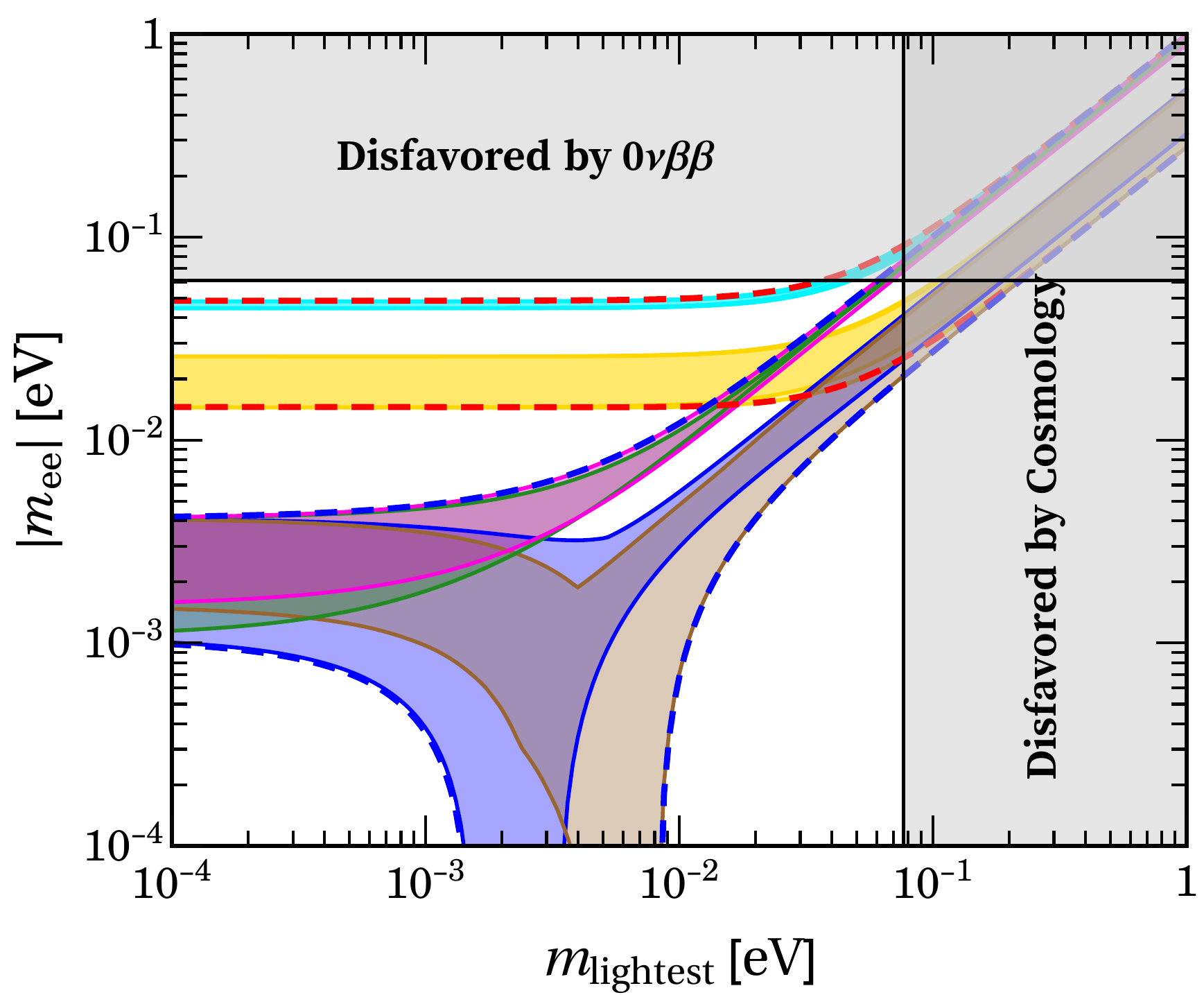}\\
~~~~~~~~~~~~\qquad\quad\includegraphics[width=0.82\textwidth]{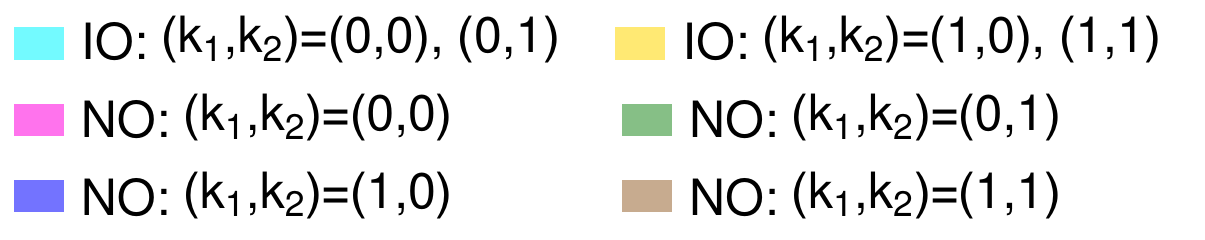}
\end{tabular}
\caption{\label{fig:0nubb_CaseI_456}The allowed regions of the effective Majorana mass $|m_{ee}|$ with respect to the lightest neutrino mass for $U_{I,4}$ (the left panel) and $U_{I,5}$ (the right panel) with $\varphi_1=\pi/3$. The red (blue) dashed lines indicate the most general allowed regions for IO (NO) neutrino mass spectrum obtained by varying the mixing parameters over their 3$\sigma$ ranges~\cite{Esteban:2016qun}. The present most stringent upper bound $|m_{ee}|<0.061$ eV from KamLAND-ZEN~\cite{KamLAND-Zen:2016pfg} and EXO-200~\cite{Albert:2017owj} is shown by horizontal grey band. The vertical grey exclusion band denotes the current sensitivity of cosmological data from the Planck collaboration~\cite{Ade:2015xua}.}
\end{figure}

\begin{table}[t!]
\centering
\footnotesize
\begin{tabular}{|c |c c c c c c|}
\hline \hline
case & $\theta_{13}/^{\circ}$ & $\theta_{12}/^{\circ}$ & $\theta_{23}/^{\circ}$ & $\delta_{CP}/\pi$ & $\alpha_{21}/\pi~(\text{mod}~1)$ & $\alpha_{31}/\pi~(\text{mod}~1)$ \\
\hline
  \multirow{2}{*}{$U_{I,4}\big|_{\varphi_1=\frac{\pi}{3}}$} & \multirow{2}{*}{$8.091-8.979$} & \multirow{2}{*}{$32.989-36.031$} & \multirow{2}{*}{$40.280-45.843$} & $0-0.304 $ & $0-0.138$ & $0-0.085$ \\
     &&&&$ \oplus 1.696-2$&$ \oplus 0.863-1$& $ \oplus 0.915-1$\\
\hline
  \multirow{2}{*}{$U_{I,5}\big|_{\varphi_1=\frac{\pi}{3}}$} & \multirow{2}{*}{$8.091-8.979$} & \multirow{2}{*}{$31.435-36.031$} & \multirow{2}{*}{$45.635-51.531$} & $0-0.464 $ & $0-0.252 $ & $0-0.162 $ \\
       &&&&$ \oplus 1.536-2$&$ \oplus 0.748-1$& $\oplus 0.838-1$\\
\hline
  \multirow{2}{*}{$U_{II,1}\big|_{\varphi_4=\frac{\pi}{3}}$} & \multirow{2}{*}{$8.091-8.979$} & \multirow{2}{*}{$32.992-36.031$} & \multirow{2}{*}{$40.280-45.845$} & $0-0.303$ & $0-0.291$ & $0.150-0.361$ \\
           &&&&$ \oplus 1.697-2$&$ \oplus 0.709-1$& $ \oplus 0.639-0.850$\\
\hline
  \multirow{2}{*}{$U_{II,2}\big|_{\varphi_4=\frac{\pi}{3}}$} & \multirow{2}{*}{$8.091-8.979$} & \multirow{2}{*}{$31.435-36.031$} & \multirow{2}{*}{$45.625-51.136$} & $0-0.444$ & \multirow{2}{*}{$0-1$} & $0-0.380$ \\
             &&&&$ \oplus 1.556-2$&& $ \oplus 0.620-1$\\
\hline
  \multirow{2}{*}{$U_{III,2}\big|_{\varphi_6=0}$} & \multirow{2}{*}{$8.096-8.975$} & \multirow{2}{*}{$31.435-36.031$} & \multirow{2}{*}{$45.580-45.715$} & \multirow{2}{*}{$0-2$} & $0-0.228$ & $0-0.168$ \\
     &&&&&$ \oplus 0.772-1$& $ \oplus 0.832-1$\\
\hline
\multirow{2}{*}{$U_{III,3}\big|_{\varphi_6=0}$ } & \multirow{2}{*}{$8.091-8.979$} & \multirow{2}{*}{$31.435-33.388$} & \multirow{2}{*}{$50.140-51.531$} & $0-0.225$ & $0-0.128$ & $0-0.098$ \\
&&&&$ \oplus 1.775-2$&$ \oplus 0.872-1$& $ \oplus 0.902-1$\\
\hline
\multirow{2}{*}{$U_{IV,1}\big|_{\theta'_{\nu}=0}$} & \multirow{2}{*}{$8.862-8.979$} & \multirow{2}{*}{$35.909-36.031$} & \multirow{2}{*}{$45.696-45.715$} & $0-0.054$ & $0-0.034$ & $0-0.019$\\
&&&&$\oplus 1.946-2$& $\oplus 0.967-1$& $\oplus 0.981-1$\\
\hline \hline
\end{tabular}
\caption{\label{Tab:z2z2cp_lepton_v2}
The allowed ranges of the mixing parameters for the viable mixing patterns arising from the scenario in which the $\Delta(6n^2)$ and CP symmetries are broken down to the remnant symmetries $Z_2$ and $Z_2\times CP$ in the charged lepton and neutrino sectors respectively. Here we require all the three mixing angles are in their experimentally preferred $3\sigma$ intervals~\cite{Esteban:2016qun}.}
\end{table}

\item[~~(\uppercase\expandafter{\romannumeral2})]

$g_l=bc^xd^x$, $g_{\nu}=abc^y$, $X_{\nu}=\left\{c^{\gamma}d^{2y+2\gamma}, abc^{y+\gamma}d^{2y+2\gamma}\right\}$

The Takagi factorization matrix $\Sigma_{\nu}$ is given by
\begin{equation}
\Sigma_{\nu}=\frac{1}{\sqrt{2}}\left(
\begin{array}{ccc}
 -e^{\frac{i \pi  \gamma }{n}} &~ 0 ~& e^{\frac{i \pi  \gamma }{n}} \\
 e^{\frac{i \pi  (2 y+\gamma )}{n}} &~ 0 ~& e^{\frac{i \pi  (2 y+\gamma )}{n}} \\
 0 &~ -\sqrt{2} e^{-\frac{2 i \pi  (y+\gamma )}{n}} ~& 0
\end{array}\right)\,.
\end{equation}
Thus the $\Sigma$ matrix is of the following form
\begin{equation}
\Sigma=\frac{1}{2}\left(
\begin{array}{ccc}
 1 &~ -\sqrt{2} e^{i \varphi_{4}} ~& -1 \\
 \sqrt{2} e^{i \varphi_{3}} &~ 0 ~& \sqrt{2} e^{i \varphi_{3}} \\
 -1 &~ -\sqrt{2} e^{i \varphi_{4}} ~& 1 \\
\end{array}
\right)\,,
\end{equation}
where an overall phase is omitted, and the discrete parameters $\varphi_{3}$ and $\varphi_{4}$ determined by the residual symmetries are of the form \begin{equation}
\varphi_3=\frac{2(y-x)}{n}\pi,\qquad
\varphi_4=-\frac{3\gamma+2(x+y)}{n}\pi\,.
\end{equation}
The lepton mixing matrix is found to be
{\small\begin{equation}
\label{eq:UPMNS_caseII}
U_{II}=\frac{1}{2}\left(
\begin{array}{ccc}
 1 &~ c_{\nu}+\sqrt{2} e^{i \varphi_{4}} s_{\nu} ~& s_{\nu}-\sqrt{2} e^{i \varphi_{4}} c_{\nu} \\
 s_{l}+\sqrt{2} e^{i \delta} c_{l} &~ s_{l} c_{\nu}-\sqrt{2} (e^{i \delta} c_{l} c_{\nu}+e^{i \varphi_{4}} s_{l}s_{\nu}) ~&s_{l} s_{\nu}-\sqrt{2}(e^{i \delta} c_{l} s_{\nu} -e^{i \varphi_{4}} s_{l}c_{\nu} )  \\
 c_{l}-\sqrt{2} e^{i \delta} s_{l} &~ c_{l}c_{\nu}+\sqrt{2} (e^{i \delta} s_{l}c_{\nu} -e^{i \varphi_{4}}c_{l} s_{\nu}) ~& c_{l} s_{\nu}+\sqrt{2} (e^{i \delta} s_{l} s_{\nu}+ e^{i \varphi_{4}} c_{l}c_{\nu}) \\
\end{array}
\right)\,,
\end{equation}}
where $\delta=-2\delta_{l}+\varphi_{3}$.
The parameters $\varphi_{4}$ can take following discrete values
\begin{equation}
\varphi_{4}~(\text{mod}~2\pi)=0,\frac{1}{n}\pi,\frac{2}{n}\pi,...,\frac{2n-1}{n}\pi \,.
\end{equation}
We can check that $U_{II}$ satisfy the following symmetry properties,
\begin{eqnarray}
\nonumber
U_{II}(\varphi_{4}+\pi,\theta_{l},\theta_{\nu},\delta)&=&U_{II}(\varphi_{4},\theta_{l},\pi-\theta_{\nu},\delta)\text{diag}(1,-1,1)\,,\\
\nonumber
U_{II}(\varphi_{4},\theta_{l},\theta_{\nu},\pi+\delta)&=&\text{diag}(1,1,-1)U_{II}(\varphi_{4},\pi-\theta_{l},\theta_{\nu},\delta)\,.
\end{eqnarray}
Therefore the fundamental regions of $\varphi_4$ and $\delta$ are $0\leq\varphi_4<\pi$ and $0\leq\delta<\pi$. From Eq.~\eqref{eq:UPMNS_caseII} we know that the fixed elements of PMNS matrix is $\frac{1}{2}$ in this case. In order to be compatible with
experimental data, the element $\frac{1}{2}$ should be $(21),(22),(31)$ or $(32)$ entries of the mixing matrix. Consequently we have four phenomenologically viable forms of the lepton mixing matrix in this case,
\begin{eqnarray}
\nonumber&&U_{II,1}=P_{12}U_{II}\,,\quad U_{II,2}=P_{12}U_{II}P_{12}\,,\\
&&U_{II,3}=P_{23}P_{12}U_{II}\,,\quad U_{II,4}=P_{23}P_{12}U_{II}P_{12}\,.
\end{eqnarray}
Since $U_{II,3}$ and $U_{II,4}$ are related to $U_{II,1}$ and $U_{II,2}$ through the exchange of the second and third rows of the PMNS matrix. Thus, they lead to the same reactor and solar mixing angles, while the atmospheric one changes from $\theta_{23}$ to $\pi/2-\theta_{23}$ and the Dirac changes from $\delta_{CP}$ to $\pi+\delta_{CP}$. Hence it is sufficient to only consider the mixing patterns $U_{II,1}$ and $U_{II,2}$. For the group $\Delta(6\cdot2^2)\cong S_{4}$, the parameter $\varphi_{4}$ can be $0$ or $\pi/2$, and the mixing pattern $U_{II}$ for
$\varphi_{4}=0$ and $\varphi_{4}=\pi/2$ correspond to the cases of \textbf{Group C} and \textbf{Group D} of Ref.~\cite{Penedo:2017vtf} exactly.
Then we turn to the next flavor group $\Delta(6\cdot3^2)=\Delta(54)$, the discrete parameter $\varphi_{4}$ can take the values of $0$, $\pi/3$ and $2\pi/3$. By applying the equivalence criterion derived in subsection~\ref{subsec:criterion_Z2xZ2xCP_lepton}, we find that $U_{II}$ for $\varphi_{4}=0$ gives the same mixing pattern as $U_{I}$ with $\varphi_{1}=\pi/3$, and the mixing matrix $U_{II}$ for $\varphi_{4}=\pi/3$ and $\varphi_{4}=2\pi/3$ are equivalent. For the mixing matrices $U_{II,1}$ and $U_{II,2}$ with $\varphi_{4}=\pi/3$, the predicted ranges of the mixing angles and CP phases are summarized in table~\ref{Tab:z2z2cp_lepton_v2}. We notice that $U_{II,1}$ and $U_{I,4}$  give similar allowed intervals of the mixing angles $\theta_{ij}$ and $\delta_{CP}$, this is because the fixed element $1/2$ is the (21) entry in both cases. The allowed regions of $\theta_{ij}$ and $\delta_{CP}$ are also similar for $U_{II,2}$ and $U_{I,5}$, and the fixed element $1/2$ is the (22) entry.
The correlations between different mixing parameters are plotted in figure~\ref{fig:U_II_1_correlations} and figure~\ref{fig:U_II_2_correlations}. As previous cases, we see that $\theta_{23}$ is correlated with CP violation phases, and there are also
strong correlations among the CP phases. We plot the allowed regions of $|m_{ee}|$ in figure~\ref{fig:0nubb_CaseII_12}.

\begin{figure}[t!]
\centering
\begin{tabular}{ >{\centering\arraybackslash} m{8.0cm} >{\centering\arraybackslash} m{8.0cm} }
\includegraphics[width=0.48\textwidth]{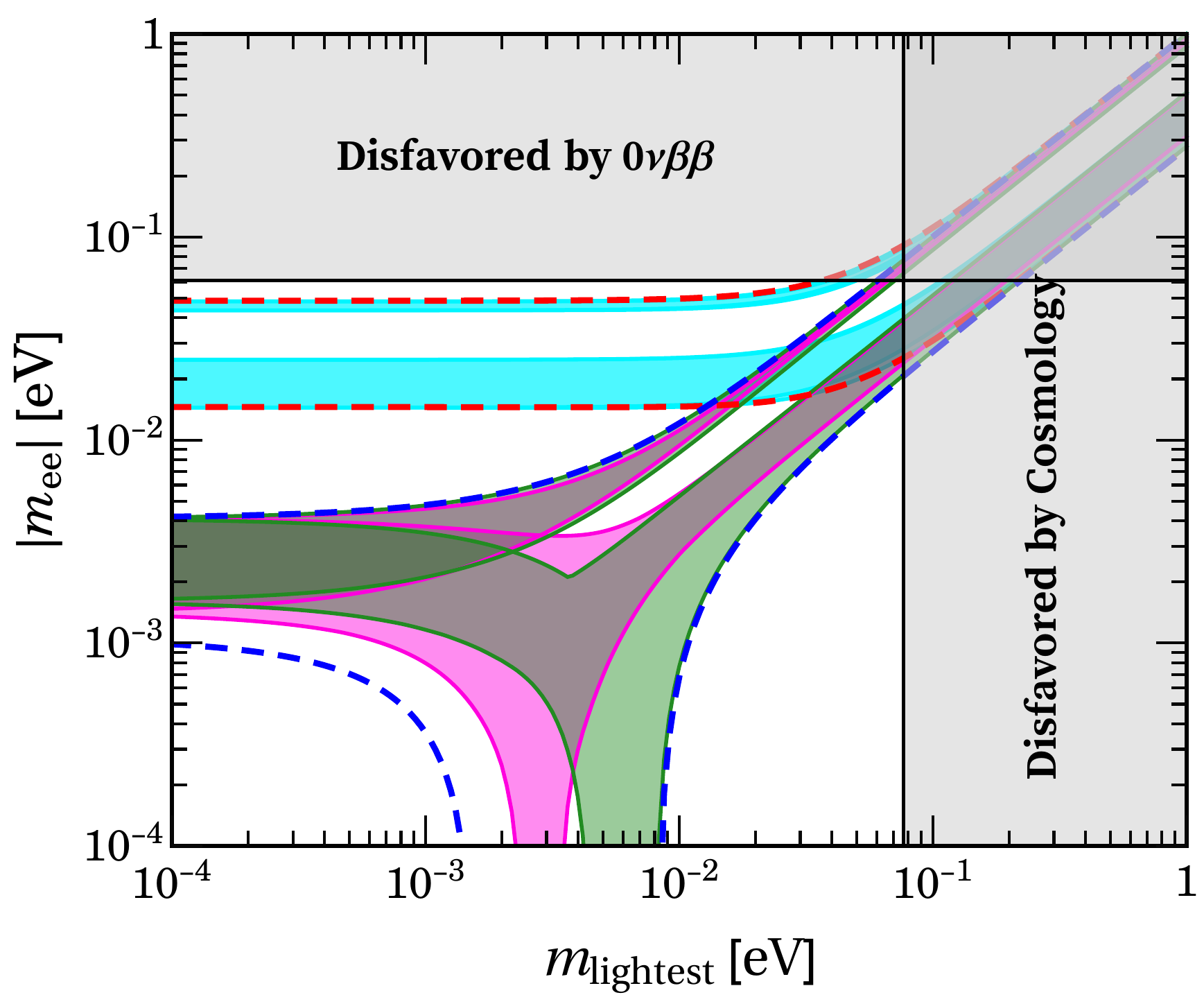}~~&
\includegraphics[width=0.48\textwidth]{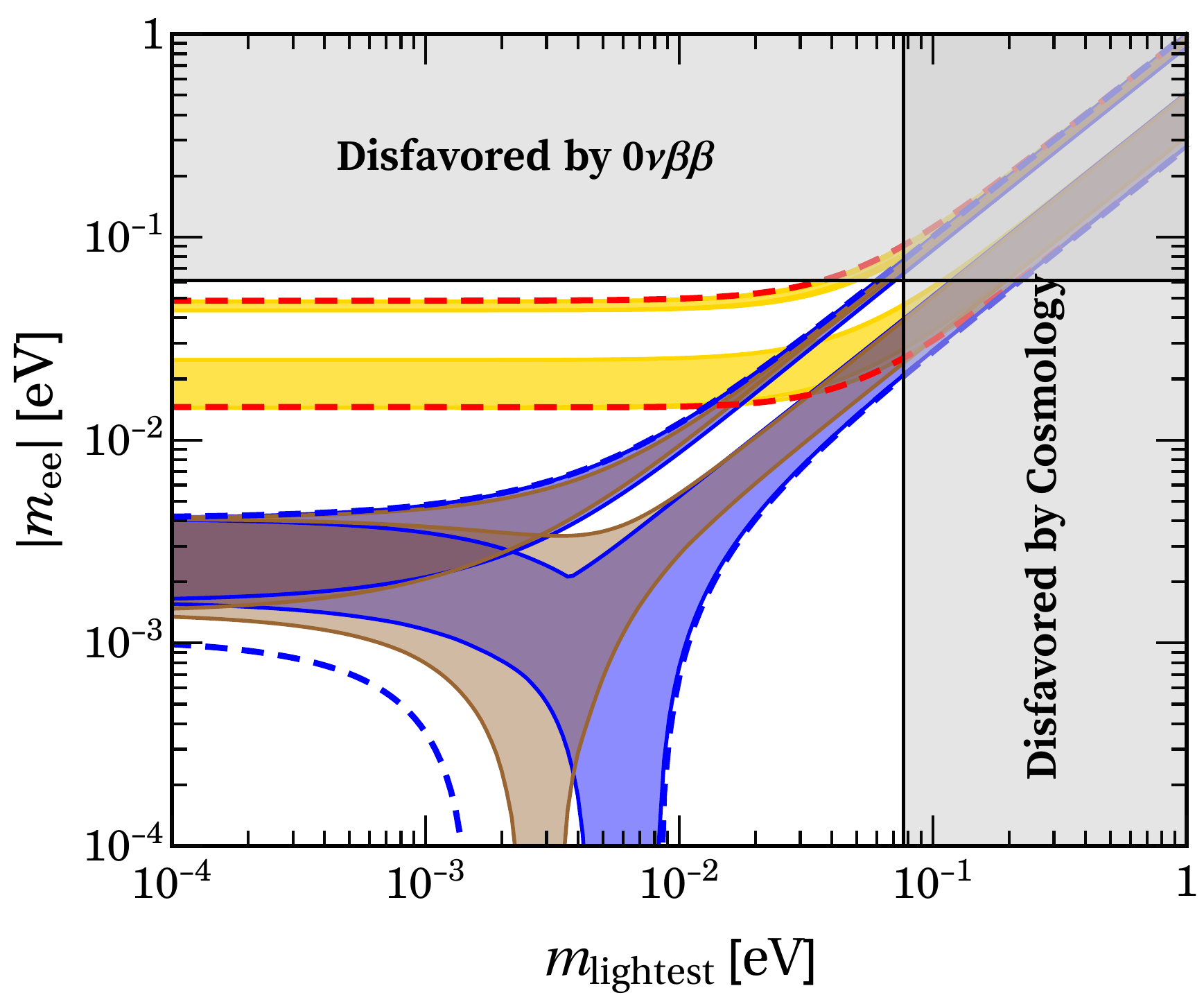}\\
\includegraphics[width=0.48\textwidth]{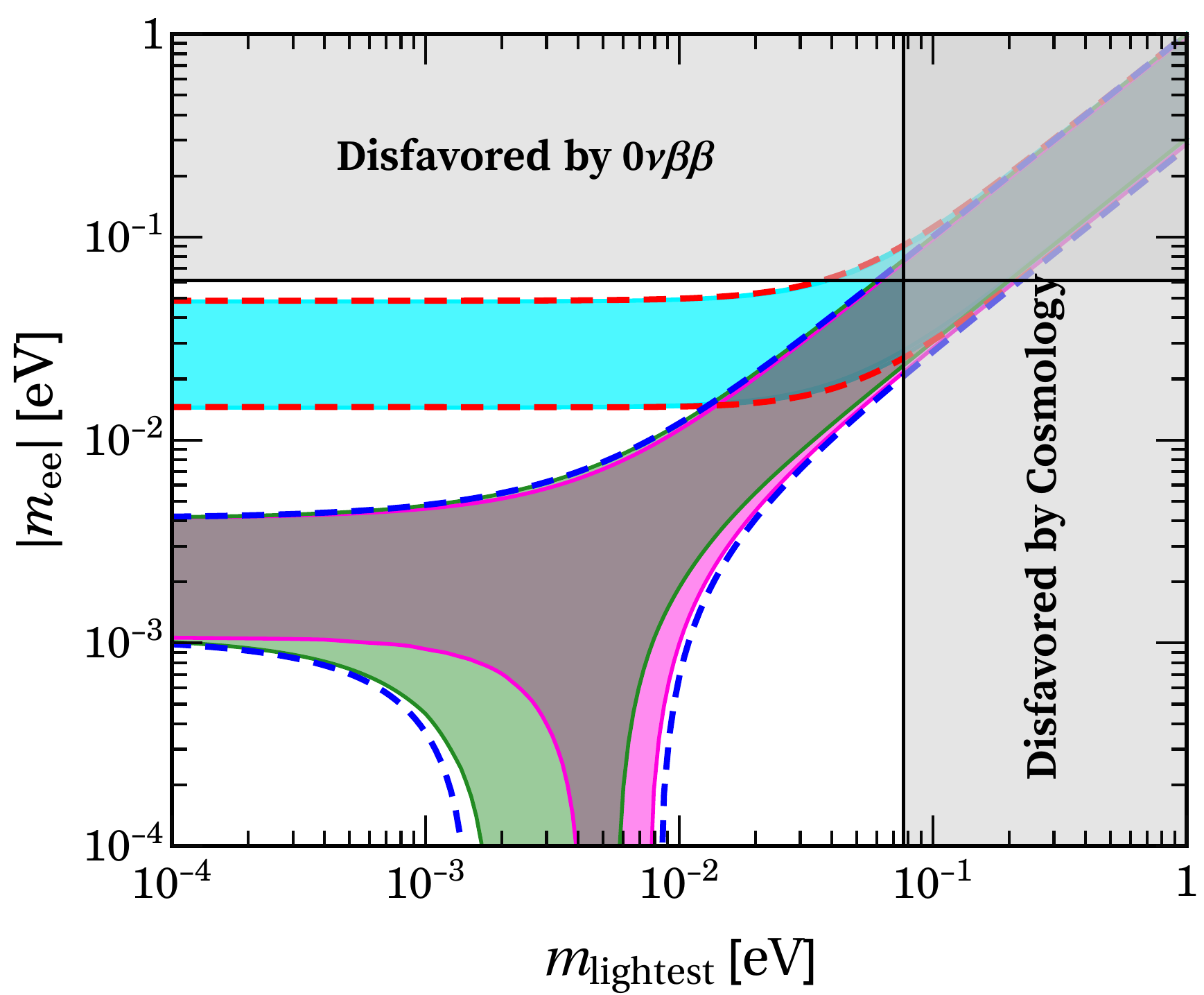}~~&
\includegraphics[width=0.48\textwidth]{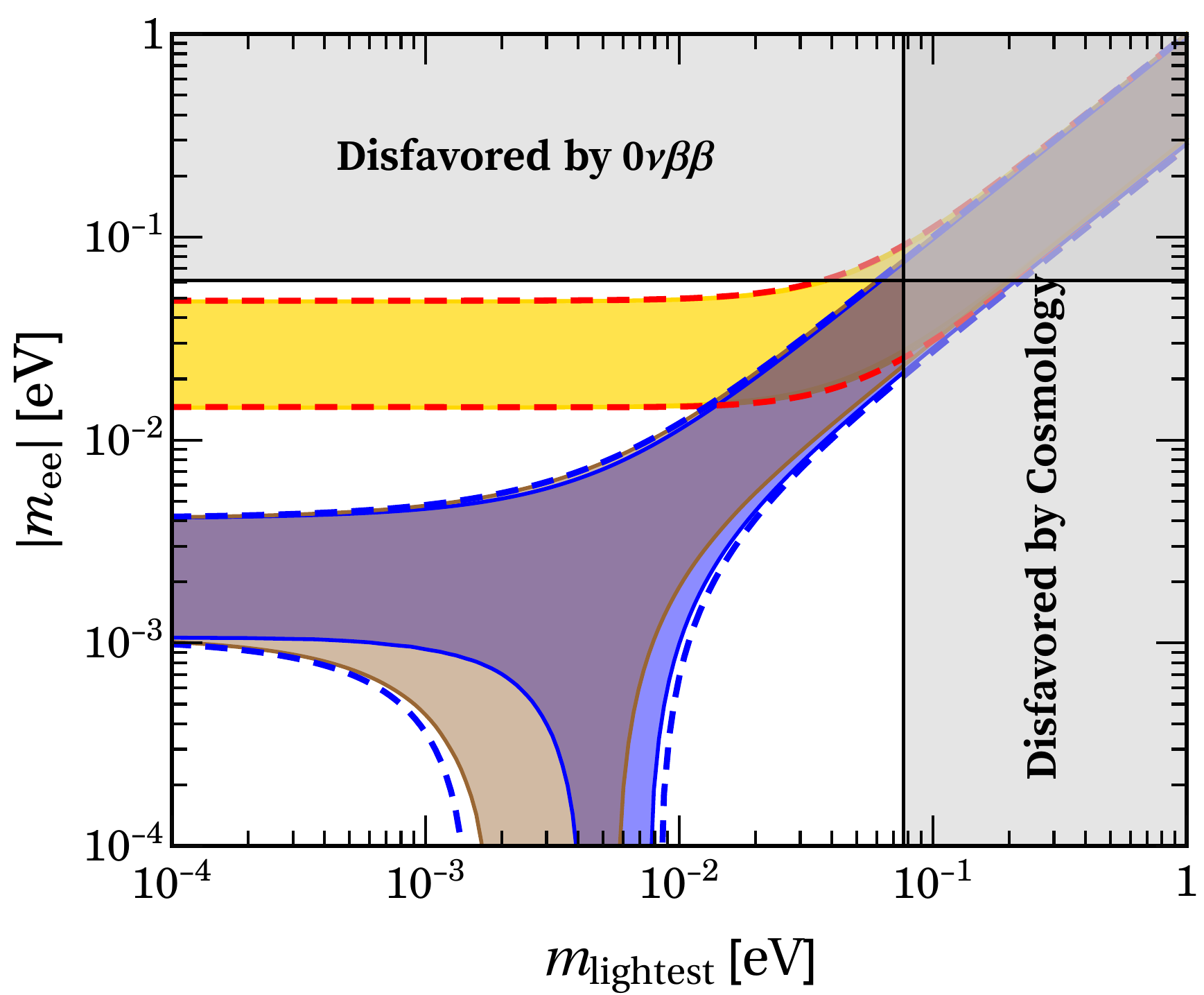}
\end{tabular}
\caption{\label{fig:0nubb_CaseII_12}The allowed regions of the effective Majorana mass $|m_{ee}|$ with respect to the lightest neutrino mass for the mixing pattern $U_{II}$ with $\varphi_{4}=\pi/3$. We adopt the same conventions as figure~\ref{fig:0nubb_CaseI_456}. The top row and bottom rows are for $U_{II,1}$ and $U_{II,2}$ respectively. The left panels correspond to $(k_{1},k_{2})=(0,0),~(0,1)$, and the right panels correspond to $(k_{1},k_{2})=(1,0),~(1,1)$.}
\end{figure}

\item[~~(\uppercase\expandafter{\romannumeral3})]
$g_l=bc^xd^x$, $g_\nu=c^{n/2}$, $X_{\nu}=\left\{c^{\gamma}d^{\rho}\right\}$

The group index $n$ should be even in order for the group to have a $Z_2$ generating element $c^{n/2}$. In the same fashion as previous cases, the matrix $\Sigma$ reads as
\begin{equation}\label{eq:sigma_III}
\Sigma=\frac{1}{\sqrt{2}}\left(
\begin{array}{ccc}
 1 & 0 & -e^{-i \varphi_{6}} \\
 0 & -\sqrt{2} e^{i \varphi_{5}} & 0 \\
 1 & 0 & e^{-i \varphi_{6}} \\
\end{array}\right)\,,
\end{equation}
with
\begin{equation}\label{eq:varphi_III}
\varphi _{5}=\frac{2 \rho - \gamma}{n}\pi, \qquad  \varphi _{6}=-\frac{2 x+\gamma +\rho }{n}\pi\,.
\end{equation}
As a consequence, the lepton mixing matrix takes the following form
\begin{equation}\label{eq:PMNS_III}
U_{III}=\frac{1}{\sqrt{2}}\left(
\begin{array}{ccc}
 -e^{i \varphi_{6}} &~ s_{\nu} ~& c_{\nu} \\
 e^{i \varphi_{6}} s_{l} &~ s_{l} s_{\nu}- \sqrt{2} e^{i \delta} c_{l} c_{\nu} ~& s_{l}c_{\nu} +\sqrt{2} e^{i \delta} c_{l} s_{\nu} \\
 e^{i \varphi_{6}} c_{l} &~ c_{l} s_{\nu}+\sqrt{2} e^{i \delta} s_{l}c_{\nu}  ~& c_{l} c_{\nu}-\sqrt{2} e^{i \delta} s_{l} s_{\nu}\\
\end{array}
\right)\,,
\end{equation}
where the parameter $\delta=2\delta_{l}+\varphi_{5}+\varphi_{6}$.
We see that the value of $\varphi_{5}$ is irrelevant, and the contribution of $\varphi_6$ is only to shift the Majorana phases. The parameter $\varphi_{6}$ can take following discrete values
\begin{equation}
\varphi_{6}~(\text{mod}~2\pi)=0,\frac{1}{n}\pi,\frac{2}{n}\pi,...,\frac{2n-1}{n}\pi \,.
\end{equation}
We can check that $U_{III}$ has the following symmetry properties:
\begin{eqnarray}
\nonumber && U_{III}(\varphi_{6},\theta_{l},\theta_{\nu},\pi+\delta)=U_{III}(\varphi_{6},\theta_{l},\pi-\theta_{\nu},\delta)\text{diag}(1,1,-1)\,,\\
\nonumber&&U_{III}(\varphi_{6}+\pi,\theta_{l},\theta_{\nu},\delta)=U_{III}(\varphi_{6},\theta_{l},\theta_{\nu},\delta)\text{diag}(-1,1,1)\,, \\
\nonumber && U_{III}(\varphi_{6}+\frac{\pi}{2},\theta_{l},\theta_{\nu},\delta)=U_{III}(\varphi_{6},\theta_{l},\theta_{\nu},\delta)\text{diag}(i,1,1)\,,\\
&&U_{III}(\pi-\varphi_{6},\theta_{l},\theta_{\nu},\pi-\delta)=U^{*}_{III}(\varphi_{6},\theta_{l},\pi-\theta_{\nu},\delta)
\text{diag}(-1,1,-1)\,.
\end{eqnarray}
Therefore the parameters $\varphi_6$ and $\delta$ can be limited in the ranges of $0\leq\varphi_6<\pi/2$ and $0\leq\delta<\pi$. One element of the mixing matrix is $-e^{i \varphi_{6}}/\sqrt{2}$ whose module is $1/\sqrt{2}$, it should be the  $(22),(23),(32)$ or $(33)$ entries of the PMNS matrix be compatible with experimental data. Hence the permutations of rows and columns give rise to four viable mixing patterns:
\begin{eqnarray}
\nonumber&&U_{III,1}=P_{12}U_{III}P_{12}\,,\quad U_{III,2}=P_{12}U_{III}P_{13}\,,\\
&&U_{III,3}=P_{23}P_{12}U_{III}P_{12}\,,\quad U_{III,4}=P_{23}P_{12}U_{III}P_{13}\,,
\end{eqnarray}
We see that $U_{III,3}$ and $U_{III,4}$ are related to $U_{III,1}$ and $U_{III,2}$ through the permutation of the second and the third rows respectively. Moreover, the predictions for a generic nonzero $\varphi_6$ can be read from those of $\varphi_6=0$, since they lead to the same mixing angles and Dirac CP phase $\delta_{CP}$ while the Majorana phase $\alpha_{21}$ $(\alpha_{31})$ differs by $2\varphi_6$ for the mixing patterns $U_{III,1}$ and $U_{III,3}$ ($U_{III,2}$ and $U_{III,4}$). Hence it is sufficient to focus on the mixing patterns $U_{III,1}$ and $U_{III,2}$ with $\varphi_{6}=0$. For the $S_4$ group with the index $n=2$, the value of $\varphi_6$ can only be $0$ in the fundamental interval $0\leq\varphi_6<\pi/2$. We find the concerned residual symmetry is exactly the \textbf{Group B} case of Ref.~\cite{Penedo:2017vtf}.

Then we perform a numerical analysis. Regarding the mixing matrix $U_{III, 2}$ with $\varphi_{6}=0$, both $\theta_{13}$ and $\theta_{12}$ can approximately take any values within their $3\sigma$ regions. By contrast, the allowed range of $\theta_{23}$ is quite narrow and it is close to $45^{\circ}$, namely, $\theta_{23}\in[45.580^{\circ}, 45.715^{\circ}]$. The reason is that the absolute value of the $(23)$ element is $1/\sqrt{2}$ and consequently the sum rule $\sin^2\theta_{23}\cos^2\theta_{13}=1/2$ is fulfilled in this case. No prediction for the Dirac CP phase $\delta_{CP}$ can be extracted and it can be any value between $0$ and $2\pi$ while the Majorana phases are constrained to be around $0/\pi$ with $\alpha_{21} (\text{mod}~\pi)\in[0,0.228\pi]\cup[0.772\pi, \pi]$ and $\alpha_{31}(\text{mod}~\pi)\in[0,0.168\pi]\cup[0.832\pi, \pi]$. The three CP violation phases are correlated as shown in figure~\ref{fig:U_III_2_correlations}.

For the mixing matrix $U_{III,3}$ with $\varphi_{6}=0$,
nearly the $3\sigma$ region of the reactor mixing angle $\theta_{13}$ can be reproduced, the solar angle is predicted to lie in the narrow range $\theta_{12}\in[31.435^{\circ}, 33.388^{\circ}]$, and the atmospheric mixing angle $\theta_{23}$ belongs to the second octant $50.140^{\circ}\leq\theta_{23}\leq51.531^{\circ}$. Furthermore, this case has clear and interesting prediction for the Dirac phase $\delta_{CP}\in[0, 0.225\pi]\cup[1.775\pi, 2\pi)$.
As regards the Majorana CP phases,
they are found close to the CP conserved limit, i.e., $\alpha_{21}(\text{mod}~\pi)\in[0, 0.128\pi]\cup[0.872\pi, \pi]
$ and $\alpha_{31}(\text{mod}~\pi)\in[0, 0.098\pi]\cup[0.902\pi, \pi]$. The correlations among the mixing parameters are displayed in figure~\ref{fig:U_III_1_correlations}, it is obvious that mixing parameters are strongly correlated with each other.
The predictions for efective Majorana mass $|m_{ee}|$ characterizing the neutrinoless double beta decay are shown in figure~\ref{fig:0nubb_CaseIII_1}. We see that effective mass $|m_{ee}|$ is predicted to be around the borders of the generic case for IO. For NO mass spectrum, $|m_{ee}|$ can be strongly suppressed to be smaller than $10^{-4}$ eV because of strong cancellations in certain intervals of $m_{\mathrm{lightest}}$.

\begin{figure}[hptb!]
\centering
\begin{tabular}{ >{\centering\arraybackslash} m{8.0cm} >{\centering\arraybackslash} m{8.0cm} }
  \includegraphics[width=0.48\textwidth]{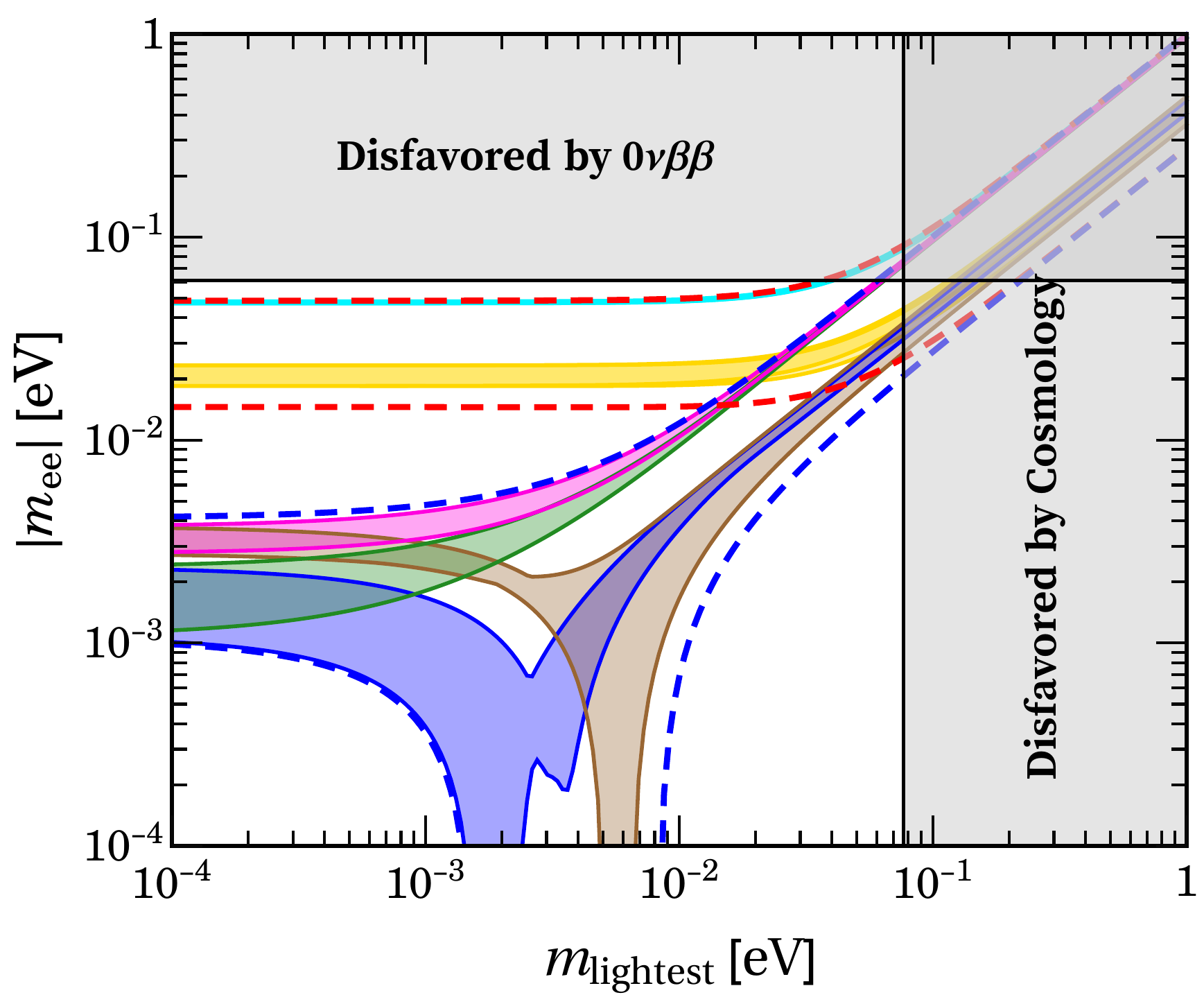}~~&
  \includegraphics[width=0.48\textwidth]{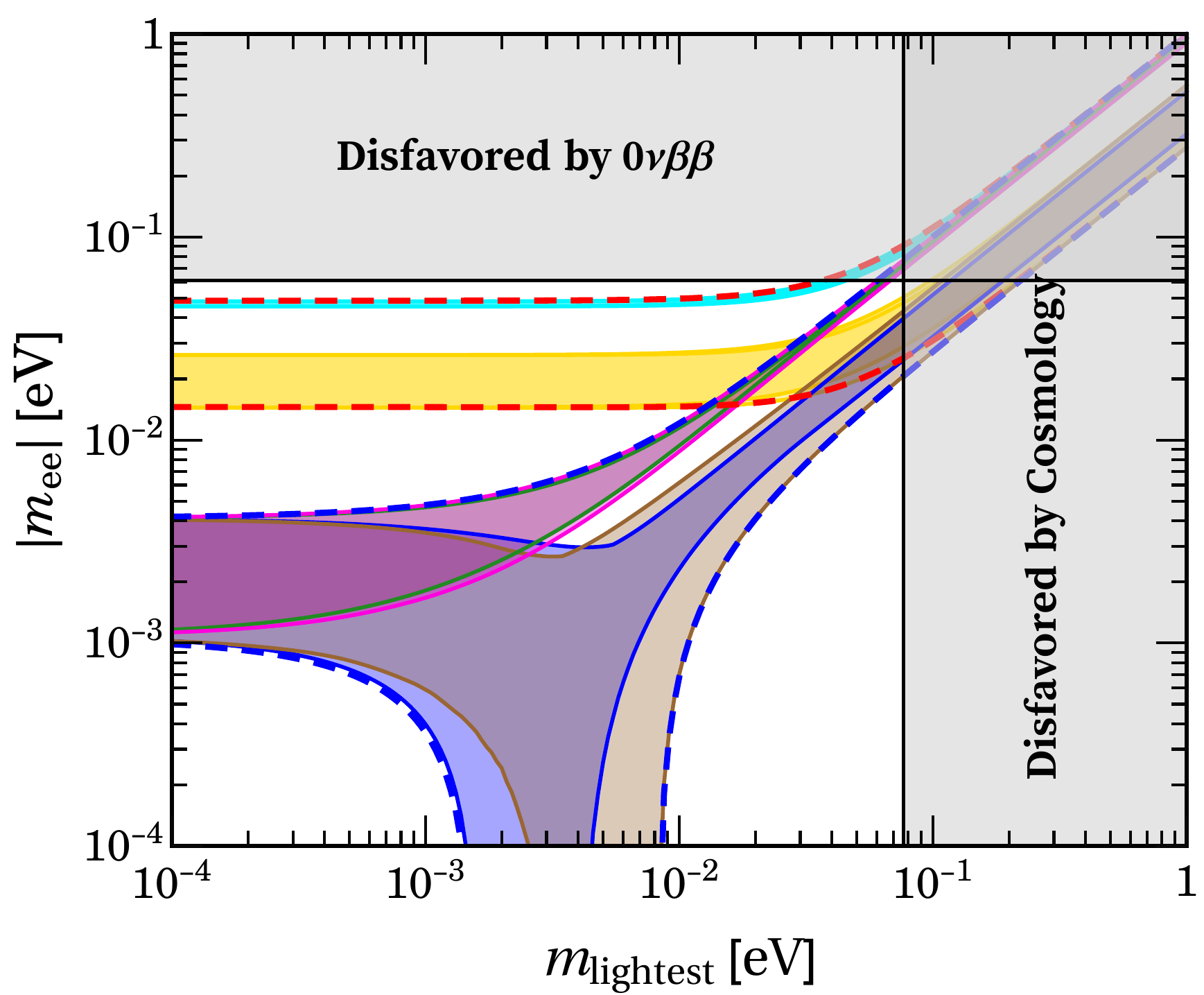}
\end{tabular}
\caption{\label{fig:0nubb_CaseIII_1}The allowed regions of the effective Majorana mass $|m_{ee}|$ with respect to the lightest neutrino mass for the mixing pattern $U_{III}$ with $\varphi_{6}=0$. We adopt the same conventions as figure~\ref{fig:0nubb_CaseI_456}. The left panel is for $U_{III,1}$ and the right is for $U_{III,2}$.}
\end{figure}

\item[~~(\uppercase\expandafter{\romannumeral4})]

$g_l=bc^xd^x$, $g_{\nu}=c^{n/2}$, $X_{\nu}=\left\{abc^{\gamma}d^{2\gamma}\right\}$

In this case, the residual flavor symmetry $Z^{g_{\nu}}_2=Z^{c^{n/2}}_2$ requires that the group index $n$ has to be even.
The $\Sigma$ matrix is of the following form
\begin{equation}\label{eq:sigma_V}
\Sigma=\frac{1}{2}\left(
\begin{array}{ccc}
 \sqrt{2}e^{-i \varphi_{7}} &~ -i e^{i \varphi_{8}} ~& -e^{i \varphi_{8}} \\
 0 &~ -i \sqrt{2} ~& \sqrt{2} \\
 \sqrt{2}e^{-i \varphi_{7}} &~ i e^{i \varphi_{8}} ~& e^{i \varphi_{8}} \\
\end{array}\right)\,,
\end{equation}
up to an overall nonphysical phase, and the discrete parameters $\varphi_{7}$ and $\varphi_{8}$ are given by
\begin{equation}\label{eq:varphi_V}
\varphi_{7}=\frac{3\gamma}{n}\pi, \qquad \varphi_{8}=\frac{2x}{n}\pi\,.
\end{equation}
Then we can read off the lepton mixing matrix
\begin{equation}\label{eq:PMNS_IV}
U_{IV}=\frac{1}{2}\left(
\begin{array}{ccc}
 1 &~ 1 ~& -\sqrt{2} e^{i \theta^{\prime}_{\nu}} \\
 s_{l}+\sqrt{2} e^{i \delta} c_{l} &~ s_{l}-\sqrt{2} e^{i \delta} c_{l} ~& \sqrt{2} e^{i \theta^{\prime}_{\nu}} s_{l} \\
 c_{l}-\sqrt{2} e^{i \delta} s_{l} &~ c_{l}+\sqrt{2} e^{i \delta} s_{l} ~& \sqrt{2} e^{i \theta^{\prime}_{\nu}} c_{l} \\
\end{array}\right)\,,
\end{equation}
where the parameters $\delta$ and $\theta^{\prime}_{\nu}$ are defined as
\begin{equation}
\delta =2\delta_{l}-2\theta_{\nu}-\varphi_{8}, \quad \theta^{\prime}_{\nu} = -\theta_{\nu}-\varphi_{7}-\varphi_{8}\,.
\end{equation}
Thus the discrete parameters $\varphi_{7}$ and $\varphi_{8}$ can be absorbed into the continuous parameters. We see that the mixing matrix $U_{IV}$ is independent of group index $n$, consequently this mixing pattern will appear in the discussion of any $\Delta(6n^{2})$ group. For the flavor group $S_4$ with $n=2$, $U_{IV}$ coincides with the mixing matrix of \textbf{Group A} in~\cite{Penedo:2017vtf}.
Furthermore, we see that the following identities are fulfilled
\begin{eqnarray}
\nonumber&&U_{IV}(\theta_{l}+\pi,\theta^{\prime}_{\nu},\delta)=\text{diag}(1,-1,-1)U_{IV}
(\theta_{l},\theta^{\prime}_{\nu},\delta)\,,\\
\nonumber&&U_{IV}(\theta_{l},\theta^{\prime}_{\nu}+\frac{\pi}{2},\delta)=U_{IV}(\theta_{l}
,\theta^{\prime}_{\nu},\delta)\text{diag}(1,1,i)\,, \\
\label{eq:PMNS_symmetry2_V}&&U_{IV}(\theta_{l},\theta^{\prime}_{\nu}, \pi+\delta)=\text{diag}(1,1,-1)U_{IV}
(\pi-\theta_{l},\theta^{\prime}_{\nu},\delta)\,.
\end{eqnarray}
Notice that the contribution of the free parameter $\theta'_{\nu}$ is to shift the Majorana phase $\alpha_{31}$ by $2\theta'_{\nu}$.
From Eq.~\eqref{eq:PMNS_IV} we see that the magnitude of one row of $U_{IV}$ is $(1/2, 1/2, 1/\sqrt{2})$, and consequently only two mixing patterns compatible with data can be obtained,
\begin{equation}
U_{IV,1}=P_{12}U_{IV}\,,\quad U_{IV,2}=P_{23}P_{12}U_{V}\,.
\end{equation}
We can extract the following results for the neutrino mixing angles
\begin{eqnarray}
\nonumber&&\sin^2\theta_{13}=\frac{1}{2}\sin^2\theta_{l},~~\quad \sin^2\theta_{12}=\frac{1}{2}-\frac{\sqrt{2}\sin2\theta_{l}\cos\delta}{3+\cos2\theta_{l}}\,,\\
&&\sin^2\theta_{23}=\frac{2}{3+\cos2\theta_{l}} ~~\text{for}~~U_{IV,1},\qquad \sin^2\theta_{23}=\frac{1+\cos2\theta_{l}}{3+\cos2\theta_{l}} ~~\text{for}~~U_{IV,2}\,.
\end{eqnarray}
The atmospheric and reactor mixing angles are related as
\begin{eqnarray}
\nonumber&&\cos^2\theta_{13}\sin^2\theta_{23}=\frac{1}{2}~~~\text{for}~~U_{IV,1},\\
&&\cos^2\theta_{13}\cos^2\theta_{23}=\frac{1}{2}~~~\text{for}~~U_{IV,2}\,.
\end{eqnarray}
Furthermore, we see that $\theta_{12}$ and $\theta_{13}$ fulfill the following inequality
\begin{equation}
\frac{1}{2}-\tan\theta_{13}\sqrt{1-\tan^2\theta_{13}} \leq\sin^2\theta_{12}\leq\frac{1}{2}+\tan\theta_{13}\sqrt{1-\tan^2\theta_{13}}\,. \end{equation}
Considering the $3\sigma$ allowed region $0.01981\leq\sin^2\theta_{13}\leq0.02436$ from the latest global data fit~\cite{Esteban:2016qun}, we find
\begin{eqnarray}
0.344\leq\sin^2\theta_{12}\leq0.656,\qquad \left\{
\begin{array}{cc}
0.510\leq\sin^2\theta_{23}\leq0.512   ~~&  \text{for}~U_{IV,1}\,, \\[4pt]
0.488\leq\sin^2\theta_{23}\leq0.490   ~~& \text{for}~U_{IV,2}\,.
\end{array}
\right.
\end{eqnarray}
We see that both $\theta_{12}$ and $\theta_{23}$ lie in the experimentally preferred $3\sigma$ range although $\theta_{12}$ is around its $3\sigma$ upper bound 0.346 given in~\cite{Esteban:2016qun}. We report the variation regions of all the mixing angles and CP phases in table~\ref{Tab:z2z2cp_lepton_v2} for the case of $\theta'_{\nu}=0$.

The forthcoming reactor neutrino oscillation experiments, such as JUNO~\cite{An:2015jdp} and RENO-50~\cite{Kim:2014rfa}, expect to make very precise measurements of the solar neutrino mixing angle $\theta_{12}$. They will be capable of reducing the error of $\theta_{12}$ to about $0.1^{\circ}$ or around $0.3\%$. Future long baseline experiments DUNE~\cite{Acciarri:2016crz,Acciarri:2015uup,Strait:2016mof,Acciarri:2016ooe}, T2HK~\cite{Kearns:2013lea,Abe:2014oxa}, T2HKK~\cite{Abe:2016ero} can make very precise measurements of the oscillation parameters $\theta_{12}$, $\theta_{23}$ and $\delta_{CP}$. Therefore future neutrino facilities have the potential to test the above predictions for mixing angles and $\delta_{CP}$, or to rule them out entirely. Furthermore, we expect that a more ambitious facility such as the neutrino factory~\cite{Geer:1997iz,DeRujula:1998umv,Bandyopadhyay:2007kx} could provide a more stringent tests of our approach. A quantitative discussion of whether and how the upcoming neutrino experiments can exclude certain mixing patterns predicted above deserves a dedicate full work of its own, it is beyond the scope of present work.

\end{description}

\section{\label{sec:quark_flavor_mixing_Z2Z2CP}Quark flavor mixing from flavor and CP symmetries breaking to residual symmetries $Z_{2}$ and $Z_{2}\times CP$}

In this section, we extend the approach of predicting lepton flavor mixing in section~\ref{sec:Lepton_flavor_mixing_Z2Z2CP} to the quark sector. Analogously we assume that the three generations of left-handed quark doublets transform as a faithful irreducible triplet $\mathbf{3}$ of the flavor symmetry group $G_{f}$. The residual symmetries of the up and down quark sectors are denoted as $G_{u}$ and $G_{d}$ respectively. Firstly we consider the scenario that the flavor and CP symmetries are broken down to $G_{u}=Z_{2}^{g_{u}}$ and $G_{d}=Z_{2}^{g_{d}}\times X_{d}$ with $g_{u}^{2}=g_{d}^{2}=1$. As a consequence, the up quark mass matrix $m_{U}$ and the down quark mass matrix $m_{D}$ are invariant under the action of $G_u$ and $G_d$, i.e.
\begin{eqnarray}
\nonumber&~~~~~\rho_{\mathbf{3}}^{\dag}(g_{u})m_{U}^{\dagger}m_{U}\rho_{\mathbf{3}}(g_{u})=m_{U}^{\dagger}m_{U},\\
&\rho_{\mathbf{3}}^{\dag}(g_{d})m_{D}^{\dagger}m_{D}\rho_{\mathbf{3}}(g_{d})=m_{D}^{\dagger}m_{D},~~~~~
X_{d}^{\dagger}m_{D}^{\dagger}m_{D}X_{d}=(m_{D}^{\dagger}m_{D})^{*}\,.
\end{eqnarray}
In the same manner as section~\ref{sec:Lepton_flavor_mixing_Z2Z2CP}, we find that the residual symmetry constrains the diagonalization matrices $U_{u}$ and $U_{d}$ of $m_{U}^{\dagger}m_{U}$ and $m_{D}^{\dagger}m_{D}$ are of the following form
\begin{equation}
U_{u}=\Sigma_{u}U_{23}^{\dagger}(\theta_{u},\delta_{u})P_{u}^{T}Q_{u}^{\dagger},~~
U_{d}=\Sigma_{d}S_{23}(\theta_{d})P_{d}Q_{d}\,,
\end{equation}
where $\Sigma_{u}$ and $\Sigma_{d}$ are unitary and they satisfy
\begin{eqnarray}
\nonumber&\Sigma_{u}^{\dagger}\rho_{\mathbf{3}}(g_{u})\Sigma_{u}=\pm\text{diag}(1,-1,-1)\,,\\
&\Sigma_{d}^{\dagger}\rho_{\mathbf{3}}(g_{d})\Sigma_{d}=\pm\text{diag}(1,-1,-1),~~~X_{d}=\Sigma_{d}\Sigma_{d}^{T}\,.
\end{eqnarray}
Therefore the quark mixing CKM matrix is determined to be
\begin{equation}
\label{eq:CKM_form_1}
V_{CKM}=U_{u}^{\dagger}U_{d}=Q_{u}P_{u}U_{23}(\theta_{u},\delta_{u})\Sigma_{u}^{\dagger}\Sigma_{d}S_{23}(\theta_{d})P_{d}Q_{d}\,,
\end{equation}
where $P_{u}$ and $P_{d}$ are permutation matrices,
$Q_{u}$ and $Q_{d}$ are arbitrary diagonal phases matrices and they can be absorbed by redefining quark fields. The range of variation of the free parameters $\theta_{u}$, $\delta_{u}$ and $\theta_{d}$ can be taken to be $[0,\pi/2]$, $[0,\pi)$ and $[0,\pi)$, respectively.

For an alternative scenario in which the quark sector residual symmetries are $G_{u}=Z_{2}^{g_{u}}\times X_{u}$ and $G_{d}=Z_{2}^{g_{d}}$, the unitary transformations $U_{u}$ and $U_{d}$ would be
\begin{equation}
U_{u}=\Sigma_{u}S_{23}(\theta_{u})P_{u}Q_{u},~~U_{d}=\Sigma_{d}U_{23}^{\dagger}(\theta_{d},\delta_{d})P_{d}^{T}Q_{d}^{\dagger}\,,
\end{equation}
where
\begin{equation}
\Sigma_{u}^{\dagger}\rho_{\mathbf{3}}(g_{u})\Sigma_{u}=\pm\text{diag}(1,-1,-1)\,,~~~
\Sigma_{u}\Sigma_{u}^{T}=X_{u}\,,~~~
\Sigma_{d}^{\dagger}\rho_{\mathbf{3}}(g_{d})\Sigma_{d}=\pm\text{diag}(1,-1,-1)\,.
\end{equation}
Consequently the CKM mixing matrix is given as
\begin{equation}
\label{eq:CKM_form_2}
V_{CKM}=Q_{u}^{\dagger}P_{u}^{T}S_{23}^{T}(\theta_{u})\Sigma_{u}^{\dagger}\Sigma_{d}U_{23}^{\dagger}(\theta_{d},\delta_{d})P_{d}^{T}Q_{d}^{\dagger}\,,
\end{equation}
with $\theta_{d}\in[0,\pi/2]$ and $\theta_{u},\delta_{d}\in[0,\pi)$. In both scenarios, one element of the CKM matrix is fixed to be certain constant which is the (11) entry of $\Sigma_{u}^{\dagger}\Sigma_{d}$. We note that for a certain pair of residual symmetries $Z_{2}$ and $Z_{2}\times CP$, the CKM matrix of $G_{u}=Z_{2}$ and $G_{d}=Z_{2}\times CP$ is related to the CKM matrix of $G_{u}=Z_{2}\times CP$ and $G_{d}=Z_{2}$ by a hermitian conjugate up to the redefinition of continuous parameters.

\subsection{\label{subsec:criterion_Z2xZ2xCP_quark}The criterion for the equivalence of two quark mixing patterns}

Similar to the lepton flavor mixing studied in section~\ref{sec:Lepton_flavor_mixing_Z2Z2CP}, two different residual symmetries in quark sector may lead to the same mixing patterns. Following similar procedure presented in section~\ref{subsec:criterion_Z2xZ2xCP_lepton}, we can
find the sufficient and necessary condition under which the CKM matrices predicted by two residual symmetries are equivalent. For two residual symmetries $\{Z^{g_u}_2, Z^{g_d}_2\times X_d\}$ and $\{Z^{g'_u}_2, Z^{g'_d}_2\times X'_d\}$, the resulting CKM mixing matrices are denoted as $V_{CKM}$ and $V'_{CKM}$ respectively with
\begin{eqnarray}
\nonumber&&V_{CKM}=Q_{u}P_{u}U_{23}(\theta_{u},\delta_{u})\Sigma_{u}^{\dagger}\Sigma_{d}S_{23}(\theta_{d})P_{d}Q_{d},\\
&&V'_{CKM}=Q'_{u}P'_{u}U_{23}(\theta'_{u},\delta'_{u})\Sigma_{u}^{'\dagger}\Sigma'_{d}S_{23}(\theta_{d}')P'_{d}Q'_{d}\,.
\end{eqnarray}
Obviously the fixed element has to be equal if the two mixing patterns are equivalent, and we assume it is the (11) entry of the CKM mixing matrix. Without loss of generality we could choose $P_{u}=P_{d}=P'_{u}=P'_{d}=1$, consequently $V_{CKM}$ and $V'_{CKM}$ become
\begin{eqnarray}
\nonumber&&V_{CKM}=Q_{u}U_{23}(\theta_{u},\delta_{u})VS_{23}(\theta_{d})Q_{d},~~~V\equiv\Sigma_{u}^{\dagger}\Sigma_{d}\,,\\
&&V'_{CKM}=Q'_{u}U_{23}(\theta'_{u},\delta'_{u})V'S_{23}(\theta_{d}')Q'_{d},~~~V'\equiv\Sigma_{u}^{'\dagger}\Sigma'_{d}\,.
\end{eqnarray}
The constant matrices $V$ and $V'$ are generically represented as follows
\begin{equation}
\label{eq:quark_2}
V=\left( \begin{array}{ccc}
a_{1} &a_{2} &a_{3}\\
a_{4} &a_{5} &a_{6}\\
a_{7} &a_{8} &a_{9} \end{array}
\right),~~
V'=\left( \begin{array}{ccc}
b_{1} &b_{2} &b_{3}\\
b_{4} &b_{5} &b_{6}\\
b_{7} &b_{8} &b_{9} \end{array}
\right)\,.
\end{equation}
The equivalence of $V_{CKM}$ and $V'_{CKM}$ entails
\begin{equation}
a_1=b_1\,,
\end{equation}
which can be taken to be positive real numbers. In the same fashion as in the lepton sector, we can rewrite the matrices $V_{CKM}$ and $V'_{CKM}$ into
\begin{eqnarray}
\nonumber&&V_{CKM}=\widetilde{Q}_{u}U_{23}(\widetilde{\theta}_{u},\widetilde{\delta}_{u})\widetilde{V}S_{23}(\widetilde{\theta}_{d})\widetilde{Q}_{d}\,,\\
\label{eq:CKM_simp}&&V'_{CKM}=\widetilde{Q}'_{u}U_{23}(\widetilde{\theta}'_{u},\widetilde{\delta}'_{u})\widetilde{V}'S_{23}(\widetilde{\theta}'_{d})\widetilde{Q}'_{d}\,,
\end{eqnarray}
where
\begin{eqnarray}
\nonumber\hskip-0.5in&&\widetilde{Q}_{u}=Q_{u}\text{diag}(1, e^{i(\delta_{1}+\delta_{2})}, e^{i(\delta_{2}-\delta_{1})}),~~~U_{23}(\widetilde{\theta}_{u},\widetilde{\delta}_{u})=\text{diag}(1, e^{-i\delta_1}, e^{i\delta_1})
U_{23}(\theta_{u},\delta_{u})
U^{\dagger}_{23}(\theta_{u}^{c},\delta_{u}^{c})\,,\\
\nonumber\hskip-0.5in&&\widetilde{V}=\text{diag}(1, e^{-i\delta_2}, e^{-i\delta_2})U_{23}(\theta_{u}^{c},\delta_{u}^{c})VS_{23}(\theta_{d}^{c})\text{diag}(1, e^{i\delta_3}, e^{i\delta_3})\,\\
\label{eq:std_form_trans1}\hskip-0.5in&&S_{23}(\widetilde{\theta}_{d})=S^{T}_{23}(\theta_{d}^{c})S_{23}(\theta_{d})=S_{23}(\theta_{d}-\theta_{d}^{c}),~~~~\widetilde{Q}_{d}=\text{diag}(1, e^{-i\delta_3}, e^{-i\delta_3})Q_{d}\,,
\end{eqnarray}
with
\begin{eqnarray}
\nonumber&&\cos\theta_{u}=|e^{i\widetilde{\delta}_{u}}\cos\theta_{u}^{c}\cos\widetilde{\theta}_{u}-e^{-i\widetilde{\delta}_{u}}\sin\theta_{u}^{c}\sin\widetilde{\theta}_{u}|,\\
\nonumber&&\sin\theta_{u}=|e^{i\widetilde{\delta}_{u}}\sin\theta_{u}^{c}\cos\widetilde{\theta}_{u}+e^{-i\widetilde{\delta}_{u}}\cos\theta_{u}^{c}\sin\widetilde{\theta}_{u}|,\\
\nonumber&&\varphi_{1}=\text{arg}[e^{i\delta_{u}^{c}}(e^{i\widetilde{\delta}_{u}}\cos\theta_{u}^{c}\cos\widetilde{\theta}_{u}-e^{-i\widetilde{\delta}_{u}}\sin\theta_{u}^{c}\sin\widetilde{\theta}_{u})],\\
\nonumber&&\varphi_{2}=\text{arg}[e^{-i\delta_{u}^{c}}(e^{i\widetilde{\delta}_{u}}\sin\theta_{u}^{c}\cos\widetilde{\theta}_{u}+e^{-i\widetilde{\delta}_{u}}\cos\theta_{u}^{c}\sin\widetilde{\theta}_{u})],\\
\nonumber&&\delta_{u}=\frac{\varphi_{1}-\varphi_{2}}{2},\\
\nonumber&&\delta_{1}=-\frac{\varphi_{1}+\varphi_{2}}{2},\\
\label{eq:std_form_trans2}&&\theta_{d}=\theta_{d}^{c}+\widetilde{\theta}_{d}\,.
\end{eqnarray}
Similar expressions for $\widetilde{Q}'_{u}$, $U_{23}(\widetilde{\theta}'_{u},\widetilde{\delta}'_{u})$, $\widetilde{V}'$, $S_{23}(\widetilde{\theta}'_{d})$ and $\widetilde{Q}'_{d}$ can be found by replacing all the parameters with the primed ones in Eqs.~(\ref{eq:std_form_trans1}, \ref{eq:std_form_trans2}). We can choose the following values for
$\delta_2$, $\theta^{c}_{u}$, $\delta^{c}_{u}$,  $\theta^{c}_{d}$ and $\delta_3$,
\begin{eqnarray}
\nonumber&&~~~\delta^{c}_{u}=\frac{\text{arg}(a_{7})-\text{arg}(a_{4})}{2},~~
\delta_{2}=\frac{\text{arg}(a_{4})+\text{arg}(a_{7})}{2}\,,\\
\nonumber&&
\sin\theta^{c}_{u}=\frac{|a_{7}|-|a_{4}|}{\sqrt{2(|a_4|^2+|a_7|^2)}},
~~\cos\theta^{c}_{u}=\frac{|a_{4}|+|a_{7}|}{\sqrt{2(|a_4|^2+|a_7|^2)}}
\,,\\
&&\cot2\theta_{d}^{c}=\frac{2\Re(a_{2}a^{*}_{3})}{|a_{2}|^{2}-|a_{3}|^{2}},~~~\delta_{3}=-\text{arg}\left(a_2\cos\theta^{c}_d-a_3\sin\theta^{c}_d\right)\,,
\end{eqnarray}
then $\widetilde{V}$ is transformed into the ``standard form''
\begin{equation}
\widetilde{V}=\left( \begin{array}{ccc}
a_{1} ~&~ \sqrt{\frac{1}{2}(1-a^2_1)} ~&~ \sqrt{\frac{1}{2}(1-a^2_1)}\,e^{i\rho_{3}}\\
\sqrt{\frac{1}{2}(1-a^2_1)} ~&~ \widehat{a}_{5} ~&~\widehat{a}_{6}e^{i\rho_{3}}\\
\sqrt{\frac{1}{2}(1-a^2_1)} ~&~ \widehat{a}_{6} ~&~\widehat{a}_{5}e^{i\rho_{3}} \end{array}
\right)\,,
\end{equation}
where
\begin{equation}
\rho_3=\delta_{3}+\text{arg}\left(a_3\cos\theta^{c}_d+a_2\sin\theta^{c}_d\right)\,.
\end{equation}
Analogously we can also transform $\widetilde{V}'$ into the following ``standard form''
\begin{equation}
\widetilde{V}'=\left( \begin{array}{ccc}
b_{1} ~&~ \sqrt{\frac{1}{2}(1-b^2_1)} ~&~ \sqrt{\frac{1}{2}(1-b^2_1)}\,e^{i\rho'_{3}}\\
\sqrt{\frac{1}{2}(1-b^2_1)} ~&~ \widehat{b}_{5} ~&~\widehat{b}_{6}e^{i\rho'_{3}}\\
\sqrt{\frac{1}{2}(1-a^2_1)} ~&~ \widehat{b}_{6} ~&~\widehat{b}_{5}e^{i\rho'_{3}} \end{array}
\right)\,.
\end{equation}
The unitarity of the matrices $\widetilde{V}$ and $\widetilde{V}'$ requires
\begin{eqnarray}
\nonumber&&a_{1}+\widehat{a}_{5}+\widehat{a}_{6}=0,~~~~2|\widehat{a}_{5}|^2+2|\widehat{a}_{6}|^2-a^2_1=1\,,\\
&&b_{1}+\widehat{b}_{5}+\widehat{b}_{6}=0,~~~~~2|\widehat{b}_{5}|^2+2|\widehat{b}_{6}|^2-b^2_1=1\,.
\end{eqnarray}
The equivalence of the two mixing patterns implies that the identity
$\widetilde{V}_{CKM}=\widetilde{V}'_{CKM}$ can be fulfilled, that is to say, the corresponding solutions
$\widetilde{\theta}'_u$, $\widetilde{\delta}_{u}'$, $\widetilde{\theta}'_{d}$ as well as $\widetilde{Q}'_{u}$, $P'_{u}$, $\widetilde{Q}'_{d}$, $P'_{d}$ can be found for any given values of $\widetilde{\theta}_u$, $\widetilde{\delta}_{u}$, $\widetilde{\theta}_{d}$ and the matrices $\widetilde{Q}_{u}$, $P_{u}$, $\widetilde{Q}_{d}$, $P_{d}$, i.e.
\begin{equation}
\label{eq:quark_equiv_1}
\widetilde{Q}_{u}U_{23}(\widetilde{\theta}_{u},\widetilde{\delta}_{u})\widetilde{V}S_{23}(\widetilde{\theta}_{d})\widetilde{Q}_{d}
=\widetilde{Q}'_{u}U_{23}(\widetilde{\theta}'_{u},\widetilde{\delta}'_{u})\widetilde{V}'S_{23}(\widetilde{\theta}'_{d})\widetilde{Q}'_{d}\,,
\end{equation}
which leads to
\begin{equation}
\label{eq:quark_equiv_2}
U_{23}^{\dagger}(\widetilde{\theta}'_{u},\widetilde{\delta}'_{u})\widetilde{Q}_{U}U_{23}(\widetilde{\theta}_{u},\widetilde{\delta}_{u})\widetilde{V}S_{23}(\widetilde{\theta}_{d})\widetilde{Q}_{D}S^{T}_{23}(\widetilde{\theta}'_{d})=\widetilde{V}'
\end{equation}
where
$\widetilde{Q}_{U}\equiv\widetilde{Q}'^{\dagger}_{u}\widetilde{Q}_{u}$ and
$\widetilde{Q}_{D}\equiv\widetilde{Q}_{d}\widetilde{Q}'^{\dagger}_{d}$ are
diagonal phase matrices.  We denote the general form of
matrices $\widetilde{Q}_{U}$ and $\widetilde{Q}_{D}$ as
\begin{equation}
\label{eq:Q_phases}\widetilde{Q}_{U}=\left( \begin{array}{ccc}
 e^{i\phi_{1}} & 0 & 0 \\
 0 & e^{i\phi_{2}} & 0 \\
 0 & 0 & e^{i\phi_{3}} \end{array}
\right),~~~~
\widetilde{Q}_{D}=\left( \begin{array}{ccc}
 e^{i\varphi_{1}} & 0 & 0 \\
 0 & e^{i\varphi_{2}} & 0 \\
 0 & 0 & e^{i\varphi_{3}} \end{array} \right)\,,
\end{equation}
where $\phi_{1,2,3}$ and $\varphi_{1,2,3}$ are free.
As shown in Eq.~\eqref{eq:simplify_U23}, we can write the
matrix
$U_{23}^{\dagger}(\widetilde{\theta}'_{u},\widetilde{\delta}'_{u})\widetilde{Q}_{U}U_{23}(\widetilde{\theta}_{u},\widetilde{\delta}_{u})$
into
\begin{equation}
\label{eq:quark_equiv_3}
U_{23}^{\dagger}(\widetilde{\theta}'_{u},\widetilde{\delta}'_{u})\widetilde{Q}_{U}U_{23}(\widetilde{\theta}_{u},\widetilde{\delta}_{u})=\text{diag}(e^{i\widehat{\phi}_{1}},e^{i\widehat{\phi}_{2}},e^{i\widehat{\phi}_{3}})U_{23}(\widehat{\theta}_{u},\widehat{\delta}_{u})
\end{equation}
with
\begin{eqnarray}
\nonumber&& \widehat{\phi}_{1}= \phi_{1}\,,\\
\nonumber&&\widehat{\phi}_{2}= (\phi_{2}+\phi_{3}+\psi_{1}+\psi_{2}-2\widetilde{\delta}'_{u})/2\,,\\
\nonumber&&\widehat{\phi}_{3}= (\phi_{2}+\phi_{3}-\psi_{1}-\psi_{2}+2\widetilde{\delta}'_{u})/2\,,\\
\nonumber&&\cos\widehat{\theta}_{u}=|e^{i(\phi_{2}-\phi_{3})/2}\cos\widetilde{\theta}'_{u}\cos\widetilde{\theta}_{u}+e^{-i(\phi_{2}-\phi_{3})/2}\sin\widetilde{\theta}'_{u}\sin\widetilde{\theta}_{u}|,\\
\nonumber&&\sin\widehat{\theta}_{u}=|e^{i(\phi_{2}-\phi_{3})/2}\cos\widetilde{\theta}'_{u}\sin\widetilde{\theta}_{u}-e^{-i(\phi_{2}-\phi_{3})/2}\sin\widetilde{\theta}'_{u}\cos\widetilde{\theta}_{u}|,\\
\nonumber&&\psi_{1}=\text{arg}[e^{i\widetilde{\delta}_{u}}(e^{i(\phi_{2}-\phi_{3})/2}\cos\widetilde{\theta}'_{u}\cos\widetilde{\theta}_{u}+e^{-i(\phi_{2}-\phi_{3})/2}\sin\widetilde{\theta}'_{u}\sin\widetilde{\theta}_{u})]\,,\\
\nonumber&&\psi_{2}=\text{arg}[e^{-i\widetilde{\delta}_{u}}(e^{i(\phi_{2}-\phi_{3})/2}\cos\widetilde{\theta}'_{u}\sin\widetilde{\theta}_{u}-e^{-i(\phi_{2}-\phi_{3})/2}\sin\widetilde{\theta}'_{u}\cos\widetilde{\theta}_{u})]\,,\\
&&\widehat{\delta}_{u}=(\psi_{1}-\psi_{2})/2\,.
\end{eqnarray}
Furthermore, we find that Eq.~\eqref{eq:quark_equiv_2} admits solution for any value of $\widetilde{\theta}_{d}$ or $\widetilde{\theta}'_{d}$ if and only if the following condition is fulfilled
\begin{equation}
e^{i\varphi_{3}}=\eta e^{i\varphi_{2}},~~~\text{with}~~~\eta=\pm1\,.
\end{equation}
Thus the combination $S_{23}(\widetilde{\theta}_{d})\widetilde{Q}_{D}S^{T}_{23}(\widetilde{\theta}'_{d})$ can be rewritten as
\begin{equation}
\label{eq:quark_equiv_new_1}
S_{23}(\widetilde{\theta}_{d})\widetilde{Q}_{D}S_{23}^{T}(\widetilde{\theta}'_{d})=
 \text{diag}(e^{i\varphi_{1}},e^{i\varphi_{2}},\eta e^{i\varphi_{2}})S_{23}(\widehat{\theta}_{d})\,,
\end{equation}
where $\widehat{\theta}_{d}=\eta \widetilde{\theta}_{d}-\widetilde{\theta}'_{d}$. As a consequence, we can simplify the equivalence condition of Eq.~\eqref{eq:quark_equiv_2} into a simple form,
\begin{equation}
\label{eq:quark_equiv_new_2}
\text{diag}(e^{i\widehat{\phi}_{1}},e^{i\widehat{\phi}_{2}},e^{i\widehat{\phi}_{3}})U_{23}(\widehat{\theta}_{u},\widehat{\delta}_{u})\widetilde{V}\text{diag}(e^{i\varphi_{1}},e^{i\varphi_{2}},\eta e^{i\varphi_{2}})S_{23}(\widehat{\theta}_{d})=\widetilde{V}'\,.
\end{equation}
Concerning the $(12)$, $(13)$, $(21)$ and $(31)$ entries of the matrices on both sides, we find that Eq.~\eqref{eq:quark_equiv_new_2} leads to
\begin{eqnarray}
\label{eq:constriant_1}
&&e^{i(\varphi_{2}-\varphi_{1})}(\cos\widehat{\theta}_{d}-e^{i\rho_{3}}\eta\sin\widehat{\theta}_{d})=1,~~~~
e^{i(\varphi_{2}-\varphi_{1}-\rho'_{3})}(e^{i\rho_{3}}\eta\cos\widehat{\theta}_{d}+\sin\widehat{\theta}_{d})=1\,,\\
\label{eq:solve_quark_21_31}&& e^{i(\varphi_{1}+\widehat{\phi}_{2})}(e^{i\widehat{\delta}_{u}}\cos\widehat{\theta}_{u}+e^{-i\widehat{\delta}_{u}}\sin\widehat{\theta}_{u})=1,~~~~
 e^{i(\varphi_{1}+\widehat{\phi}_{3})}(e^{-i\widehat{\delta}_{u}}\cos\widehat{\theta}_{u}-e^{i\widehat{\delta}_{u}}\sin\widehat{\theta}_{u})=1\,,
\end{eqnarray}
which requires
\begin{eqnarray}
&&\cos\rho_{3}\sin 2\widehat{\theta}_{d}=0\,,~~~~~\cos2\widehat{\delta}_{u}\sin2\widehat{\theta}_{u}=0\,.
\end{eqnarray}
Hence we obtain the constraints
\begin{eqnarray}
&&\cos\rho_3=0,~~~\text{or}~~~\widehat{\theta}_{d}=0, \pi/2\,,\\
&&\cos2\widehat{\delta}_{u}=0,~~~\text{or}~~~\widehat{\theta}_{u}=0, \pi/2\,.
\end{eqnarray}
Consequently from Eq.~\eqref{eq:solve_quark_21_31} we can determine the values of $\widehat{\phi}_{2}$ and $\widehat{\phi}_{3}$ as
\begin{eqnarray}
\nonumber&&e^{i\widehat{\phi}_{2}}=e^{-i(\widehat{\delta}_{u}+\varphi_{1})},~~~~
e^{i\widehat{\phi}_{3}}=e^{i(\widehat{\delta}_{u}-\varphi_{1})},~~\text{for}~~\widehat{\theta}_{u}=0\,,\\
\nonumber&&e^{i\widehat{\phi}_{2}}=e^{i(\widehat{\delta}_{u}-\varphi_{1})},~~~~
 e^{i\widehat{\phi}_{3}}=-e^{-i(\widehat{\delta}_{u}+\varphi_{1})},~~\text{for}~~\widehat{\theta}_{u}=\pi/2\,,\\
&&e^{i\widehat{\phi}_{2}}=e^{i(\kappa_{2}\widehat{\theta}_{u}-\widehat{\delta}_{u}-\varphi_{1})},~~~
e^{i\widehat{\phi}_{3}}=e^{i(\kappa_{2}\widehat{\theta}_{u}+\widehat{\delta}_{u}-\varphi_{1})},~~\text{for}~~\cos2\widehat{\delta}_{u}=0\,,
\end{eqnarray}
where $\kappa_{2}=-i e^{2i\widehat{\delta}_{u}}$ is either $+1$ or $-1$. In the following, we shall discuss the constraints arising form the (22), (23), (32) and $(33)$ entries in Eq.~\eqref{eq:quark_equiv_new_2} for all possible cases.

\begin{itemize}
\item{$\widehat{\theta}_{d}=0$}

Plugging this value of $\widehat{\theta}_{d}$ into Eq.~\eqref{eq:constriant_1}, we have
\begin{equation}
\label{eq:elemnts_constriant_1}
e^{i\varphi_{2}}=e^{i\varphi_{1}},~~~~\eta e^{i(\rho'_{3}-\rho_{3})}=1\,.
\end{equation}
Then the equivalence of these two mixing patterns requires
\begin{eqnarray}
\nonumber&&\widehat{a}_{5}=\widehat{b}_{5},~~~~\widehat{a}_{6}=\widehat{b}_{6},~~\text{for}~~\widehat{\theta}_{u}=0\,,\\
\nonumber&&\widehat{a}_{5}=\widehat{b}_{6},~~~\widehat{a}_{6}=\widehat{b}_{5},~~\text{for}~~\widehat{\theta}_{u}=\pi/2\,,\\
&&\tan\widehat{\theta}_{u}=i\kappa_{2}\frac{\widehat{b}_{5}-\widehat{a}_{5}}{\widehat{a}_{6}-\widehat{b}_{5}},~~~\text{for}~~\cos2\widehat{\delta}_{u}=0\,.
\end{eqnarray}
We notice that the unitarity of $\widetilde{V}$ and $\widetilde{V}'$ implies the combination $i(\widehat{b}_{5}-\widehat{a}_{5})/(\widehat{a}_{6}-\widehat{b}_{5})$ is real. Hence $V_{CKM}$ and $V'_{CKM}$ would be essentially the same mixing pattern of the condition $\eta e^{i(\rho'_{3}-\rho_{3})}=1$ is fulfilled.

\item{$\widehat{\theta}_{d}=\pi/2$}

For this value of $\widehat{\theta}_{d}$, Eq.~\eqref{eq:constriant_1} leads to
\begin{equation}
\label{eq:elemnts_constriant_2}
e^{i\varphi_{2}}=e^{i(\varphi_{1}+\rho'_{3})},~~~-\eta e^{i(\rho'_{3}+\rho_{3})}=1\,.
\end{equation}
Solving the equivalence condition of Eq.~\eqref{eq:quark_equiv_new_2}, we find
\begin{eqnarray}
\nonumber&&\widehat{a}_{5}=\widehat{b}_{6},~~~\widehat{a}_{6}=\widehat{b}_{5},~~\text{for}~~\widehat{\theta}_{u}=0\,,\\
\nonumber&&\widehat{a}_{5}=\widehat{b}_{5},~~~~\widehat{a}_{6}=\widehat{b}_{6},~~\text{for}~~\widehat{\theta}_{u}=\pi/2\,,\\
&&\tan\widehat{\theta}_{u}=i\kappa_{2}\frac{\widehat{a}_{6}-\widehat{b}_{5}}{\widehat{b}_{5}-\widehat{a}_{5}},~~~\text{for}~~\cos2\widehat{\delta}_{u}=0\,.
\end{eqnarray}

\item{$\cos\rho_{3}=0$ }

In this case, from Eq.~\eqref{eq:constriant_1} we obtain
\begin{equation}
\label{eq:elemnts_constriant_3}
e^{i\varphi_{2}}=e^{i(\varphi_{1}+\eta\kappa_{1}\widehat{\theta}_{d})},~~~~e^{i\rho'_{3}}=i\eta\kappa_{1},~~\text{with}~~e^{i\rho_{3}}=i\kappa_{1}\,.
\end{equation}
where $\kappa_{1}=\pm 1$. In the same fashion as previous case, we find the equivalence condition of Eq.~\eqref{eq:quark_equiv_new_2} requires
\begin{eqnarray}
\nonumber&&\tan\widehat{\theta}_{d}=i\eta\kappa_{1}\frac{\widehat{b}_{5}-\widehat{a}_{5}}{\widehat{a}_{6}-\widehat{b}_{5}},~~\text{for}~~\widehat{\theta}_{u}=0\,,\\
\nonumber&&\tan\widehat{\theta}_{d}=i\eta\kappa_{1}\frac{\widehat{a}_{6}-\widehat{b}_{5}}{\widehat{b}_{5}-\widehat{a}_{5}},~~\text{for}~~\widehat{\theta}_{u}=\pi/2\,,\\
\nonumber&&\tan(\eta\kappa_1\widehat{\theta}_{d}+\kappa_2\widehat{\theta}_{u})=i\frac{\widehat{b}_{5}-\widehat{a}_{5}}{\widehat{a}_{6}-\widehat{b}_{5}},~~~\text{for}~~\cos2\widehat{\delta}_{u}=0\,.
\end{eqnarray}

\end{itemize}
We summarize that the two CKM mixing matrices for $\widetilde{V}$ and $\widetilde{V}'$ in Eq.~\eqref{eq:CKM_simp} would give the same mixing pattern, if the following conditions are satisfied,
\begin{equation}
\eta e^{i(\rho'_{3}-\rho_{3})}=1,~~\text{or}~~-\eta e^{i(\rho'_{3}+\rho_{3})}=1,~~\text{or}~~e^{i\rho_{3}}=i\kappa_{1},~~e^{i\rho'_{3}}=i\eta\kappa_{1}\,,
\end{equation}
where $\eta,\kappa_{1}=\pm1$. The above conditions for the equivalence of the two mixing patterns in this scenario can be compactly written as
\begin{equation}
\label{eq:equiv_quark_1}e^{2i(\rho'_{3}-\rho_{3})}=1,~~\text{or}~~ e^{2i(\rho'_{3}+\rho_{3})}=1\,.
\end{equation}
For the scenario of residual symmetry $G_{u}=Z_{2}^{g_{u}}\times X_{u}$ and $G_{d}=Z_{2}^{g_{d}}$ in the quark sector, the quark mixing matrix is given by Eq.~\eqref{eq:CKM_form_2}. The equivalence condition can be derived in the same manner. In fact, we can easily determine whether two distinct residual symmetries generate the same quark mixing pattern by taking the hermitian conjugate conjugate of the mixing matrix and applying the criterion of Eq.~\eqref{eq:equiv_quark_1}.

\subsection{\label{sec:quark_z2z2cp}Examples of quark mixing patterns from $\Delta(6n^{2})$ and CP symmetries}

It is well-known that the CKM matrix has been quite precisely measured by the $B$ factories. The global fit results for the magnitudes of all nine CKM elements are \cite{Patrignani:2016xqp},
\begin{equation}
\label{eq:abs_VCKM}|V_{CKM}|=\begin{pmatrix}
0.97434^{+0.00011}_{-0.00012} ~&~0.22506\pm0.00050 ~&~ 0.00357\pm0.00015 \\
0.22492\pm0.00050~&~0.97351\pm0.00013 ~&~0.0411 \pm 0.0013\\
0.00875^{+0.00032}_{-0.00033} ~&~0.0403\pm0.0013 ~&~ 0.99915\pm0.00005
\end{pmatrix}\,.
\end{equation}
The full fit values of quark mixing angles and Jarlskog invariant, given by the UTfit collaboration~\cite{Bona:2005vz,Bona:2007vi,utfit:2014}, read as
\begin{eqnarray}
\label{eq:full_fit}
\nonumber && \sin\theta^{q}_{12}=0.22497\pm0.00069, \qquad \sin\theta^{q}_{23}=0.04229\pm0.00057\,, \\
&& \sin\theta^{q}_{13}=0.00368\pm0.00010, \qquad J^{q}_{CP}=(3.115\pm0.093)\times10^{-5}\,.
\end{eqnarray}
In this section, we shall investigate the possible quark mixing patterns which arise form the breaking of $\Delta(6n^2)$ and CP into $Z_{2}^{g_{u}}$ and $Z_{2}^{g_{d}}\times X_{d}$ in the up quark and down quark sectors respectively. We find that agreement with the experimental data can be achieved if the residual symmetries are $g_{u}=bc^{x}d^{x}$, $g_{d}=bc^{y}d^{y}$ and $X_{d}=\{c^{\rho}d^{-2y-\rho},bc^{y+\rho}d^{-y-\rho}\}$ where $x,y,\rho=0,1,\ldots, n-1$. Using the master formula of Eq.~\eqref{eq:CKM_form_1}, the CKM matrix is determined to be
\begin{equation}
\label{eq:U_CKM_Z2Z2CP_form}
V_{CKM}=\left(\begin{array}{ccc}
\cos\varphi_{1} &~ s_{d}\sin \varphi_{1}~& -c_{d}\sin \varphi_{1}\\
s_{u}\sin \varphi_{1}&~ c_{u} c_{d}e^{i \delta} -s_{u} s_{d}\cos \varphi_{1}~& c_{u} s_{d}e^{i \delta} +c_{d} s_{u}\cos \varphi_{1} \\
c_{u}\sin \varphi_{1}&~ -c_{d} s_{u}e^{i \delta} -c_{u} s_{d}\cos \varphi_{1}~& -s_{u} s_{d}e^{i \delta}+c_{u} c_{d}\cos \varphi_{1}
\end{array}
\right)\,,
\end{equation}
up to permutations of rows and columns, where the parameters $c_{u}$, $c_{d}$, $s_{u}$, $s_{d}$ and $\delta$ are defined as
\begin{equation}\label{eq:define_sc}
c_{u}=\cos\theta_{u}, \quad c_{d}=\cos\theta_{d}, \quad s_{u}=\sin\theta_{u}, \quad s_{d}=\sin\theta_{d}, \quad \delta= 2\delta_{u}-\varphi_{2}\,.
\end{equation}
The discrete parameters $\varphi_{1}$ and $\varphi_{2}$ are given by
\begin{equation}
\varphi_{1}=\frac{x-y}{n}\pi,~~\varphi_{2}=\frac{x+3(y+\rho)}{n}\pi\,.
\end{equation}
We see that the CKM mixing matrix in Eq.~\eqref{eq:U_CKM_Z2Z2CP_form} depends not only on three continuous parameters $\theta_{u}$, $\theta_{d}$ and $\delta$ but also on the discrete parameter $\varphi_{1}$ whose value is determined by the choice of the residual symmetry. The value of another discrete parameter $\varphi_{2}$ is irrelevant since it can be absorbed into the continuous free parameter $\delta_{u}$.
Moreover, the parameters $\varphi_{1}$ can take the following discrete values
\begin{equation}
\varphi_{1}~(\text{mod}~2\pi)=0,\frac{1}{n}\pi,\frac{2}{n}\pi,...,\frac{2n-1}{n}\pi\,.
\end{equation}
The matrix $V_{CKM}$ has the following symmetry properties,
\begin{eqnarray}
\nonumber
V_{CKM}(\varphi_{1}, \pi+\theta_{u},\theta_{d},\delta)&=&\text{diag}(1,-1,-1)V_{CKM}(\varphi_{1},\theta_{u},\theta_{d},\delta)\\
\nonumber V_{CKM}(\varphi_{1},\pi-\theta_{u},\theta_{d},\delta)&=&\text{diag}(1,1,-1)V_{CKM}(\varphi_{1},\theta_{u},\theta_{d},\delta-\pi)\,,\\
\nonumber
V_{CKM}(\varphi_{1},\theta_{u}, \pi+\theta_{d},\delta)&=&V_{CKM}(\varphi_{1},\theta_{u},\theta_{d},\delta)\text{diag}(1,-1,-1)\\
\nonumber
V_{CKM}(\pi+\varphi_{1},\theta_{u},\theta_{d},\delta)&=&V_{CKM}(\varphi_{1},\theta_{u},\pi-\theta_{d},\delta)\text{diag}(-1,-1,1)\\
\nonumber
V_{CKM}(\pi-\varphi_{1},\theta_{u},\theta_{d},\delta)&=&\text{diag}(-1,1,1)V_{CKM}(\varphi_{1},\theta_{u},\pi-\theta_{d},\delta)\text{diag}(1,-1,1)\\
\nonumber
V_{CKM}(\varphi_{1},\theta_{u},\theta_{d}, \pi+\delta)&=&V_{CKM}(\varphi_{1},\theta_{u},\pi-\theta_{d},\delta)\text{diag}(1,1,-1)\\
V_{CKM}(\varphi_{1},\theta_{u},\theta_{d},\pi-\delta)&=&V_{CKM}^{*}(\varphi_{1},\theta_{u},\pi-\theta_{d},\delta)\text{diag}(1,1,-1)\,,
\end{eqnarray}
where the above diagonal matrices can be absorbed by the quark fields. Consequently the parameter $\varphi_1$ can be limited in the range of $0\leq\varphi_1\leq\pi/2$, and the free parameters $\theta_{u}$, $\theta_{d}$ and $\delta$ take values in the range of $0\leq\theta_{u}\leq\pi/2$, $0\leq\theta_{d}<\pi$ and $0\leq\delta<\pi$ respectively.
Furthermore, we see that the residual symmetry fixes one element of the CKM mixing matrix is $\cos\varphi_{1}$, and the 36 possible permutations of rows and columns give rise to nine independent mixing patterns
\begin{equation}
\label{eq:CKM_caseI}
\begin{array}{lll}
V_{CKM,1}=V_{CKM}, ~~&~~ V_{CKM,2}=V_{CKM}P_{12},~~&~~ V_{CKM,3}=V_{CKM}P_{13}\,, \\
V_{CKM,4}=P_{12}V_{CKM}, ~~&~~ V_{CKM,5}=P_{12}V_{CKM}P_{12},~~&~~ V_{CKM,6}=P_{12}V_{CKM}P_{13}\,, \\
V_{CKM,7}=P_{23}P_{12}V_{CKM}, ~~&~~ V_{CKM,8}=P_{23}P_{12}V_{CKM}P_{12}, ~~&~~ V_{CKM,9}=P_{23}P_{12}V_{CKM}P_{13}\,.
\end{array}
\end{equation}

For another scheme where the $\Delta(6n^{2})$ and CP symmetries
are broken down to $Z_{2}^{g_{u}}\times X_{u}$ in the up quark sector and $Z_{2}^{g_{d}}$ in the down quark sector, the experimental data on quark mixing can be accommodated if $g_{u}=bc^{x}d^{x}$, $X_{u}=\{c^{\rho}d^{-2x-\rho},bc^{x+\rho}d^{-x-\rho}\}$ and $g_{d}=bc^{y}d^{y}$. From Eq.~\eqref{eq:CKM_form_2} we can obtain the corresponding form of the CKM mixing matrix
\begin{equation}
 \label{eq:U_CKM_Z2Z2CP_form_2}
V'_{CKM}=\left(\begin{array}{ccc}
 \cos\varphi'_{1} &~ c'_{d}\sin \varphi'_{1}~& -s'_{d}\sin \varphi'_{1}\\
 c'_{u}\sin \varphi'_{1}&~ s'_{u} s'_{d}e^{i \delta'} -c'_{u} c'_{d}\cos \varphi'_{1}~& s'_{u} c'_{d}e^{i \delta'} +s_{d} c_{u}\cos \varphi'_{1}\\
 s'_{u}\sin \varphi'_{1}&~ -s_{d} c_{u}e^{i \delta'} -s'_{u} c'_{d}\cos \varphi'_{1}~& -c'_{u} c'_{d}e^{i \delta'}+s'_{u} s'_{d}\cos \varphi'_{1} \\
\end{array}
\right)\,,
\end{equation}
where the permutation matrices and phases matrices are neglected for simplicity. We see that $V'_{CKM}$ depends on three free continuous parameters $\theta_{u}'$, $\theta_{d}'$, $\delta'$ and one discrete parameter $\varphi_{1}'$ with
\begin{equation}\label{eq:define_sc_2}
c'_{u}=\cos\theta'_{u}, \quad c'_{d}=\cos\theta'_{d}, \quad s'_{u}=\sin\theta'_{u}, \quad s'_{d}=\sin\theta'_{d}, \quad \delta'=\varphi'_{2}-2\delta'_{d}\,,
\end{equation}
and
\begin{equation}
\varphi'_{1}=\frac{x-y}{n}\pi,~~~~\varphi'_{2}=\frac{y+3(x+\rho)}{n}\pi\,.
\end{equation}
Moreover we find that the two CKM matrices in Eq.~\eqref{eq:U_CKM_Z2Z2CP_form} and Eq.~\eqref{eq:U_CKM_Z2Z2CP_form_2} are related as follows
\begin{equation}
V'_{CKM}(\varphi'_{1},\theta'_{u},\theta'_{d},\delta')=V_{CKM}(\varphi'_{1},\pi/2-\theta'_{u},\pi/2-\theta_{d}',\delta')\,.
\end{equation}
Consequently $V_{CKM}$ and $V'_{CKM}$ lead to the same mixing pattern and it is sufficient to only consider the first scheme where the quark mixing matrix is determined to be $V_{CKM}$.

We scan all possible values of discrete parameters $\varphi_{1}$ for each value of group index $n$ with $n\leq 40$. The continuous parameters $\theta_{u}$, $\delta_{u}$ and $\theta_{d}$ freely vary between $0$ and $\pi$. The cases that can give a good fit to the experimental data are summarized in table~\ref{Tab:numerical_result_quark_z2z2cp}, where we list the values of $n$, $\varphi_{1}$ and the resulting predictions for $\sin\theta_{ij}^{q}$ and $J_{CP}$ at certain benchmark values of $\theta_{u}$, $\delta_{u}$, $\theta_{d}$. For the $\Delta(6n^2)$ groups with $n\leq40$, we find six permutations $V_{CKM,1}$, $V_{CKM,2}$, $V_{CKM,4}$, $V_{CKM,5}$, $V_{CKM,6}$, and $V_{CKM,8}$ can describe the experimental data of CKM matrix shown in Eq.~\eqref{eq:full_fit}. The $6$ possible matrices are $V_{CKM,1}$, $V_{CKM,2}$, $V_{CKM,4}$, $V_{CKM,5}$, $V_{CKM,6}$, and $V_{CKM,8}$. Furthermore, for the mixing patterns $V_{CKM,2}$ and $V_{CKM,4}$, the smallest value of index $n$ which can accommodate the experimental data is $n=7$.

For the matrix $V_{CKM,2}$, the fixed element $\cos\varphi_{1}$ is $(12)$ entry of CKM matrix and the expressions of mixing parameters can be extracted as follows
\begin{eqnarray}
\nonumber
&&\sin^{2}\theta_{13}^{q}=\sin^{2}\varphi_{1}\cos^{2}\theta_{d}\,,\\
 \nonumber
&&\sin^{2}\theta_{12}^{q}=\frac{\cos^{2}\varphi_{1}}{1-\sin^{2}\varphi_{1}\cos^{2}\theta_{d}}\,,\\
 \nonumber
&&\sin^{2}\theta_{23}^{q}=\frac{2\cos^{2}\theta_{u} \sin^{2}\theta_{d}+2\sin^{2}\theta_{u}\cos^{2}\theta_{d}\cos^{2}\varphi_{1}-\cos\varphi_{1}\cos\delta \sin2\theta_{u}\sin2\theta_{d}}{2-2\sin^{2}\varphi_{1}\cos^{2}\theta_{d}}\,,\\
\label{eq:U_CKM2_expression}
 &&J_{CP}=\frac{1}{8}\sin\varphi_{1}\sin2\varphi_{1}\sin\delta \sin2\theta_{u} \sin2\theta_{d}\,.
\end{eqnarray}
For the parameter values
\begin{equation}
\varphi_{1}=3\pi/7,~~\theta_{u}=0.51357\pi,~~\theta_{d}=0.49880\pi,~~\delta=0.36887\pi\,,
\end{equation}
we obtain
\begin{eqnarray}
\nonumber
&&\sin\theta_{13}^{q}=0.00368,~~~\sin\theta_{12}^{q}=0.22252\,,\\
&&\sin\theta_{23}^{q}=0.04229,~~~J^{q}_{CP}=3.115\times 10^{-5}\,.
\end{eqnarray}
It is remarkable that the predicted values $\sin\theta_{13}^{q}$, $\sin\theta_{23}^{q}$ and $J^{q}_{CP}$ coincide with the best fitting values given by the UTfit collaboration~\cite{Bona:2005vz,Bona:2007vi,utfit:2014}.
The mixing angle $\sin\theta_{12}^{q}$ is about $1\%$ larger than its measured value and this small discrepancy can be easily resolved in a model with small corrections.

For the mixing matrix $V_{CKM,4}$, its $(21)$ element is $\cos\varphi_{1}$.
The quark flavor mixing angles and CP invariant are determined to be of the following form
\begin{eqnarray}
\nonumber
&&\sin^{2}\theta_{13}^{q}=\cos^{2}\theta_{d} \sin^{2}\theta_{u}\cos^{2}\varphi_{1}+\cos^{2}\theta_{u}\sin^{2}\theta_{d}-\frac{1}{2}\sin2\theta_{u}\sin2\theta_{d}\cos\varphi_{1}\cos\delta\,,\\
\nonumber
&&\sin^{2}\theta_{12}^{q}=1-\frac{\sin^{2}\theta_{u} \sin^{2}\varphi_{1}}{1-\cos^{2}\theta_{d} \sin^{2}\theta_{u}\cos^{2}\varphi_{1}-\cos^{2}\theta_{u}\sin^{2}\theta_{d}+\frac{1}{2}\sin2\theta_{u}\sin2\theta_{d}\cos\varphi_{1}\cos\delta}\,,\\
\nonumber
&&\sin^{2}\theta_{23}^{q}=\frac{\cos^{2}\theta_{d} \sin^{2}\varphi_{1}}{1-\cos^{2}\theta_{d} \sin^{2}\theta_{u}\cos^{2}\varphi_{1}-\cos^{2}\theta_{u}\sin^{2}\theta_{d}+\frac{1}{2}\sin2\theta_{u}\sin2\theta_{d}\cos\varphi_{1}\cos\delta}\,,\\
&&J_{CP}^{q}=\frac{1}{8}\sin\varphi_{1}\sin2\varphi_{1}\sin\delta \sin2\theta_{u} \sin2\theta_{d}\,.
\end{eqnarray}
The quark mixing angles and CP violation phase can be in accordance
with experimental data for $\varphi_{1}=3\pi/7$, e.g.,
\begin{eqnarray}
\nonumber&&\theta_{u}=0.50370\pi,~~~\theta_{d}=0.48619\pi,~~~\delta=0.09426\pi\,,\\
\nonumber
&&\sin\theta_{13}^{q}=0.00368,~~~~\sin\theta_{12}^{q}=0.22278\,,\\
&&\sin\theta_{23}^{q}=0.04232,~~~~J_{CP}^{q}=3.111\times 10^{-5}\,.
\end{eqnarray}
We notice that $\sin\theta_{13}^{q}$, $\sin\theta_{23}^{q}$ and $J_{CP}^{q}$ are in the experimentally preferred ranges while $\sin\theta_{12}^{q}$ is a bit larger than the best fit value given in Eq.~\eqref{eq:full_fit}. Since the leading order predictions generically receive subleading corrections in a concrete model such that we expect the current data can be reproduced.

\begin{table}[t!]
\centering
\footnotesize
\begin{tabular}{|c|c|c|c|c|c|c|c|c|c|}
\hline
\hline
 & $n$ & $\varphi_{1}$ & $\theta_{u}/\pi$ & $\theta_{d}/\pi$  & $\delta/\pi$ & $\sin\theta_{13}^{q}$ & $\sin\theta_{12}^{q}$ & $\sin\theta_{23}^{q}$ & $J^{q}_{CP}/10^{-5}$ \\
\hline
$V_{CKM,1}$ & $14,28$ & $\pi/14$ & $0.48756$ & $0.49474$  & $0.49910$ & $0.00368$ & $0.22249$ & $0.04229$ & $3.115$ \\
  \hline
$V_{CKM,2}$ & $7,14,21,28,35$ & $3\pi/7$ & $0.51357$  & $0.49880$  & $0.36887$ & $0.00368$ & $0.22252$ & $0.04229$ & $3.115$ \\
  \hline
$V_{CKM,4}$ & $7,14,21,28,35$ & $3\pi/7$ & $0.50370$ & $0.48619$  & $0.09426$ & $0.00368$ & $0.22278$ & $0.04232$ & $3.111$ \\
  \hline
$V_{CKM,5}$ & $27$ & $2\pi/27$ & $0.55799$  & $0.55903$ & $0.99407$ & $0.00368$ & $0.22680$ & $0.04252$ & $3.112$  \\
  \hline
\multirow{3}{*}{$V_{CKM,6}$}& $35$ & $17\pi/35$ & $0.00117$ & $0.42775$ & $0.33034$ & $0.00368$ & $0.22497$ & $0.04487$ & $3.115$  \\
  \cline{2-10}
& $37$ & $18\pi/37$ & $0.00117$ & $0.42775$  & $0.36426$ & $0.00368$ & $0.22497$ & $0.04244$ & $3.115$  \\
 \cline{2-10}
& $39$ & $19\pi/39$ & $0.00117$ & $0.42776$ & $0.40925$  & $0.00368$ & $0.22497$ & $0.04027$ & $3.115$ \\
 \hline
\multirow{2}{*}{$V_{CKM,8}$}
& $37$ & $18\pi/37$ & $0.92771$  & $0.00276$ & $0.12567$ & $0.00369$ & $0.22497$ & $0.043162$ & $3.104$ \\
\cline{2-10}
& $39$ & $19\pi/39$ & $0.92771$  & $0.00359$ & $0.10108$ & $0.00369$ & $0.22499$ & $0.04165$ & $3.104$ \\ \hline\hline
\end{tabular}
\caption{\label{Tab:numerical_result_quark_z2z2cp} Numerical results of the
quark mixing parameters for the residual symmetries $G_{u}=Z_{2}^{g_{u}}$ and $G_{d}=Z_{2}^{g_{d}}\times X_{d}$ with $g_{u}=bc^{x}d^{x}$, $g_{d}=bc^{y}d^{y}$ and $X_{d}=\{c^{\rho}d^{-2y-\rho},bc^{y+\rho}d^{-y-\rho}\}$, where we focus on the $\Delta(6n^2)$ group with $n\leq40$. We display quark mixing angles $\sin\theta_{ij}^{q}$ and CP invariant $J_{CP}^{q}$ which are in accordance with the experimental data for certain values of $\theta_{u}$, $\delta_{u}$, $\theta_{d}$ and $\varphi_{1}$.}
\end{table}

In this work, we are eager to know whether it is possible to describe quark and lepton flavor mixing structures from a common flavor group $\Delta(6n^{2})$ and CP symmetry. In section~\ref{sec:lepton_z2z2cp} we have studied the lepton mixing patterns which can be obtained from the breaking of flavor group $\Delta(6n^{2})$ and CP into residual symmetries $Z_{2}^{g_{l}}$ and $Z_{2}^{g_{\nu}}\times X_{\nu}$ in charged lepton sector and neutrino sector respectively. We have focused on the smaller $\Delta(6n^2)$ group with $n=3,4$ in section~\ref{sec:lepton_z2z2cp}. However, the index has to be at least $n=7$ in order to explain the experimental data on quark mixing. We now assume that the flavor group $\Delta(6\cdot 7^{2})$ and CP symmetries are broken down to $Z_{2}^{g_{l}}$ in charged lepton sector and $Z_{2}^{g_{\nu}}\times X_{\nu}$ in neutrino sector. After considering all possible residual symmetries,
we find that only case I and case II can lead to possible mixing patterns in agreement with experimental data. The numerical predictions for lepton mixing angles and CP phases are summarized in
in table~\ref{Tab:U_{I}_n=7_and_U_{II}_n=7}.

\begin{figure}[t!]
\centering
\includegraphics[width=0.98\textwidth]{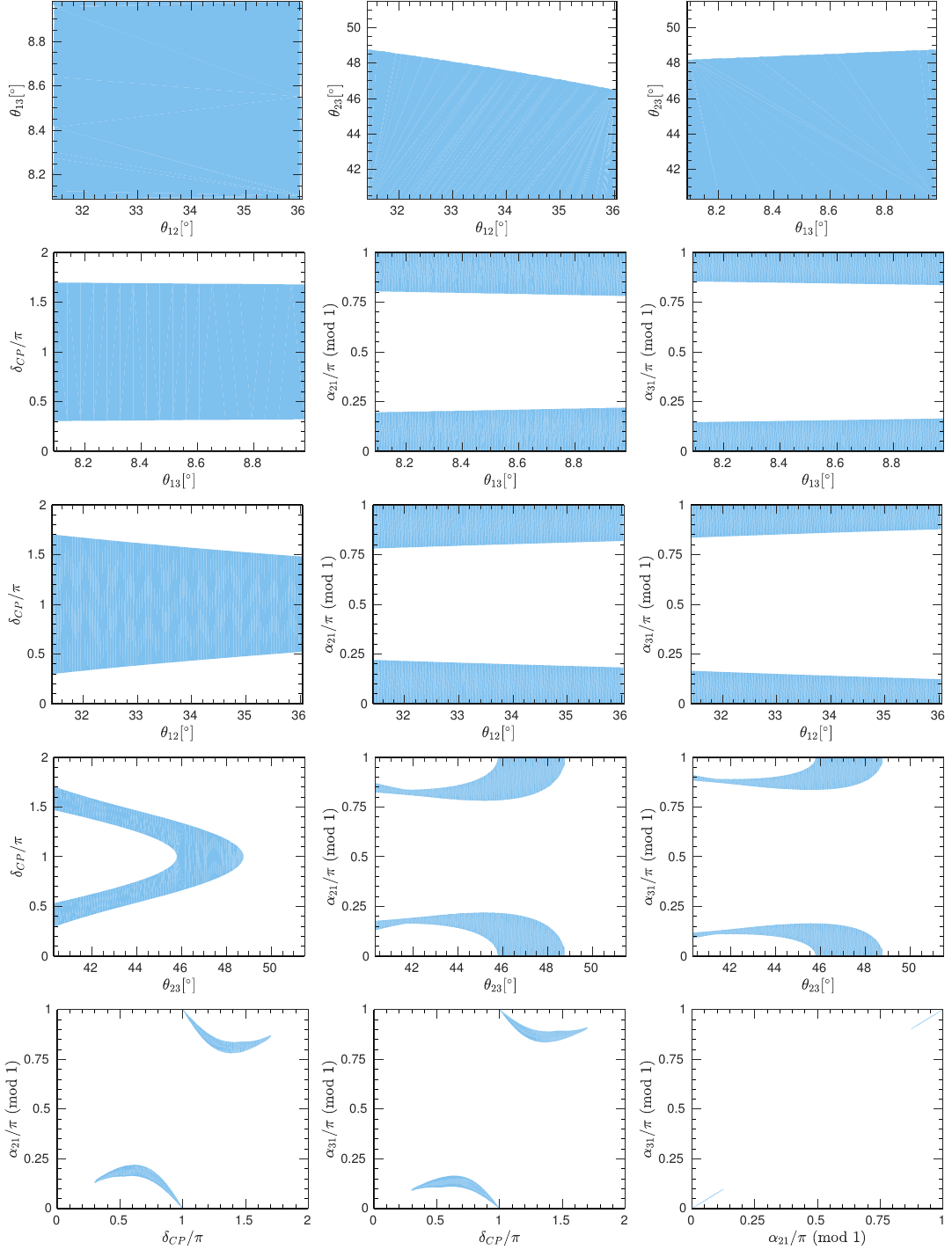}
\caption{\label{Fig:lepton_case15_n7}
Correlations between different mixing parameters for the mixing pattern $U_{I, 5}$ with $\varphi_1=2\pi/7$, where the residual symmetry is  $\{G_{l},G_{\nu},X_{\nu}\}=\{Z_{2}^{bc^{x}d^{x}},Z_{2}^{bc^{y}d^{y}},(c^{\delta}d^{-2y-\delta},bc^{y+\delta}d^{-y-\delta})\}$, the three lepton mixing angles are required to be compatible with the experimental data at $3\sigma$ level~\cite{Esteban:2016qun}.}
\end{figure}

As an example, for the mixing pattern $U_{I,5}$ with $\varphi_{1}=2\pi/7$, the mixing angles $\theta_{13}$ and $\theta_{12}$ can approximately take any value within their allowed $3\sigma$ ranges while $\theta_{23}$ lies in the range $\theta_{23}\in[40.280^{\circ}, 48.768^{\circ}]$. The Dirac CP violation phase $\delta_{CP}$ can vary from $0.299\pi$ to $1.701\pi$. However, the Majorana phases $\alpha_{21}$ and $\alpha_{31}$ are determined to be around $0$ and $\pi$, i.e., $\alpha_{21}\in[0, 0.219\pi]\cup[0.781\pi,\pi]$, $\alpha_{31}\in[0, 0.165\pi]\cup[0.835\pi,\pi]$. The correlations between different mixing parameters of this case are displayed in figure~\ref{Fig:lepton_case15_n7}. We can see that the atmospheric mixing angle $\theta_{23}$ and $\delta_{CP}$ are strongly correlated and there are peculiar correlations between three CP violation phases.

\begin{table}[!hbp]
\centering
\footnotesize
\begin{tabular}{|c |c| c c c c c c|}
\hline
\hline
 & $\varphi_{1}$ &$\theta_{13}/^{\circ}$ & $\theta_{12}/^{\circ}$ & $\theta_{23}/^{\circ}$ & $\delta_{CP}/\pi$ & $\alpha_{21}/\pi(\text{mod}~1)$ & $\alpha_{31}/\pi(\text{mod}~1)$ \\
\hline
  \multirow{2}{*}{$U_{I,5}$} & \multirow{2}{*}{$2\pi/7$} &\multirow{2}{*}{$8.091-8.979$} & \multirow{2}{*}{$31.435-36.031$} & \multirow{2}{*}{$40.280-48.768$} & \multirow{2}{*}{$0.299-1.701$} & $0-0.219$ & $0-0.165$ \\
    &&&&&&$\oplus 0.781-1$&$\oplus 0.835-1$\\
\hline
\multirow{2}{*}{$U_{I,6}$}& \multirow{2}{*}{$2\pi/7$} &\multirow{2}{*}{$8.095-8.979$} & \multirow{2}{*}{$31.435-36.031$} & \multirow{2}{*}{$50.860-50.967$} & \multirow{2}{*}{$0-2 $} & $0-0.181$ & $0-0.133$ \\
    &&&&&&$\oplus 0.819-1$&$\oplus 0.867-1$\\
  \hline
  \hline
   & $\varphi_{4}$ &$\theta_{13}/^{\circ}$ & $\theta_{12}/^{\circ}$ & $\theta_{23}/^{\circ}$ & $\delta_{CP}/\pi$ & $\alpha_{21}/\pi(\text{mod}~1)$ & $\alpha_{31}/\pi(\text{mod}~1)$ \\
\hline
  \multirow{11}{*}{$U_{II,1}$} & \multirow{2}{*}{0}&\multirow{2}{*}{$8.091-8.979$} & \multirow{2}{*}{$33.010-36.031$} & \multirow{2}{*}{$40.280-45.828$} & $0-0.304$ & $0-0.138$ & $0-0.085$ \\
         &&&&& $\oplus 1.696-2 $ & $\oplus 0.862-1$ & $\oplus 0.915-1$\\
\cline{2-8}
                 & \multirow{3}{*}{$\frac{\pi}{7}$} & \multirow{3}{*}{$8.091-8.979$} & \multirow{3}{*}{$32.993-36.031$} & \multirow{3}{*}{$40.280-45.844$} & \multirow{1}{*}[-5pt]{$0-0.304$} & $0-0.183$ & $0-0.158$ \\
     &&&&& \multirow{2}{*}[-1pt]{$\oplus 1.696-2 $} & $\oplus 0.204-0.480$ & $\oplus 0.229-0.400$\\
     &&&&&&$\oplus 0.907-1$ & $\oplus 0.987-1$\\
  \cline{2-8}
                 & \multirow{3}{*}{$\frac{2\pi}{7}$} & \multirow{3}{*}{$8.091-8.979$} & \multirow{3}{*}{$33.000-36.031$} & \multirow{3}{*}{$40.280-45.848$} & \multirow{1}{*}[-5pt]{$0-0.303$} & $0-0.252$ & \multirow{1}{*}[-5pt]{$0.093-0.286$} \\
     &&&&& \multirow{2}{*}[-1pt]{$\oplus 1.697-2 $} & $\oplus 0.570-0.878$ & \multirow{2}{*}[-1pt]{$\oplus 0.537-0.730$}\\
     &&&&&&$\oplus 0.945-1$&\\
  \cline{2-8}
                 & \multirow{3}{*}{$\frac{3\pi}{7}$} &\multirow{3}{*}{$8.091-8.979$} & \multirow{3}{*}{$33.995-36.031$} & \multirow{3}{*}{$40.280-45.862$} & \multirow{1}{*}[-5pt]{$0-0.304$} & \multirow{1}{*}[-5pt]{$0-0.505$} & $0-0.131$ \\
        &&&&& \multirow{2}{*}[-1pt]{$\oplus 1.697-2 $} & \multirow{2}{*}[-1pt]{$\oplus  0.971-1$} & $\oplus  0.345-0.636$\\
     &&&&&&&$\oplus  0.841-1$\\
  \hline
\hline
\multirow{8}{*}{$U_{II,2}$}
                 & \multirow{2}{*}{0}&\multirow{2}{*}{$8.091-8.979$} & \multirow{2}{*}{$31.435-36.031$} & \multirow{2}{*}{$45.641-51.531$} & $0-0.463$ & $0-0.252$ & $0-0.162$ \\
        &&&&& $\oplus 1.537-2 $ & $\oplus 0.748-1$ & $\oplus 0.838-1$\\
  \cline{2-8}
                 & \multirow{2}{*}{$\frac{\pi}{7}$} & \multirow{2}{*}{$8.091-8.979$} & \multirow{2}{*}{$31.435-36.031$} & \multirow{2}{*}{$45.642-51.531$} & $0-0.464$ & $0-0.783$ & \multirow{2}{*}{$0.178-0.822 $} \\
        &&&&& $\oplus 1.536-2$ & $\oplus 0.830-1$ &\\
  \cline{2-8}
                 & \multirow{2}{*}{$\frac{2\pi}{7}$} & \multirow{2}{*}{$8.091-8.979$} & \multirow{2}{*}{$31.435-36.031$} & \multirow{2}{*}{$45.633-51.531$} & $0-0.464$ & $0-0.509$ & $0.099-0.488$ \\
        &&&&& $\oplus 1.536-2$ & $\oplus 0.667-1$ &$\oplus 0.512-0.901$\\
\cline{2-8}
                 & \multirow{2}{*}{$\frac{3\pi}{7}$} &\multirow{2}{*}{$8.091-8.979$} & \multirow{2}{*}{$34.125-36.031$} & \multirow{2}{*}{$45.621-47.050$} & $0-0.212$ & $0-0.561$ & $0-0.166$ \\
        &&&&& $\oplus 1.788-2$ & $\oplus 0.963-1$ &$\oplus 0.834-1$\\
\hline
\hline
\end{tabular}
\caption{\label{Tab:U_{I}_n=7_and_U_{II}_n=7}
The ranges of the mixing parameters for the viable cases of $U_{I}$ and $U_{II}$ with the group index $n=7$, where the constraints imposed are the experimental values at $3\sigma$ for the mixing angles~\cite{Esteban:2016qun}. }
\end{table}

\clearpage

\clearpage

\section{\label{sec:Lepton_flavor_mixing_OneCP}Lepton flavor mixing from single residual CP transformation in the neutrino sector}

We have discussed both lepton and quark mixing patterns which arise from the residual symmetries $Z_{2}$ and $Z_{2}\times CP$ in previous sections. The resulting mixing matrices depend on three free real parameters. In Ref.~\cite{Lu:2016jit}, we have explored another proposal in which the charged lepton and neutrino mass matrices are invariant under the action of a residual abelian subgroup $G_{l}$ and a single CP transformation $X_{\nu}$ respectively. The lepton mixing matrix would depend on three free real parameters as well. The general form of the PMNS matrix in this scheme has been given in~\cite{Lu:2016jit} as follow
\begin{equation}
\label{eq:U_PMNS_onecp}
U_{PMNS}=P_{l}U_{l}^{\dagger}\Sigma_{\nu}O_{3}Q_{\nu}\,,
\end{equation}
where $P_{l}$ is a three dimensional permutation matrix and $Q_{\nu}$ is a diagonal matrix with entries $\pm1$ and $\pm i$. The unitary transformation $U_{l}$ diagonalizes the representation matrix of the generator of $G_{l}$,
\begin{equation}
U^{\dagger}_{l}\rho_{\mathbf{3}}(g_{l})U_{l}=\rho^{\text{diag}}_{\mathbf{3}}(g_{l})\,,
\end{equation}
where $\rho^{\text{diag}}_{\mathbf{3}}(g_{l})$ is a diagonal phase matrix.  $\Sigma_{\nu}$ is the Takagi factorization matrix of $X_{\nu}$ and it satisfies
\begin{equation}
X_{\nu}=\Sigma_{\nu}\Sigma_{\nu}^{T}\,.
\end{equation}
Moreover, $O_{3}$ in Eq.~\eqref{eq:U_PMNS_onecp} is a generic $3\times3$ real orthogonal matrix, and we shall parameterize $O_{3}$ as
\begin{equation}
\label{eq:O33_form}
O_{3}=\left(\begin{array}{ccc}
 1 & 0 & 0 \\
 0 & \cos\theta_{1} & \sin\theta_{1} \\
                       0 & -\sin\theta_{1} & \cos\theta_{1}\\ \end{array} \right)
 \left(\begin{array}{ccc}
 \cos\theta_{2} & 0 & \sin\theta_{2} \\
 0 & 1 & 0 \\
                       -\sin\theta_{2} & 0 & \cos\theta_{2}\\ \end{array} \right)
 \left(\begin{array}{ccc}
 \cos\theta_{3} & \sin\theta_{3} & 0 \\
 -\sin\theta_{3} & \cos\theta_{3} & 0 \\
 0 & 0 & 1\\ \end{array} \right)\,,
\end{equation}
where the variation range of the free parameters $\theta_{1,2,3}$ can
be taken to be $[0,\pi)$. Let us consider two distinct residual symmetries $(G_{l}, X_{\nu})$ and $(G'_{l}, X'_{\nu})$, the lepton mixing matrices read
\begin{equation}
U_{PMNS}=P_{l}U_{l}^{\dagger}\Sigma_{\nu}O_{3}Q_{\nu},~~~~U'_{PMNS}=P'_{l}U'^{\dagger}_{l}\Sigma'_{\nu}O'_{3}Q'_{\nu}\,.
\end{equation}
The necessary and sufficient condition that $U_{PMNS}$ and $U'_{PMNS}$ describe the same lepton mixing pattern is~\cite{Lu:2016jit}
\begin{equation}
UU^{T}=Q_{L}P_{L}U'U'^{T}P^{T}_{L}Q_{L}\,,
\end{equation}
where $U\equiv U_{l}^{\dagger}\Sigma_{\nu}$, $U'\equiv U'^{\dagger}_{l}\Sigma'_{\nu}$, $P_{L}$ is a permutation matrix and $Q_{L}$
is an arbitrary phase matrix.

The possible mixing patterns which arise from the breaking of $\Delta(6n^{2})$ flavor group and CP symmetry into a single remnant CP transformation in neutrino sector and an abelian subgroup in the charged lepton sector have been studied~\cite{Li:2017zmk}, the analytical expressions of the mixing matrix and the mixing parameters have been presented in~\cite{Li:2017zmk}.
In order to study the mixing patterns comprehensively for all admitted $G_{l}$ and $X_{\nu}$, it is sufficient to only consider six types of residual symmetries~\cite{Li:2017zmk}. In the following, we shall briefly review the possible lepton mixing matrices which can be obtained from $\Delta(6n^2)$ and CP in this approach.

\begin{itemize}[labelindent=-0.5em, leftmargin=2.2em]

\item[(I)]{$G_{l}=\braket{c^{s}d^{t}},X_{\nu}=c^{x}d^{y}$}

In this case the lepton mixing matrix is of the form~\cite{Li:2017zmk}
\begin{equation}
U'_{I}=O_{3}Q_{\nu}\,,
\end{equation}
where the row permutation $P_{l}$ can be absorbed into the orthogonal matrix $O_3$. Obviously the Dirac CP phase is trivial for this mixing pattern, and consequently it is disfavored by the latest experimental evidence for maximal $\delta_{CP}\sim3\pi/2$~\cite{Abe:2017uxa,Adamson:2017gxd}.

\item[(II)]{$G_{l}=\braket{c^{s}d^{t}},X_{\nu}=bc^{x}d^{-x}$}

The lepton mixing matrix is determined to be
\begin{equation}
U'_{II}=\frac{1}{\sqrt{2}}\left(\begin{array}{ccc}
0 & -i & 1\\
\sqrt{2} & 0 & 0 \\
0 & i & 1 \\ \end{array} \right) O_{3}Q_{\nu}\,,
\end{equation}
up to possible row permutations. The matrix $U_{III}$ satisfies
\begin{equation}
P_{13}U'_{II}(\theta_{1},\theta_{2},\theta_{3})=U'_{II}(-\theta_{1},\theta_{2},-\theta_{3})\text{diag}(1,-1,1)\,.
\end{equation}
Hence all possible permutations lead to three independent mixing matrices,
\begin{equation}
U'_{II,1}=U'_{II},\qquad U'_{II,2}=P_{12}U'_{II},\qquad U'_{II,3}=P_{23}U'_{II}\,.
\end{equation}
We find that $U'_{II,1}$ and $U'_{II,3}$ predict $\tan\theta_{13}=\cos\theta_{23}$ and $\tan\theta_{13}=\sin\theta_{23}$ respectively such that the experimental data~\cite{Esteban:2016qun} of the mixing angles $\theta_{13}$ and $\theta_{23}$ can not be accommodated simultaneously. For the mixing matrix $U'_{II,2}$, the lepton mixing parameters are given by
\begin{eqnarray}
\nonumber&&\sin^2\theta_{13}=\sin ^2\theta_{2}, \quad \sin^2\theta_{12}=\sin ^2\theta_{3} , \quad \sin^2\theta_{23}=\frac{1}{2}\,,\\
\label{eq:mixing_para_caseII}&&J_{CP}=\frac{1}{8}\cos \theta_{2}\sin 2\theta_{2} \sin 2\theta_{3}, \quad |\sin\delta|=1\,.
\end{eqnarray}
Notice that both $\theta_{23}$ and $\delta_{CP}$ are maximal.

\item[(III)]{$G_{l}=\braket{bc^{s}d^{t}},~X_{\nu}=abc^{x}d^{2x}$}

For this case, the PMNS mixing matrix is given by
\begin{equation}
U'_{III}=\frac{1}{2}\left(\begin{array}{ccc}
 -\sqrt{2} & -i & 1 \\
 \sqrt{2} & -i & 1 \\
 0 & i\sqrt{2} & \sqrt{2} \\ \end{array} \right) O_{3}Q_{\nu}\,,
\end{equation}
which fulfills the equality
\begin{equation}
P_{12}U'_{III}(\theta_{1},\theta_{2},\theta_{3})=U'_{III}(\theta_{1},-\theta_{2},-\theta_{3})\text{diag}(-1,1,1)\,.
\end{equation}
As a result, the six possible row permutations lead to three independent mixing patterns which can be chosen as
\begin{equation}
U'_{III,1}=U'_{III}\,,~~~~ U'_{III,2}=P_{23}U'_{III}\,,~~~~ U'_{III,3}=P_{13}U'_{III}\,.
\end{equation}

\item[(IV)]{$G_{l}=\braket{bc^{s}d^{t}},~X_{\nu}=c^{x}d^{y}$}

Up to possible permutation of rows, the lepton mixing matrix is determined to be
\begin{equation}
\label{eq:onecp_lepton_PMNS_5}
U'_{IV}=\frac{1}{\sqrt{2}}\left(\begin{array}{ccc}
1 & 0 & -e^{i\rho_{1}} \\
1 & 0 & e^{i\rho_{1}} \\
0 & \sqrt{2} & 0 \\ \end{array} \right)O_{3}Q_{\nu}\,.
\end{equation}
where $\rho_{1}=(x+y+s+t)\pi/ n$ and it can take the following discrete values
\begin{equation}
\rho_{1}~(\text{mod}~2\pi)=0,\frac{1}{n}\pi,\frac{2}{n}\pi,\cdots,\frac{2n-1}{n}\pi\,.
\end{equation}
The mixing matrix $U'_{IV}$ has the following properties
\begin{eqnarray}
\nonumber P_{12}U'_{IV}(\rho_{1},\theta_{1},\theta_{2},\theta_{3})&=&U'_{IV}(\rho_{1}+\pi,\theta_{1},\theta_{2},\theta_{3})\,,\\
\nonumber U'_{IV}(\rho_{1}+\pi,\theta_{1},\theta_{2},\theta_{3})&=&U'_{IV}(\rho_{1},-\theta_{1},-\theta_{2},\theta_{3})\text{diag}(1,1,-1)\,,\\
U'_{IV}(\pi-\rho_{1},\theta_{1},\theta_{2},\theta_{3})&=&\text{diag}(-e^{-i\rho_{1}},e^{-i\rho_{1}},1)U'_{IV}(\rho_{1},\theta_{1}',\theta_{2}',\theta_{3}')\text{diag}(1,1,-1)\,,
\end{eqnarray}
where the parameters $\theta'_{1,2,3}$ fulfill $O_{3}(\theta_{1}',\theta_{2}',\theta_{3}')=P_{13}O_{3}(-\theta_{1},-\theta_{2},\theta_{3})$. As a consequence, the parameter $\rho_{1}$ can be limited in the range of $0\leq \rho_{1} \leq \frac{\pi}{2}$ and we only need to consider the following three row permutations of $U'_{IV}$,
\begin{equation}
U'_{IV,1}=U'_{IV},\qquad U'_{IV,2}=P_{23}U'_{IV},\qquad U'_{IV,3}=P_{13}U'_{IV}\,.
\end{equation}

\item[(V)]{$G_{l}=\braket{ac^{s}d^{t}},~X_{\nu}=bc^{x}d^{-x}$}

In this case, the lepton mixing matrix take the following form
\begin{equation}
U'_{V}=\sqrt{\frac{2}{3}}\left(\begin{array}{ccc}
\frac{e^{i\rho_{2}}}{\sqrt{2}} & 0 & 1 \\
-\frac{e^{i\rho_{2}}}{\sqrt{2}} & \cos\frac{\pi}{6} & \sin\frac{\pi}{6} \\
\frac{e^{i\rho_{2}}}{\sqrt{2}} & \cos\frac{\pi}{6} & -\sin\frac{\pi}{6} \\ \end{array} \right) O_{3}Q_{\nu}\,,
\end{equation}
with $\rho_{2}=-(3x+s-2t)\pi/n$. All of the six possible row permutations lead to the same mixing pattern because $U'_{V}$ has the following symmetry properties,
\begin{eqnarray}
\nonumber P_{23}U'_{V}(\rho_{2},\theta_{1},\theta_{2},\theta_{3})&=&\text{diag}(1,-1,-1)U'_{V}(\rho_{2},-\theta_{1},\theta_{2},-\theta_{3})\text{diag}(1,-1,1)\,,\\
\nonumber P_{12}P_{23}U'_{V}(\rho_{2},\theta_{1},\theta_{2},\theta_{3})&=&\text{diag}(1,-1,-1)U'_{V}(\rho_{2},\theta_{1}-\frac{2\pi}{3},\theta_{2},\theta_{3})\,,\\
P_{13}P_{23}U'_{VI}(\rho_{2},\theta_{1},\theta_{2},\theta_{3})&=&\text{diag}(-1,-1,1)U'_{V}(\rho_{2},\theta_{1}+\frac{2\pi}{3},\theta_{2},\theta_{3})\,.
\end{eqnarray}
In addition, we find that $U'_{V}$ satisfies
\begin{eqnarray}
U'_{V}(\rho_{2}+\pi,\theta_{1},\theta_{2},\theta_{3})&=&U'_{V}(\rho_{2},\theta_{1},-\theta_{2},-\theta_{3})\text{diag}(-1,1,1)\,,\\
U'_{V}(\pi-\rho_{2},\theta_{1},\theta_{2},\theta_{3})&=&U'^{*}_{V}(\rho_{2},\theta_{1},-\theta_{2},-\theta_{3})\text{diag}(-1,1,1)\,.
\end{eqnarray}
As a result, the fundamental region of $\rho_{2}$ can be chosen to be $[0,\pi)$.

\item[(VI)]{$G_{l}=\braket{ac^{s}d^{t}},~X_{\nu}=c^{x}d^{y}$}

Up to possible permutation of rows, the lepton PMNS mixing matrix reads
\begin{equation}
U'_{VI}=\frac{1}{\sqrt{3}}\left(\begin{array}{ccc}
1 ~&~ e^{i\rho_{3}} ~&~ e^{i\rho_{4}} \\
1 ~&~ \omega^{2}e^{i\rho_{3}} ~&~ \omega e^{i\rho_{4}} \\
1 ~&~ \omega e^{i\rho_{3}} ~&~ \omega^{2} e^{i\rho_{4}} \\ \end{array}\right)O_{3}Q_{\nu}\,,
\end{equation}
with
\begin{equation}
\rho_{3}=\frac{2(y+t)-x}{n}\pi,~~\rho_{4}=\frac{x+y+2s}{n}\pi\,.
\end{equation}
The parameters $\rho_{3}$ and $\rho_{4}$ can take the following discrete values
\begin{equation}
\rho_{3},\rho_{4}~(\text{mod}~2\pi)=0,\frac{1}{n}\pi,\frac{2}{n}\pi,\cdots,\frac{n-1}{n}\pi\,.
\end{equation}
The mixing matrix $U'_{VI}$ has the following properties
\begin{eqnarray}
\nonumber U'_{VI}(\rho_{3}+\pi,\rho_{4},\theta_{1},\theta_{2},\theta_{3})&=& U'_{VI}(\rho_{3},\rho_{4},-\theta_{1},\theta_{2},-\theta_{3})\text{diag}(1,-1,1)\,, \\
\nonumber U'_{VI}(\rho_{3},\rho_{4}+\pi,\theta_{1},\theta_{2},\theta_{3})&=& U'_{VI}(\rho_{3},\rho_{4},-\pi-\theta_{1},\theta_{2},-\theta_{3})\text{diag}(1,-1,1)\,, \\
\nonumber P_{23}U'_{VI}(\rho_{3},\rho_{4},\theta_{1},\theta_{2},\theta_{3})&=&U'_{VI}(\rho_{4},\rho_{3},\pi/2-\theta_{1},\theta_{2},-\theta_{3})\text{diag}(1,-1,1)\,, \\
U'^{*}_{VI}(\rho_{3},\rho_{4},\theta_{1},\theta_{2},\theta_{3})&=&P_{23}U'_{VI}(\pi-\rho_{3},\pi-\rho_{4},\theta_{1}+\pi,\theta_{2},\theta_{3})\,.
\end{eqnarray}
Consequently the parameters $\rho_{3}$ and $\rho_{4}$ can be limited in the range $0\leq\rho_3,\rho_4<\pi$ without loss of generality, and three out of the six possible row permutations are independent if all possible values of $\rho_{3}$ and $\rho_{4}$ are considered,
\begin{equation}
U'_{VI,1}=U'_{VI}\,,~~~ U'_{VI,2}=P_{12}U'_{VI}\,, ~~~ U'_{VI,3}=P_{23}P_{12}U'_{VI}\,.
\end{equation}

As regards the predictions for the mixing parameters for each cases, we refer the reader to~\cite{Li:2017zmk}.

\end{itemize}

\begin{figure}[t!]
\centering
\includegraphics[width=0.98\textwidth]{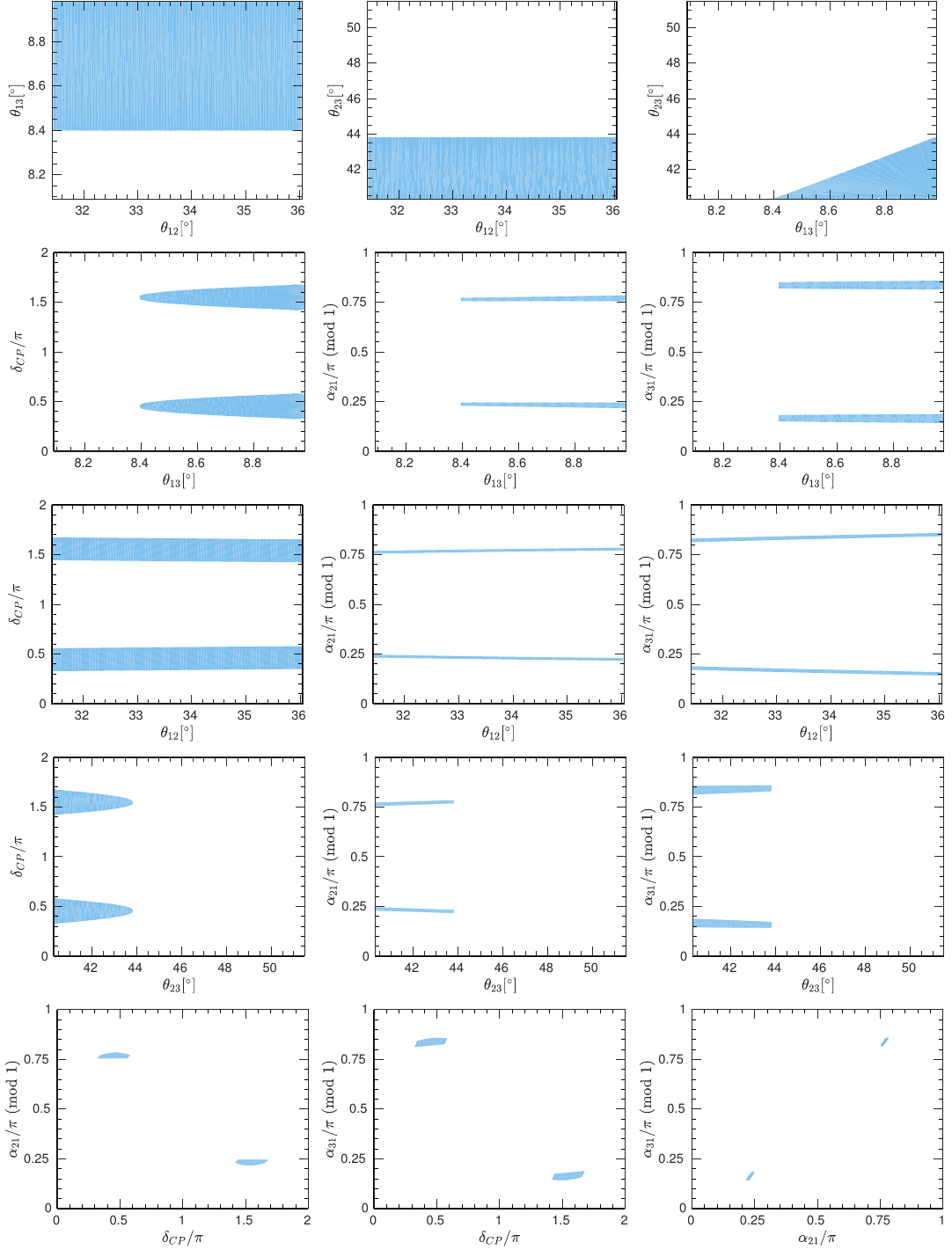}
\caption{
Correlations between different mixing parameters for the mixing pattern $U'_{IV, 1}$ with $\rho_1=\pi/7$,
where the $\Delta(6n^2)$ and CP symmetries are broken to the residual symmetry $G_{l}=\braket{bc^{s}d^{t}},~X_{\nu}=c^{x}d^{y}$,
the three lepton mixing angles are required to be compatible with the experimental data at $3\sigma$ level~\cite{Esteban:2016qun}.
}
\label{fig:onecp_lepton_example}
\end{figure}

In the following, we shall focus on the $\Delta(6\cdot 7^2)=\Delta(294)$ flavor group. We shall perform a numerical analysis for all above cases of PMNS mixing matrices by treating the free parameters $\theta_{1,2,3}$ as random numbers in the range of $[0,\pi]$ and all possible values of the discrete parameters for $n=7$ will be considered.
The reason why we take $n=7$ is that the value of index $n$ of $\Delta(6n^{2})$ has to be at least $n=7$ in order to simultaneously describe both quark and lepton flavor mixings if a single CP is preserved.
The analysis of quark flavor mixing with single residual CP is given in section~\ref{sec:onecp_quark}. The predictions for the three mixing angles $\theta_{13}$, $\theta_{12}$ and $\theta_{23}$ as well as CP violating phases $\delta_{CP}$, $\alpha_{21}$ and $\alpha_{31}$ are studied.
In table~\ref{table:onecp_lepton_123} and table~\ref{table:onecp_lepton_4}, we summarize the allowed ranges of the mixing parameters for all the phenomenological viable cases which can be obtained from the $\Delta(294)$ group. In particular, we notice that for $U'_{IV,1}$ with $\rho_{1}=\pi/7$ the Dirac CP phase $\delta_{CP}$ is predicted to be around $\pi/2$ and $3\pi/2$ which are favored by the present experimental data~\cite{Abe:2017uxa,Adamson:2017gxd}.
The explicit form of this mixing matrix has been given in Eq.~\eqref{eq:onecp_lepton_PMNS_5}, then we can straightforwardly extract the mixing parameters and find
\begin{eqnarray}
 &&\sin^{2}\theta_{13}=\frac{1}{2}(\sin^{2}\theta_{2}+\cos^{2}\theta_{1}\cos^{2}\theta_{2}-\cos\theta_{1}\sin2\theta_{2}\cos\rho_{1})\,,\\
 &&\sin^{2}\theta_{12}=\sin^{2}\theta_{13}+\frac{\sin2\theta_{3}\sin\theta_{1}(\cos\theta_{2}\cos\rho_{1}+\cos\theta_{1}\sin\theta_{2})+\sin^{2}\theta_{1}\cos2\theta_{3}}{2-\sin^{2}\theta_{2}-\cos^{2}\theta_{1}\cos^{2}\theta_{2}+\cos\theta_{1}\sin2\theta_{2}\cos\rho_{1}}\,,\\
 &&\sin^{2}\theta_{23}=1-\frac{2\sin^{2}\theta_{1}\cos^{2}\theta_{2}}{2-\sin^{2}\theta_{2}-\cos^{2}\theta_{1}\cos^{2}\theta_{2}+\cos\theta_{1}\sin2\theta_{2}\cos\rho_{1}}\,,\\
 &&J_{CP}=-\frac{1}{4}\sin\theta_{1}\cos\theta_{2}\sin\rho_{1}[\sin2\theta_{1}\sin\theta_{2}\cos2\theta_{3}+\sin2\theta_{3}(\cos^{2}\theta_{1}-\sin^{2}\theta_{1}\sin^{2}\theta_{2})]\,.
 \end{eqnarray}
The correlations among the mixing angles and CP phases are presented in figure \ref{fig:onecp_lepton_example}. From figure~\ref{fig:onecp_lepton_example} and table~\ref{table:onecp_lepton_123} we see that the approximately full $3\sigma$ region of $\theta_{12}$ can be achieved, the reactor mixing angle $\theta_{13}$ lies in the interval $[8.396^{\circ}, 8.979^{\circ}]$, and the atmospheric angle $\theta_{23}\in[40.281^{\circ}, 43.806^{\circ}]$ is predicted to be in the first octant. For the CP violation phases, the Dirac CP phase $\delta_{CP}$ is around $0.5\pi$ and $1.5\pi$ which are favored by present experimental data~\cite{Abe:2017uxa,Adamson:2017gxd}. The Majorana phases are strongly constrained, and they are determined to be in the ranges $\alpha_{21}\, (\text{mod} \pi)\in[0.216, 0.246]\cup[0.754, 0.784]$ and $\alpha_{31}\, (\text{mod} \pi) \in [0.142, 0.188]\cup[0.812, 0.858]$.

\begin{figure}[hptb!]
\centering
\includegraphics[width=0.6\textwidth]{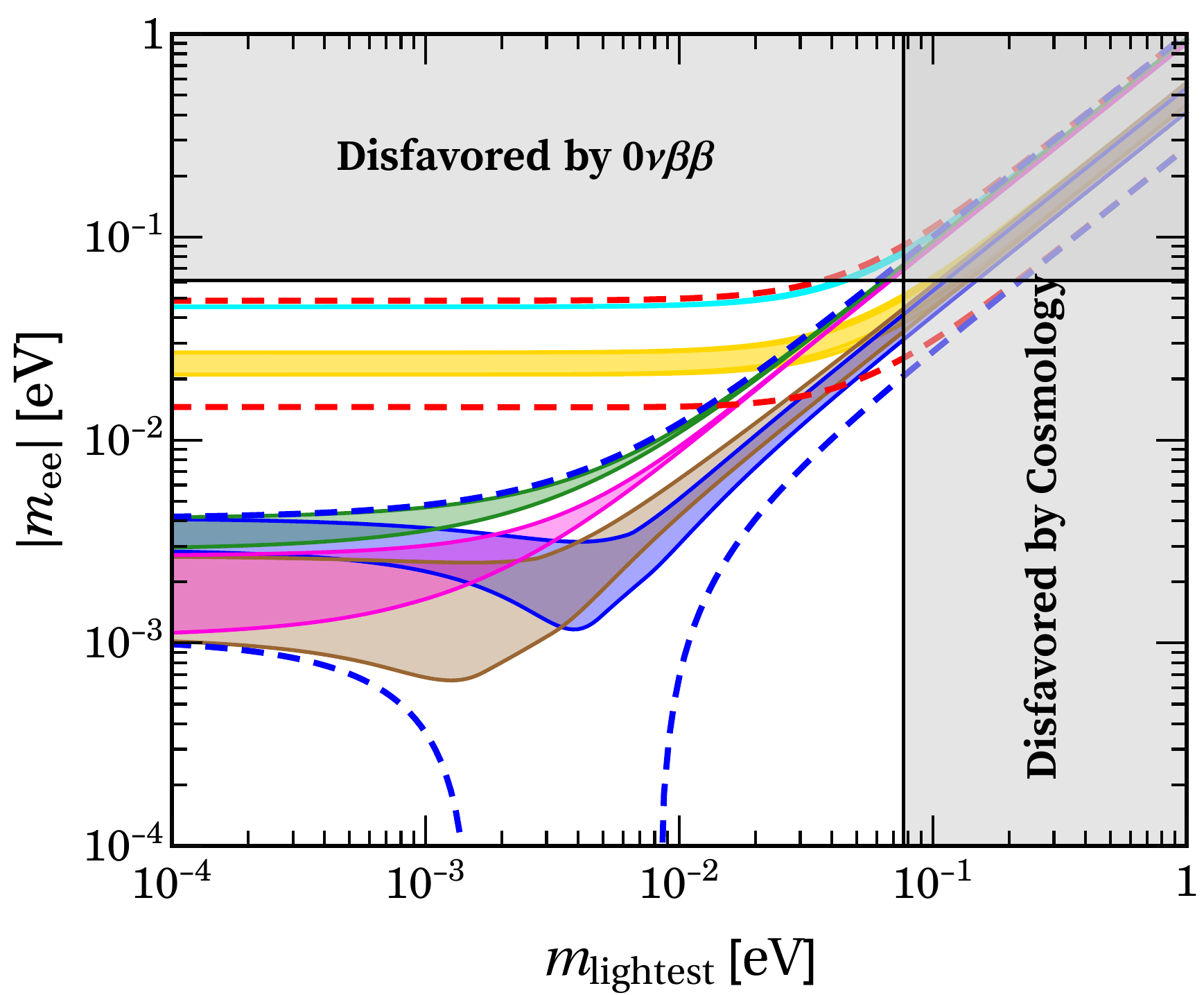}
\caption{\label{fig:caseV1_z2z2cp_nldb}The allowed regions of the effective Majorana mass $|m_{ee}|$ with respect to the lightest neutrino mass for the mixing pattern $U'_{IV,1}$ with $\rho_{1}=\pi/7$. Here we adopt the same conventions as figure~\ref{fig:0nubb_CaseI_456}.}
\end{figure}

Furthermore, we can obtain predictions for the neutrinoless double beta decay effective Majorana mass $|m_{ee}|$ as a function of the lightest neutrino mass $m_{\text{lightest}}$. We display the attainable values
of $|m_{ee}|$ for the mixing pattern $U'_{IV,1}$ with $\rho_{1}=\pi/7$ in figure~\ref{fig:caseV1_z2z2cp_nldb}. Notice that there is no cancellation in $|m_{ee}|$ for any values of $m_{\text{lightest}}$ in the case
of NO, and thus $|m_{ee}|$ has a lower bound $|m_{ee}|\geq6.543\times 10^{-4}$ eV.

\begin{table}[!hbp]
\centering
\footnotesize
\begin{tabular}{|c |c| c c c c c c|}
  \hline
  \hline
  \multicolumn{2}{|c|}{case} &$\theta_{13}/^{\circ}$ & $\theta_{12}/^{\circ}$ & $\theta_{23}/^{\circ}$ & $\delta_{CP}/\pi$ & $\alpha_{21}/\pi(\text{mod}~1)$ & $\alpha_{31}/\pi(\text{mod}~1)$ \\
  \hline
 \multicolumn{2}{|c|}{$U'_{II,2}$} &$8.091-8.979$ & $31.435-36.031$ & $45$ & $0.5,1.5$ & $0$ & $0$ \\
 \hline
\multicolumn{2}{|c|}{\multirow{2}{*}{ $U'_{III,1}$ }} &\multirow{2}{*}{$8.091-8.979$} & \multirow{2}{*}{$31.435-36.031$} & \multirow{2}{*}{$49.525-51.530$} & $0-0.757$ & $0.125-0.299$ & \multirow{2}{*}{$0.169-0.831$} \\
  \multicolumn{2}{|c|}{}  &&&& $\oplus 1.243-2$ & $\oplus 0.701-0.876$ & \\
\hline
\hline
case & $\rho_{1}$ &$\theta_{13}/^{\circ}$ & $\theta_{12}/^{\circ}$ & $\theta_{23}/^{\circ}$ & $\delta_{CP}/\pi$ & $\alpha_{21}/\pi~(\text{mod}~1)$ & $\alpha_{31}/\pi~(\text{mod}~1)$ \\
\hline
  \multirow{4}{*}{$U'_{IV,1}$ }
     & 0 & $8.091-8.979$ & $31.435-36.031$ & $40.282-51.530$ & $0,1$ & $0$ & $0$ \\
\cline{2-8}
& \multirow{2}{*}{$\frac{\pi}{7}$} & \multirow{2}{*}{$8.396-8.979$} & \multirow{2}{*}{$31.435-36.031$} & \multirow{2}{*}{$40.281-43.806$} & $0.326-0.580$ & $0.216-0.246$ & $0.142-0.188$ \\
     &&&&& $\oplus 1.421-1.675 $ & $\oplus 0.754-0.784$ & $\oplus 0.812-0.858$\\
  \hline
\multirow{11}{*}{$U'_{IV, 3}$}
                & 0 & $8.091-8.979$ & $31.435-36.031$ & $40.282-53.531$ & $0,1$ & $0$ & $0$ \\
\cline{2-8}
& \multirow{4}{*}{$\frac{\pi}{7}$} & \multirow{4}{*}{$8.091-8.979$} & \multirow{4}{*}{$31.435-36.031$} & \multirow{4}{*}{$40.281-51.529$} & $0.143-0.147$ & \multirow{4}{*}{$0$} & \\
     &&&&& $\oplus 0.853-0.857$ & &$0.286-0.294 $\\
     &&&&& $\oplus 1.143-1.147$ & &$\oplus 0.706-0.714 $\\
     &&&&& $\oplus 1.853-1.857$ & &\\
  \cline{2-8}
& \multirow{4}{*}{$\frac{2\pi}{7}$} & \multirow{4}{*}{$8.091-8.979$} & \multirow{4}{*}{$31.435-36.031$} & \multirow{4}{*}{$40.282-51.530$} & $0.286-0.296$ & \multirow{4}{*}{$0$} & \\
     &&&&& $\oplus 0.704-0.714$ & & $0.407-0.429 $ \\
     &&&&& $\oplus1.286-1.296$ & & $\oplus 0.571-0.593$\\
     &&&&& $\oplus1.704-1.714$ & &\\
 \cline{2-8}
& \multirow{2}{*}{$\frac{3\pi}{7}$} &\multirow{2}{*}{$8.091-8.979$} & \multirow{2}{*}{$31.435-36.031$} & \multirow{2}{*}{$40.280-51.429$} & $0.429-0.571$ & \multirow{2}{*}{$0$} & $0-0.143$ \\
        &&&&& $\oplus 1.429-1.571$ & &$\oplus 0.857-1$\\
\hline
  \hline
case & $\rho_{2}$ &$\theta_{13}/^{\circ}$ & $\theta_{12}/^{\circ}$ & $\theta_{23}/^{\circ}$ & $\delta_{CP}/\pi$ & $\alpha_{21}/\pi~(\text{mod}~1)$ & $\alpha_{31}/\pi~(\text{mod}~1)$ \\
\hline
\multirow{9}{*}{$U'_{V,1}$}
     & 0 & $8.091-8.979$ & $31.435-36.031$ & $40.281-51.530$ & $1$ & $0$ & $0$ \\
\cline{2-8}
& \multirow{3}{*}{$\frac{\pi}{7}$} & \multirow{3}{*}{$8.091-8.979$} & \multirow{3}{*}{$31.435-36.031$} & \multirow{3}{*}{$40.283-51.530$} & $0-0.016$ & \multirow{1}{*}[-5pt]{$0.283-0.308$} & $0-0.037$ \\
&&&&& $\oplus 0.760-1.032 $ & \multirow{2}{*}[-1pt]{$\oplus 0.692-0.717$} & $\oplus 0.567-0.620$\\
&&&&& $\oplus 1.778-2 $ & & $\oplus 0.982-1$\\
\cline{2-8}
& \multirow{3}{*}{$\frac{2\pi}{7}$} & \multirow{3}{*}{$8.091-8.979$} & \multirow{3}{*}{$31.435-36.031$} & \multirow{3}{*}{$40.281-51.529$} & $0-0.046$ & \multirow{1}{*}[-5pt]{$0.369-0.442$} & \multirow{1}{*}[-5pt]{$0-0.193$} \\
&&&&& $\oplus 0.399-1.086 $ & \multirow{2}{*}[-1pt]{$\oplus 0.558-0.631$} & \multirow{2}{*}[-1pt]{$\oplus 0.946-1$}\\
&&&&& $\oplus 1.443-2 $ & & \\
 \cline{2-8}
& \multirow{2}{*}{$\frac{3\pi}{7}$} & \multirow{2}{*}{$8.092-8.979$} & \multirow{2}{*}{$32.310-36.031$} & \multirow{2}{*}{$40.284-51.530$} & \multirow{2}{*}{$0-2$} & $0-0.203$ & $0-0.599$ \\
&&&&& & $\oplus 0.797-1$ & $\oplus 0.873-1$ \\
\hline\hline
\end{tabular}
\caption{
  The ranges of the mixing parameters for the mixing patterns $U'_{II,2},U'_{III,1},U'_{IV,1}, U'_{IV, 3}$ and $U'_{V,1}$ with the group index   $n=7$, where the constraints imposed are the experimental values at $3\sigma$ for the mixing angles~\cite{Esteban:2016qun}.
  \label{table:onecp_lepton_123}}
\end{table}

\begin{table}[!hbp]
\centering
\footnotesize
\begin{tabular}{|c |c| c| c c c c c c|}
\hline
  \hline
  case & $\rho_{3}$ & $\rho_{4}$ &$\theta_{13}/^{\circ}$ & $\theta_{12}/^{\circ}$ & $\theta_{23}/^{\circ}$ & $\delta_{CP}/\pi$ & $\alpha_{21}/\pi~(\text{mod}~1)$ & $\alpha_{31}/\pi~(\text{mod}~1)$ \\
\hline
  \multirow{17}{*}{$U'_{VI,1}$}
     & $0$ & $0$ & $8.091-8.979$ & $31.435-36.031$ & $45$ & $0.5,1.5$ & $0$ & $0$ \\
\cline{2-9}
    & \multirow{3}{*}{$0$} & \multirow{3}{*}{$\frac{\pi}{7}$} & \multirow{3}{*}{$8.091-8.979$} & \multirow{3}{*}{$31.435-36.031$} & \multirow{3}{*}{$40.280-51.531$} & \multirow{1}{*}[-5pt]{$0.246-0.689$} & \multirow{1}{*}[-5pt]{$0.243-0.308$} & $0-0.094$ \\
     &&&&&& \multirow{2}{*}[-1pt]{$\oplus 1.196-1.785 $} & \multirow{2}{*}[-1pt]{$\oplus 0.692-0.757$} & $\oplus 0.577-0.741$\\
       &&&&&& & & $\oplus 0.996-1$\\
\cline{2-9}
    & \multirow{2}{*}{$0$} & \multirow{2}{*}{$\frac{2\pi}{7}$} & \multirow{2}{*}{$8.091-8.979$} & \multirow{2}{*}{$31.435-36.031$} & \multirow{2}{*}{$40.281-51.530$} & $0.046-0.712$ & $0.369-0.453$ & $0-0.253$ \\
     &&&&&& $\oplus 0.993-1.775 $ & $\oplus 0.547-0.631$ & $\oplus 0.987-1$\\
 \cline{2-9}
& \multirow{3}{*}{$0$} & \multirow{3}{*}{$\frac{3\pi}{7}$} & \multirow{3}{*}{$8.091-8.979$} & \multirow{3}{*}{$33.131-36.031$} & \multirow{3}{*}{$40.282-51.530$} & $0-0.144$ & \multirow{1}{*}[-5pt]{$0-0.176$} & \multirow{1}{*}[-5pt]{$0-0.624$} \\
     &&&&&& $\oplus 0.265-1.075 $ & \multirow{2}{*}[-1pt]{$\oplus 0.824-1$} & \multirow{2}{*}[-1pt]{$\oplus 0.943-1$}\\
       &&&&&& $\oplus 1.332-2 $ & & \\
 \cline{2-9}
    & \multirow{3}{*}{$\frac{\pi}{7}$} & \multirow{3}{*}{$\frac{\pi}{7}$} & \multirow{3}{*}{$8.091-8.979$} & \multirow{3}{*}{$31.435-36.031$} & \multirow{3}{*}{$40.281-51.531$} & \multirow{1}{*}[-5pt]{$0.216-0.805$} & \multirow{1}{*}[-5pt]{$0.243-0.308$} & $0-0.004$ \\
     &&&&&& \multirow{2}{*}[-1pt]{$\oplus 1.311-1.754 $} & \multirow{2}{*}[-1pt]{$\oplus 0.692-0.757$} & $\oplus 0.259-0.423$\\
       &&&&&& & & $\oplus 0.906-1$\\
\cline{2-9}
   & \multirow{2}{*}{$\frac{2\pi}{7}$} & \multirow{2}{*}{$\frac{2\pi}{7}$} & \multirow{2}{*}{$8.091-8.979$} & \multirow{2}{*}{$31.435-36.031$} & \multirow{2}{*}{$40.280-51.531$} & $0.226-1.009$ & $0.369-0.453$ & $0-0.013$ \\
     &&&&&& $\oplus 1.289-1.953 $ & $\oplus 0.547-0.631$ & $\oplus 0.747-1$\\
 \cline{2-9}
& \multirow{3}{*}{$\frac{3\pi}{7}$} & \multirow{3}{*}{$\frac{3\pi}{7}$} & \multirow{3}{*}{$8.091-8.979$} & \multirow{3}{*}{$33.128-36.031$} & \multirow{3}{*}{$40.282-51.531$} & $0-0.668$ & \multirow{1}{*}[-5pt]{$0-0.176$} & \multirow{1}{*}[-5pt]{$0-0.057$} \\
     &&&&&& $\oplus 0.925-1.734 $ & \multirow{2}{*}[-1pt]{$\oplus 0.824-1$} & \multirow{2}{*}[-1pt]{$\oplus 0.376-1$}\\
       &&&&&& $\oplus 1.855-2 $ & & \\
  \hline
\multirow{17}{*}{$U'_{VI,2}$}
& \multirow{2}{*}{$\frac{\pi}{7}$}  & \multirow{2}{*}{$\frac{2\pi}{7}$} & \multirow{2}{*}{$8.091-8.979$} & \multirow{2}{*}{$34.590-36.031$} & \multirow{2}{*}{$40.291-51.505$} & \multirow{2}{*}{$0.451-1.549$} & $0-0.119$ & $0-0.243$ \\
     &&&&&&& $\oplus 0.881-1$ & $\oplus 0.757-1$ \\
  \cline{2-9}
& \multirow{2}{*}{$\frac{\pi}{7}$}  & \multirow{2}{*}{$\frac{3\pi}{7}$} & \multirow{2}{*}{$8.091-8.979$} & \multirow{2}{*}{$33.193-36.031$} & \multirow{2}{*}{$40.281-42.829$} & $0-0.605$ & $0-0.173$ & $0-0.582$ \\
     &&&&&& $1.768-2$ & $0.827-1$ & $0.940-1$ \\
  \cline{2-9}
 & \multirow{2}{*}{$\frac{\pi}{7}$}  & \multirow{2}{*}{$\frac{5\pi}{7}$} & \multirow{2}{*}{$8.091-8.979$} & \multirow{2}{*}{$33.193-36.031$} & \multirow{2}{*}{$40.282-42.859$} & $0-0.234$ & $0-0.173$ & $0-0.060$ \\
     &&&&&& $\oplus 1.394-2$ & $\oplus 0.827-1$ & $\oplus 0.418-1$ \\
  \cline{2-9}
  & \multirow{2}{*}{$\frac{\pi}{7}$}  & \multirow{2}{*}{$\frac{6\pi}{7}$} & \multirow{2}{*}{$8.091-8.979$} & \multirow{2}{*}{$34.586-36.031$} & \multirow{2}{*}{$40.281-51.531$} & \multirow{2}{*}{$0.452-1.549$} & $0-0.120$ & $0-0.242$ \\
     &&&&&& & $\oplus 0.881-1$ & $\oplus 0.758-1$ \\
  \cline{2-9}
  & \multirow{2}{*}{$\frac{2\pi}{7}$}  & \multirow{2}{*}{$\frac{2\pi}{7}$} & \multirow{2}{*}{$8.091-8.979$} & \multirow{2}{*}{$31.435-36.031$} & \multirow{2}{*}{$40.281-51.531$} & $0-1.491$ & $0.164-0.326$ & $0-0.302$ \\
     &&&&&& $\oplus 1.856-2$ & $\oplus 0.674-0.836$ & $\oplus 0.752-1$ \\
  \cline{2-9}
  & \multirow{2}{*}{$\frac{2\pi}{7}$}  & \multirow{2}{*}{$\frac{3\pi}{7}$} & \multirow{2}{*}{$8.091-8.979$} & \multirow{2}{*}{$33.193-36.031$} & \multirow{2}{*}{$40.281-42.828$} & $0-0.233$ & $0-0.173$ & $0-0.060$ \\
     &&&&&& $\oplus 1.396-2$ & $\oplus 0.827-1$ & $\oplus 0.418-1$ \\
  \cline{2-9}
  & \multirow{2}{*}{$\frac{2\pi}{7}$}  & \multirow{2}{*}{$\frac{6\pi}{7}$} & \multirow{2}{*}{$8.091-8.979$} & \multirow{2}{*}{$33.195-36.031$} & \multirow{2}{*}{$40.281-42.858$} & $0-0.604$ & $0-0.173$ & $0-0.582$ \\
     &&&&&& $\oplus 1.768-2$ & $\oplus 0.827-1$ & $\oplus 0.940-1$ \\
  \cline{2-9}
  & \multirow{3}{*}{$\frac{3\pi}{7}$} & \multirow{3}{*}{$\frac{3\pi}{7}$} & \multirow{3}{*}{$8.091-8.979$} & \multirow{3}{*}{$31.435-36.031$} & \multirow{3}{*}{$40.281-42.905$} & $0-0.225$ & \multirow{1}{*}[-5pt]{$0.360-0.451$} & \multirow{1}{*}[-5pt]{$0-0.041$} \\
     &&&&&& $\oplus 1.026-1.521 $ & \multirow{2}{*}[-1pt]{$\oplus 0.549-0.640$} & \multirow{2}{*}[-1pt]{$\oplus 0.798-1$}\\
       &&&&&& $\oplus 1.707-2 $ & & \\
  \hline
  \hline
\end{tabular}
\caption{
The ranges of the mixing parameters for the mixing patterns $U'_{VI,1}$ and $U'_{VI, 2}$ with $n=7$, where the constraints imposed are the experimental values at $3\sigma$ for the mixing angles~\cite{Esteban:2016qun}.}
\label{table:onecp_lepton_4}
\end{table}

\section{\label{sec:onecp_quark}Quark flavor mixing from single residual CP transformation in up or down quark sector}

In this section, we shall investigate whether it is also possible to derive quark mixing in the same way, as presented in section~\ref{sec:Lepton_flavor_mixing_OneCP}. We assume that the parental flavor and CP symmetries are broken down to an abelian subgroup $G_{u}$ in the up quark sector and to a single residual CP transformation $X_{d}$ in the down quark sector. The residual symmetry $G_{u}$ is able to distinguish the three generations of up type quarks and its generator is denoted as $g_{u}$. The invariance of up quark mass matrix $m_{u}$ under the action of $G_{u}$ requires
\begin{equation}
\label{eq:quark_1}
\rho_{\mathbf{3}}^{\dagger}(g_{u})m_{u}^{\dagger}m_{u}\rho_{\mathbf{3}}(g_{u})=m_{u}^{\dagger}m_{u}\,.
\end{equation}
This implies that
\begin{equation}
[\rho_{\mathbf{3}}(g_{u}), m_{u}^{\dagger}m_{u}]=0\,.
\end{equation}
Since $m_{u}^{\dagger}m_{u}$ commutes with $\rho_{\mathbf{3}}(g_{u})$, they are diagonalized by the same unitary transformation.
We can change basis via the unitary transformation $\Sigma_{u}$ such that $\rho_{\mathbf{3}}(g_{u})$ is diagonalized,
\begin{equation}
\label{eq:quark_4}
\Sigma_{u}^{\dagger}\rho_{\mathbf{3}}(g_{u})\Sigma_{u}=\rho^{\text{diag}}_{\mathbf{3}}(g_{u})\,,
\end{equation}
where $\rho^{\text{diag}}_{\mathbf{3}}(g_{u})$ is a diagonal phase matrices. As the order of the up type quark masses is undefined in this approach, the unitary matrix $U_u$ which diagonalizes $m_{u}^{\dagger}m_{u}$, is uniquely determined up to permutations and phases of its column vectors,
\begin{equation}
\label{eq:U_{u}_form1}
U_{u}=\Sigma_{u}P_{u}^{\dagger}Q_{u}^{\dagger}\,,
\end{equation}
where $P_{u}$ is a generic permutation matrix and $Q_{u}$ is a diagonal phase matrix.

In down quark sector, the residual CP symmetry $X_{d}$ constrains the down quark mass matrix $m_d$ as
\begin{equation}
\label{eq:X_{d}_trans}
X_{d}^{\dagger}m_{d}^{\dagger}m_{d}X_{d}=(m_{d}^{\dagger}m_{d})^{*}\,.
\end{equation}
Consequently the diagonalization matrix $U_{d}$ of $m_{d}^{\dagger}m_{d}$ should fulfill the condition
\begin{equation}
\label{eq:U_{d}_X_{d}_relation}
U_{d}^{\dagger}X_{d}U_{d}^{*}=Q_{d}^{*2}\,,
\end{equation}
with $Q_{d}$ is an arbitrary diagonal phase matrix. The residual CP transformation $X_{d}$ is a symmetric unitary matrix, thus one can perform a Takagi factorization
\begin{equation}
X_{d}=\Sigma_{d}\Sigma_{d}^{T}\,.
\end{equation}
The residual CP transformation $X_{d}$ would enforce $U_{d}$ to be of the form
\begin{equation}
\label{eq:U_{d}_form1}
U_{d}=\Sigma_{d}O_{3} Q_{d}\,,
\end{equation}
where $O_{3}$ is a real orthogonal matrix given by Eq.~\eqref{eq:O33_form}. The Cabibb-Kobayashi-Maskawa(CKM) martrix $V_{CKM}$ is a result of the mismatch between $U_{u}$ and $U_{d}$, consequently $V_{CKM}$ is derived as
\begin{equation}
\label{eq:U_CKM_form1}
V_{CKM}=U_{u}^{\dagger}U_{d}=Q_{u}P_{u}\Sigma_{u}^{\dagger}\Sigma_{d}O_{3} Q_{d}\,.
\end{equation}
In the second scenario, the residual symmetries are a single CP transformation $X_{u}$ in up quark sector and an abelian subgroup $G_{d}$ with generator $g_d$ in the down quark sector. Following the same procedures as the above case, we can obtain the general form of $U_{u}$ and $U_{d}$ as follows,
\begin{equation}
U_{u}=\Sigma_{u}O_{3} Q_{u}^{\dagger},~~~~U_{d}=\Sigma_{d}P_{d}Q_{d}\,,
\end{equation}
where $Q_{u},Q_{d}$ are diagonal phase matrices, $P_{d}$ is a permutation matrix, $\Sigma_{u}$ is the Takagi factorization of $X_{u}$ with $X_{u}=\Sigma_{u}\Sigma_{u}^{T}$ and $\Sigma_{d}$ is a diagonalization matrix of $\rho_{\mathbf{3}}(g_{d})$. Hence the quark mixing matrix $V_{CKM}$ is determined to be
\begin{equation}
\label{eq:U_CKM_form2}
V_{CKM}=Q_{u}O_{3}^{T}\Sigma_{u}^{\dagger}\Sigma_{d}P_{d}Q_{d}\,.
\end{equation}
Notice that the diagonal phase matrices $Q_{u}$ and $Q_{d}$ in Eq.~\eqref{eq:U_CKM_form1} and Eq.~\eqref{eq:U_CKM_form2} are unphysical since they can be eliminated by rephasing the up and down quark fields.

\subsection{\label{equivalen_criterion_quark_single_CP}Equivalence condition}

In the following, we shall derive the criterion to determine whether two distinct residual symmetries give rise to the same quark mixing matrix. For two residual symmetries $\{G_{u}, X_{d}\}$ and $\{G'_{u}, X'_{d}\}$ in this scheme, the CKM mixing matrices are given by
\begin{equation}
V_{CKM}=Q_{u}P_{u}\Sigma_{u}^{\dagger}\Sigma_{d}O_{3} Q_{d},\qquad
V'_{CKM}=Q'_{u}P'_{u}\Sigma'^{\dagger}_{u}\Sigma'_{d}O'_{3} Q'_{d}\,.
\end{equation}
For any values of $O_{3}$, $P_{u}$, $Q_{u}$ and $Q_d$, if one can always find solutions for $O'_{3}$, $P'_{u}$, $Q'_{u}$ and $Q'_{d}$,
such that the equality
\begin{equation}
\label{eq:CKM_eq_single}V_{CKM}=V'_{CKM}\,,
\end{equation}
is satisfied, then $V_{CKM}$ and $V'_{CKM}$ would describe the same quark mixing pattern. To be more concrete, the condition of Eq.~\eqref{eq:CKM_eq_single} requires
\begin{equation}
\label{eq:sim_con_single}UO_{3}=Q_{U}P_{U}U'O_{3}'Q_{D}\,,
\end{equation}
where
\begin{equation}
U=U_{u}^{\dagger}\Sigma_{d},~~U'=U_{u}'^{\dagger}\Sigma'_{d}\,,~~ P_{U}=P^{T}_{u}P'_{u},~~Q_{U}= P^{T}_{u}Q_{u}^{\dagger}Q_{u}'P_{u}\,,~~ Q_{D}=Q_{d}'Q_{d}^{\dagger}\,.
\end{equation}
We multiply both sides of Eq.~\eqref{eq:sim_con_single} with their transpose, then obtain
\begin{equation}
\label{eq:CKM_eqv_cond1}
UU^{T}=Q_{U}P_{U}U'O_{3}'Q_{D}^{2}O_{3}'^{T}U'^{T}P^{T}_{U}Q_{U}\,.
\end{equation}
It is remarkable that the left hand side of Eq.~\eqref{eq:CKM_eqv_cond1} is a constant matrix while the right hand side of Eq.~\eqref{eq:CKM_eqv_cond1} contains the orthogonal matrix $O'_{3}$ which depends on three continuous free parameters. Therefore Eq.~\eqref{eq:CKM_eqv_cond1} is satisfied if and only if $Q_{D}^{2}\propto\text{diag}(1, 1, 1)$, and $Q_{D}^{2}$ can be set to be an identity matrix without loss of generality. As a consequence, the equivalence condition of the two mixing patterns in this scheme is
\begin{equation}
\label{eq:CKM_eqv_cond2}
UU^{T}=Q_{U}P_{U}U'U'^{T}P^{T}_{U}Q_{U}\,.
\end{equation}
Conversely if the condition of Eq.~\eqref{eq:CKM_eqv_cond2} is fulfilled for  a permutation matrix $P_{U}$ and a phase matrix $Q_{U}$, $V_{CKM}$ and $V'_{CKM}$ would be the same quark mixing pattern. In the same fashion, we can obtain the necessary and sufficient condition under which two residual symmetries $\{G_{d}, X_{u}\}$ and $\{G'_{d}, X'_{u}\}$ lead to the same quark mixing matrix,
\begin{equation}
\label{eq:CKM_equiv_GdXu}
U^{T}U=Q_{D}P_{D}^{T}U^{'T}U'P_{D}Q_{D}\,,
\end{equation}
where $P_{D}$ is a permutation matrix and $Q_{D}$ is a generic phase matrix.

\subsection{Possible quark mixing patterns from $\Delta(6n^{2})$ and CP symmetries}

If the flavor group $\Delta(6n^{2})$ and CP symmetry are broken down to an abelian subgroup $G_{u}$ and a single CP transformation $X_{d}$ in the up and down quark sectors respectively, the CKM mixing matrix would be of the same from as the PMNS mixing matrix for residual symmetries $G_{u}$ in the charged lepton sector and $X_{d}$ in the neutrino sector except that $Q_{\nu}$ should be replaced with the phase matrix $Q_{d}$. The mixing matrix would become the hermitian conjugate if the residual symmetries of the up and down quark mass matrices are interchanged $G_{u}\rightarrow G_{d}$ and $X_{d}\rightarrow X_{u}$. As shown in section~\ref{sec:Lepton_flavor_mixing_OneCP}, it is sufficient to consider the following six independent combinations of the residual symmetries,
\begin{eqnarray}
\nonumber
(G_{u}, X_{d}), (G_{d}, X_{u})=&&(\langle c^{\gamma}d^{\rho}\rangle, c^{x}d^{y}),~~ (\langle c^{\gamma}d^{\rho}\rangle, bc^{x}d^{-x}),~~ (\langle bc^{\gamma}d^{\rho}\rangle, c^{x}d^{y}),~~ (\langle bc^{\gamma}d^{\rho}\rangle, bc^{x}d^{-x})\,,\\
\label{eq:res_sym_quark_single}&&(\langle bc^{\gamma}d^{\rho}\rangle, abc^{x}d^{2x}), ~~ (\langle ac^{\gamma}d^{\rho}\rangle, c^{x}d^{y}),~~ (\langle ac^{\gamma}d^{\rho}\rangle, bc^{x}d^{-x})\,,
\end{eqnarray}
where $\gamma,\rho,x,y=0,1,...,n-1$.

We have analyzed all the possible residual symmetries in Eq.~\eqref{eq:res_sym_quark_single} up to $n\leq40$, the numerical results are summarized in table~\ref{Tab:numerical_result_quark_Onecp}. We find that the index $n$ has to be at least $n=7$ in order to explain the experimental data of  quark mixing in Eq.~\eqref{eq:full_fit}, and the corresponding remnant symmetry is $(X_{u}, G_{d})=(c^{x}d^{y}, \langle bc^{\gamma}d^{\rho}\rangle)$. Using the general formula of the CKM matrix in Eq.~\eqref{eq:U_CKM_form2}, the quark mixing matrix up to permutations of columns reads as
\begin{equation}
\label{eq:U_CKM_example}
V''_{CKM}=\frac{1}{\sqrt{2}}O_{3}^{T}
\left(\begin{array}{ccc}
1 & 1 & 0 \\
0 & 0 & \sqrt{2} \\
-e^{-i\varphi_{1}} & e^{-i\varphi_{1}} & 0
\end{array} \right)\,,
\end{equation}
where
\begin{equation}
\varphi_{1}=\frac{x+y+\gamma+\rho}{n}\pi\,,
\end{equation}
which can take the following discrete values
\begin{equation}
\varphi_{1} (\text{mod}~2\pi)=0,\frac{1}{n}\pi,\frac{2}{n}\pi,...,\frac{2n-1}{n}\pi\,.
\end{equation}
We can check that the matrix $V'_{CKM}$ has the following symmetry properties,
\begin{eqnarray}
\nonumber V''_{CKM}(\varphi_{1}+\pi,\theta_{1},\theta_{2},\theta_{3})&=&\text{diag}(1,1,-1)V''_{CKM}(\varphi_{1},-\theta_{1},-\theta_{2},\theta_{3})\,,\\
\nonumber V''_{CKM}(\varphi_{1},\theta_{1},\theta_{2},\theta_{3})P_{12}&=&\text{diag}(1,1,-1)V''_{CKM}(\varphi_{1},-\theta_{1},-\theta_{2},\theta_{3})\,,\\
V''_{CKM}(\pi-\varphi_{1},\theta_{1},\theta_{2},\theta_{3})&=&\text{diag}(1,1,-1)V''_{CKM}(\varphi_{1},\theta'_{1},\theta'_{2},\theta'_{3})\text{diag}(-e^{i\varphi_{1}},e^{i\varphi_{1}},1)\,,
\end{eqnarray}
where the parameters $\theta'_{1,2,3}$ fulfill $O_{3}(\theta'_{1},\theta'_{2},\theta'_{3})=P_{13}O_{3}(-\theta_{1},-\theta_{2},\theta_{3})$. As a result, the parameter $\varphi_{1}$ can be limited in the range of $0\leq \varphi_{1}\leq\frac{\pi}{2}$, and six column permutations lead to three independent mixing patterns,
\begin{equation}
V''_{CKM,1}=V''_{CKM},~~~~ V''_{CKM,2}=V''_{CKM}P_{13}, ~~~~ V''_{CKM,3}=V''_{CKM}P_{23}\,.
\end{equation}
For the mixing pattern $V''_{CKM,1}$, we can extract the following results for the mixing angles and CP invariants
\begin{eqnarray}
\nonumber
\sin^{2}\theta_{13}^{q}&=&(\cos\theta_{3}\sin\theta_{1}\sin\theta_{2}+\cos\theta_{1}\sin\theta_{3})^{2}\,,\\
\nonumber
\sin^{2}\theta_{12}^{q}&=&\frac{1}{2}+\frac{(\sin\theta_{1}\sin\theta_{3}-\cos\theta_{1}\cos\theta_{3}\sin\theta_{2})\cos\theta_{2}\cos\theta_{3}\cos\varphi_{1}}{1-(\cos\theta_{3}\sin\theta_{1}\sin\theta_{2}+\cos\theta_{1}\sin\theta_{3})^{2}}\,,\\
\nonumber
\sin^{2}\theta_{13}^{q}&=&1-\frac{\cos^{2}\theta_{2}\sin^{2}\theta_{1}}{(\cos\theta_{3}\sin\theta_{1}\sin\theta_{2}+\cos\theta_{1}\sin\theta_{3})^{2}}\,,\\
J_{CP}^{q}&=&\frac{1}{4}[\cos 2\theta_{3}\sin 2\theta_{1}\sin\theta_{2}+\sin2\theta_{3}(\cos^{2}\theta_{1}-\sin^{2}\theta_{1}\sin^{2}\theta_{2})]\cos\theta_{2}\sin\theta_{1}\sin\varphi_{1}\,.
\end{eqnarray}
For the small group index $n=7$, we find that only $V''_{CKM,1}$ can
accommodate the experimental data on CKM mixing matrix for certain values of the continuous parameters $\theta_{1,2,3}$ and the discrete parameter $\varphi_{1}$. As an example,
\begin{equation}
\varphi_{1}=\frac{\pi}{7},~~\theta_{1}=0.49172\pi,~~\theta_{2}=0.01054\pi,~~\theta_{3}=0.73906\pi\,,
\end{equation}
the quark mixing parameters are determined to be
\begin{eqnarray}
\nonumber&& \sin\theta_{12}^{q}=0.22497,~~~~ \sin\theta_{13}^{q}=0.00356\,,\\
&&\sin\theta_{23}^{q}=0.04195, ~~~~  J_{CP}^{q}=3.233\times 10^{-5}\,,
\end{eqnarray}
which is compatible with the data. The other two mixing matrices $V''^{\dagger}_{CKM,1}$ and $V''^{\dagger}_{CKM,2}$ can also be compatible with the precisely measured quark mixing, and they can be reproduced from the remnant symmetry $(G_{u}, X_{d})=(\langle bc^{\gamma}d^{\rho}\rangle, c^{x}d^{y})$.

\begin{table}[!hbp]
\centering
\footnotesize
\begin{tabular}{|c|c|c|c|c|c|c|c|c|c|}
\hline
\hline
 & $n$ & $\varphi_{1}$ & $\theta_{1}/\pi$ & $\theta_{2}/\pi$ & $\theta_{4}/\pi$ & $\sin\theta_{13}^{q}$ & $\sin\theta_{12}^{q}$ & $\sin\theta_{23}^{q}$ & $J^{q}_{CP}/10^{-5}$   \\
\hline

\multirow{18}{*}{$V''_{CKM,1}$} & $7,14,21,28,35$ & $\pi/7$ & $0.49100$ & $0.00994$ & $0.76127$ & $0.00356$ & $0.22497$ & $0.04195$ & $3.233$ \\
 \cline{2-10}
& $8,16,24,32,40$ & $\pi/8$ & $0.50988$ & $0.99068$ & $0.21264$ & $0.00375$ & $0.22497$ & $0.04248$ & $3.043$ \\
 \cline{2-10}
& $9,18,27,36$ & $\pi/9$ & $0.48975$ & $0.00913$ & $0.79731$ & $0.00390$ & $0.22497$ & $0.04293$ & $2.859$  \\
 \cline{2-10}
& $15,30$ & $2\pi/15$ & $0.50772$ & $0.98893$ & $0.27850$ & $0.00366$ & $0.22497$ & $0.04223$ & $3.138$ \\
 \cline{2-10}
& $17,34$ & $2\pi/17$ & $0.49273$ & $0.01156$ & $0.70724$ & $0.00383$ & $0.22497$ & $0.04272$ & $2.950$  \\
 \cline{2-10}
& $22$ & $3\pi/22$ & $0.50785$ & $0.98906$ & $0.27445$ & $0.00362$ & $0.22497$ & $0.04214$ & $3.170$   \\
 \cline{2-10}
& $23$ & $3\pi/23$ & $0.49031$ & $0.00943$ & $0.78213$ & $0.00369$ & $0.22497$ & $0.04231$ & $3.106$   \\
 \cline{2-10}
& $25$ & $3\pi/25$ & $0.49268$ & $0.01150$ & $0.70895$ & $0.00380$ & $0.22497$ & $0.04264$ & $2.981$  \\
 \cline{2-10}
& $26$ & $3\pi/26$ & $0.48985$ & $0.00918$ & $0.79463$ & $0.00385$ & $0.22497$ & $0.04279$ & $2.919$  \\
 \cline{2-10}
& $29$ & $4\pi/29$ & $0.49064$ & $0.00966$ & $0.77235$ & $0.00361$ & $0.22497$ & $0.04209$ & $3.185$   \\
 \cline{2-10}
& $31$ & $4\pi/31$ & $0.49026$ & $0.00940$ & $0.78359$ & $0.00370$ & $0.22497$ & $0.04236$ & $3.090$  \\
 \cline{2-10}
& $33$ & $4\pi/33$ & $0.49265$ & $0.01146$ & $0.70987$ & $0.00379$ & $0.22497$ & $0.04260$ & $2.996$ \\
 \cline{2-10}
& $35$ & $4\pi/35$ & $0.48983$ & $0.00916$ & $0.79535$ & $0.00386$ & $0.22497$ & $0.04283$ & $2.904$  \\
 \cline{2-10}
& $36$ & $5\pi/36$ & $0.49069$ & $0.00970$ & $0.77070$ & $0.00360$ & $0.22497$ & $0.04206$ & $3.195$ \\
 \cline{2-10}
&  \multirow{2}{*}{$37$} & $4\pi/37$ & $0.49291$ & $0.01179$ & $0.70129$ & $0.00393$ & $0.22497$ & $0.04303$ & $2.814$  \\
 \cline{3-10}
&  & $5\pi/37$ & $0.49050$ & $0.00956$ & $0.77651$ & $0.00364$ & $0.22497$ & $0.04217$ & $3.157$   \\
 \cline{2-10}
& $38$ & $5\pi/38$ & $0.49234$ & $0.01113$ & $0.71945$ & $0.00368$ & $0.22497$ & $0.04228$ & $3.119$  \\
 \cline{2-10}
& $39$ & $5\pi/39$ & $0.49023$ & $0.00938$ & $0.78441$ & $0.00371$ & $0.22497$ & $0.04238$ & $3.081$  \\

\hline
\multirow{18}{*}{$V''^{\dagger}_{CKM,1}$} & $16,32$ & $\pi/16$ & $0.49083$ & $0.00930$ & $0.68459$ & $0.00403$ & $0.22491$ & $0.04082$ & $3.174$   \\
 \cline{2-10}
& $17,34$ & $\pi/17$ & $0.49066$ & $0.00946$ & $0.68378$ & $0.00386$ & $0.22494$ & $0.04157$ & $3.146$   \\
 \cline{2-10}
& $18,36$ & $\pi/18$ & $0.49050$ & $0.00959$ & $0.68311$ & $0.00370$ & $0.22497$ & $0.04222$ & $3.118$   \\
\cline{2-10}
& $19,38$ & $\pi/19$ & $0.49035$ & $0.00962$ & $0.68256$ & $0.00353$ & $0.22496$ & $0.04264$ & $3.122$ \\
\cline{2-10}
& $20,40$ & $\pi/20$ & $0.49025$ & $0.00914$ & $0.68213$ & $0.00356$ & $0.22487$ & $0.04183$ & $3.252$ \\
\cline{2-10}
& $21$ & $\pi/21$ & $0.49015$ & $0.00904$ & $0.68172$ & $0.00360$ & $0.22489$ & $0.04183$ & $3.231$  \\
\cline{2-10}
& $22$ & $\pi/22$ & $0.49006$ & $0.00901$ & $0.68137$ & $0.00364$ & $0.22492$ & $0.04198$ & $3.190$ \\
\cline{2-10}
& $23$ & $\pi/23$ & $0.48997$ & $0.00901$ & $0.68106$ & $0.00367$ & $0.22495$ & $0.04218$ & $3.141$  \\
\cline{2-10}
& $24$ & $\pi/24$ & $0.48988$ & $0.00902$ & $0.68079$ & $0.00369$ & $0.22499$ & $0.04240$ & $3.088$  \\
\cline{2-10}
& $25$ & $\pi/25$ & $0.48980$ & $0.00904$ & $0.68055$ & $0.00371$ & $0.22502$ & $0.04263$ & $3.033$  \\
\cline{2-10}
  & $26$ & $\pi/26$ & $0.48972$ & $0.00906$ & $0.68034$ & $0.00373$ & $0.22505$ & $0.04286$ & $2.977$  \\
\cline{2-10}
  & $27$ & $\pi/27$ & $0.48965$ & $0.00909$ & $0.68015$ & $0.00375$ & $0.22507$ & $0.04309$ & $2.921$  \\
\cline{2-10}
& $28$ & $\pi/28$ & $0.51042$ & $0.99088$ & $0.32001$ & $0.00376$ & $0.22510$ & $0.04330$ & $2.865$   \\
\cline{2-10}
& $29$ & $\pi/29$ & $0.51048$ & $0.99086$ & $0.32016$ & $0.00377$ & $0.22512$ & $0.04351$ & $2.809$  \\
\cline{2-10}
& $33$ & $2\pi/33$ & $0.49074$ & $0.00938$ & $0.68417$ & $0.00394$ & $0.22493$ & $0.04121$ & $3.160$ \\
\cline{2-10}
& $35$ & $2\pi/35$ & $0.49058$ & $0.00953$ & $0.68343$ & $0.00378$ & $0.22496$ & $0.04191$ & $3.131$ \\
\cline{2-10}
& $37$ & $2\pi/37$ & $0.49042$ & $0.00963$ & $0.68282$ & $0.00362$ & $0.22498$ & $0.04249$ & $3.109$ \\
\cline{2-10}
& $39$ & $2\pi/39$ & $0.49029$ & $0.00928$ & $0.68235$ & $0.00352$ & $0.22488$ & $0.04202$ & $3.235$  \\
\hline
\multirow{5}{*}{$V''^{\dagger}_{CKM,2}$} & $36$ & $\pi/36$ & $0.00156$ & $0.75000$ & $0.57108$ & $0.00347$ & $0.22482$ & $0.04362$ & $3.309$  \\\cline{2-10}
& $37$ & $\pi/37$ & $0.99842$ & $0.25000$ & $0.42893$ & $0.00352$ & $0.22486$ & $0.04244$ & $3.267$ \\ \cline{2-10}
& $38$ & $\pi/38$ & $0.99838$ & $0.25286$ & $0.42895$ & $0.00356$ & $0.22489$ & $0.04229$ & $3.225$ \\
 \cline{2-10}
& $39$ & $\pi/39$ & $0.00165$ & $0.74588$ & $0.57103$ & $0.00361$ & $0.22492$ & $0.04229$ & $3.183$  \\ \cline{2-10}
& $40$ & $\pi/40$ & $0.00167$ & $0.74499$ & $0.57102$ & $0.00365$ & $0.22495$ & $0.04229$ & $3.142$  \\
\hline\hline
\end{tabular}
\caption{ \label{Tab:numerical_result_quark_Onecp}
Numerical results of the quark mixing parameters when the $\Delta(6n^2)$ abd CP symmetries are broken to an abelian subgroup and single CP in the up and down quark sectors. The mixing pattern $V''_{CKM, 1}$ can be obtained from the residual symmetry $(X_{u}, G_{d})=(c^{x}d^{y}, \langle bc^{\gamma}d^{\rho}\rangle)$, $V''^{\dagger}_{CKM, 1}$ and $V''^{\dagger}_{CKM,2}$ arise from $(G_{u}, X_{d})=(\langle bc^{\gamma}d^{\rho}\rangle, c^{x}d^{y})$. Here we focus on the $\Delta(6n^2)$ group with $n\leq40$. We display quark mixing angles $\sin\theta_{ij}^{q}$ and CP invariant $J_{CP}^{q}$ which are compatible with the experimental data for certain values of $\theta_{u}$, $\delta_{u}$, $\theta_{d}$ and $\varphi_{1}$.}
\end{table}

\section{\label{sec:conclusion}Conclusion}

Discrete flavor symmetry in combination with generalized CP has been widely exploited to predict lepton mixing angles and CP violating phases. It is well-known that the observed quark mixing pattern is drastically different from the neutrino mixing. It is intriguing to investigate whether both neutrino and quark mixing can be explained from the same discrete flavor symmetry group. In~\cite{Li:2017abz} we find that a unified interpretation of quark and lepton mixing can be achieved if the flavor and CP symmetries are broken to $Z_2\times CP$ in all the neutrino, charged lepton, up quark
and down quark sectors, or alternatively the residual symmetry of the charged lepton mass term is $Z_m$, $m\geq3$ instead of $Z_2\times CP$. In the present work, we have considered two other possible approaches to explain the patterns of quark and lepton mixing based on discrete
flavor symmetry and generalized CP.

In the first scenario, the residual symmetries of the charged lepton and neutrino mass matrices are $Z_2$ and $Z_2\times CP$ respectively. The lepton mixing matrix is predicted to be of the form Eq.~\eqref{eq:PMNS_z2z2cp}. All
mixing angles and CP phases are then expressed in terms of three free parameters $\theta_{l}$, $\delta_{l}$ and $\theta_{\nu}$. We derive the criterion to determine whether two distinct residual symmetries give rise to the same lepton mixing pattern. It is remarkable that the criterion given by Eqs.~(\ref{eq:equv_cond1_Z2Z2CP}, \ref{eq:equv_cond2_Z2Z2CP},\ref{eq:equv_cond3_Z2Z2CP}) is quite simple if we change the $\Sigma\equiv\Sigma^{\dagger}_{l}\Sigma_{\nu}$ matrix into the ``standard'' form. As an example, we analyze the lepton mixing patterns arising from the flavor group $\Delta(6n^2)$ and CP which are broken to all possible residual symmetries indicated above. Then we discuss whether it is possible to obtain the experimentally favored quark mixing in a similar fashion, assuming that the flavor and CP symmetries are broken to $Z_2$ and $Z_2\times CP$ in the up and down quark sectors. The most general form of the CKM mixing is given by Eq.~\eqref{eq:CKM_form_2} in this case. We also derive the sufficient and necessary condition for the equivalence of two quark mixing matrices. It is remarkable that the experimentally preferred quark and lepton mixing patterns can be obtained from the $\Delta(6n^2)$ flavor symmetry and generalized CP, and the smallest flavor group is $\Delta(294)$ with $n=7$.

The second scenario has an abelian subgroup and single CP transformation as residual symmetries of the charged lepton and neutrino mass matrices respectively. The lepton mixing matrix is fixed up to a real orthogonal matrix $O_{3}$ which contains three rotation angles $\theta_{1,2,3}$. We extend this approach to the quark sector. The single CP transformation can be preserved by the down (or up) quark sector, accordingly the residual symmetry of the up (or down) quark mass matrix would be an abelian subgroup. We find that the $\Delta(6n^2)$ flavor symmetry combined with CP can reproduce the drastically different texture of quark and lepton mixing in this scheme, and the smallest flavor group to achieve this is $\Delta(294)$ as well.

The neutrino mixing angles and CP phases are predicted to depend on three free parameters $\theta_{l}$, $\delta_{l}$, $\theta_{\nu}$ or $\theta_{1,2,3}$ in the above two scenarios. Detailed numerical analyses show that the values of CP violating phases are correlated with the values of the three mixing angles. In particular, there is generally strong correlation between the Dirac CP phase $\delta_{CP}$ and the atmospheric angle $\theta_{23}$. Future neutrino oscillation experiments
can significantly improve the sensitivity to $\theta_{12}$, $\theta_{23}$ and $\delta_{CP}$~\cite{An:2015jdp,Kim:2014rfa,Acciarri:2016crz,Acciarri:2015uup,Strait:2016mof,Acciarri:2016ooe,Kearns:2013lea,Abe:2014oxa,Abe:2016ero,Geer:1997iz,DeRujula:1998umv,Bandyopadhyay:2007kx}.
We expect that forthcoming neutrino facilities are able to exclude certain mixing patterns that we have identified, or provide strong evidence for their continued relevance. In addition, the next generation neutrinoless double beta decay experiments are able to probe the whole parameter space of the inverse ordering mass spectrum. Thus the predicted mixing patterns for inverted ordering could be excluded or confirmed independently of neutrino oscillation.

In the present work, we have considered the set-up in which the residual symmetries of the quark and lepton sectors are of the same structure. Alternatively the lepton mixing can be understood in the semi-direct approach~\cite{Feruglio:2012cw,Ding:2013hpa,Ding:2013bpa,Feruglio:2013hia,Li:2013jya,Ding:2013nsa,Ding:2014hva,Ding:2014ssa,Li:2014eia,
Hagedorn:2014wha,Ding:2014ora,Branco:2015hea,Branco:2015gna,Li:2015jxa,DiIura:2015kfa,Ballett:2015wia,Ding:2015rwa,Li:2016ppt} while the residual symmetry of quark sector is kept intact.
Then all the lepton mixing angles and CP violation phases are expressed in term of one single real parameter $\theta$ and the CKM mixing matrix still depends on three parameters.
Finally we would like to mention that the symmetry breaking patterns discussed in the present work provide new starting points for building models which can explain quark and lepton mixing simultaneously, and the $\Delta(294)$ flavor symmetry looks particularly interesting.

\section*{Acknowledgements}

This work is supported by the National Natural Science Foundation of China under Grant No.11522546.

\vskip 2cm

\bibliographystyle{utphys}

\bibliography{quark_lepton_D6nsq}

\begin{figure}[t!]
\centering
\includegraphics[width=0.98\textwidth]{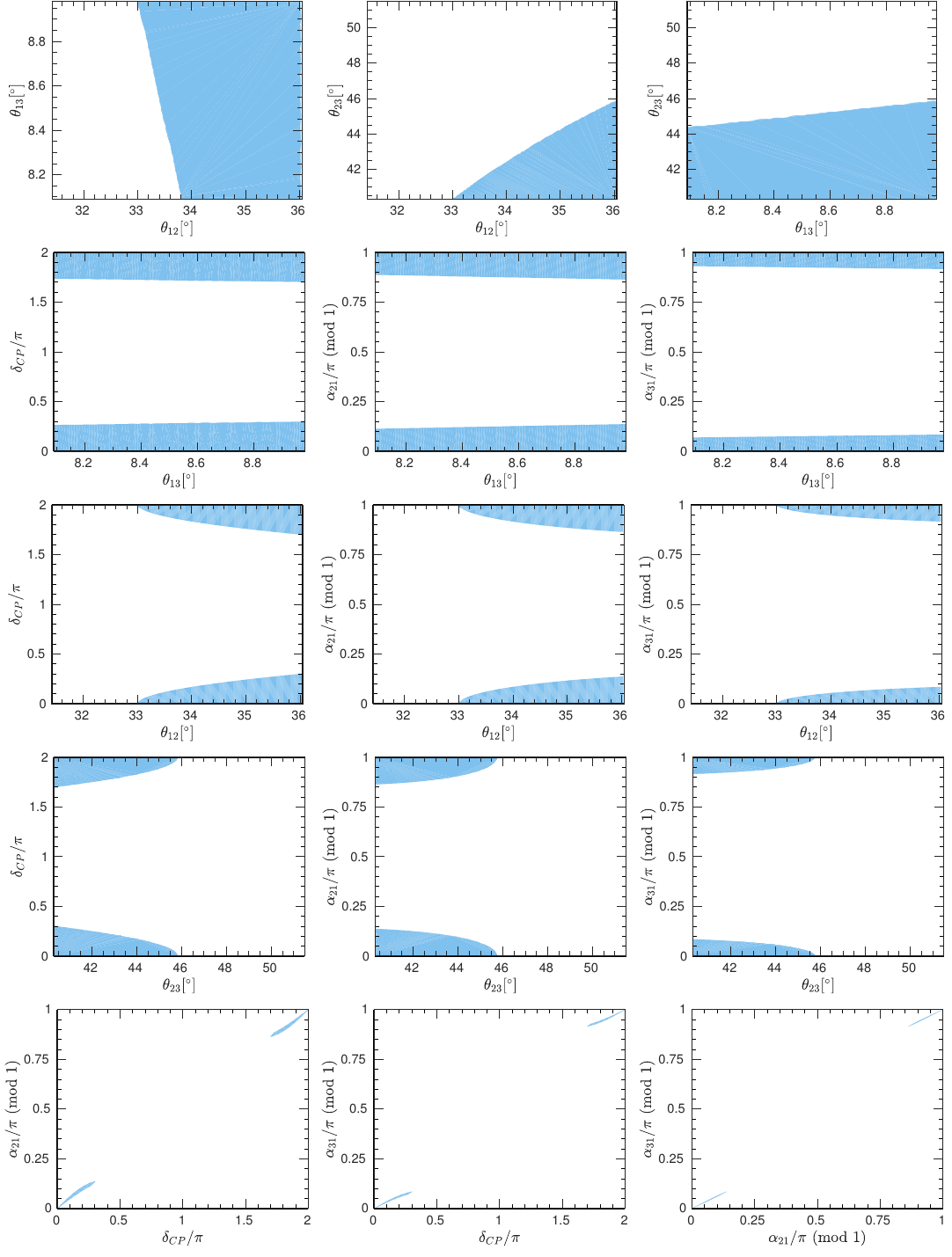}
\caption{\label{fig:U_I_4_correlations}Correlations between different mixing parameters for the mixing pattern $U_{I,4}$ with $\varphi_1=\pi/3$, where the residual symmetry is  $(G_{l},G_{\nu},X_{\nu})=(Z_{2}^{bc^{x}d^{x}},Z_{2}^{bc^{y}d^{y}}, \{c^{\rho}d^{-2y-\rho},bc^{y+\rho}d^{-y-\rho}\})$, and the three lepton mixing angles are required to be compatible with the experimental data at $3\sigma$ level~\cite{Esteban:2016qun}.
}
\end{figure}

\begin{figure}[t!]
\centering
\includegraphics[width=0.98\textwidth]{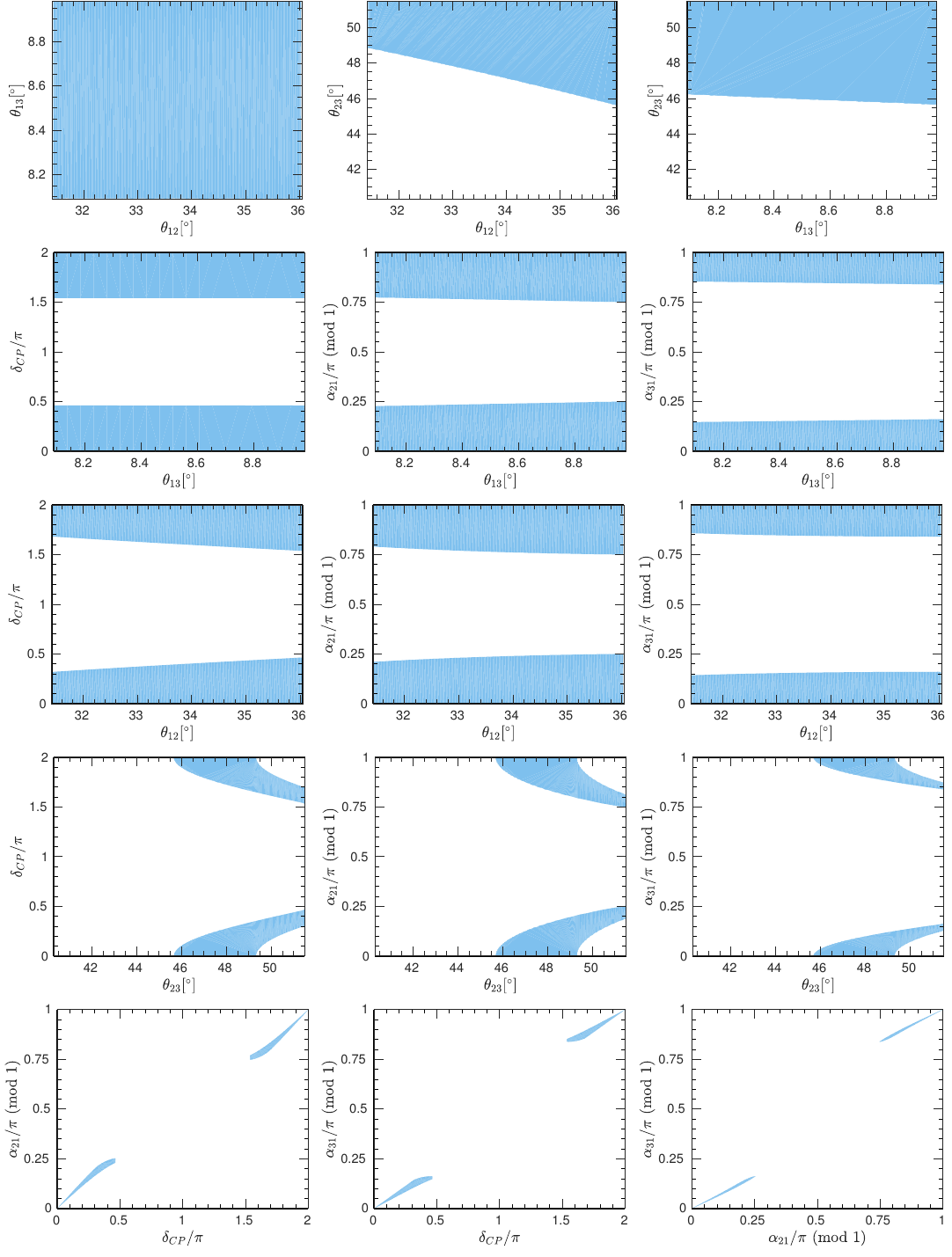}
\caption{\label{fig:U_I_5_correlations}
Correlations between different mixing parameters for the mixing pattern $U_{I, 5}$ with $\varphi_1=\pi/3$, where the residual symmetry is  $(G_{l},G_{\nu},X_{\nu})=(Z_{2}^{bc^{x}d^{x}},Z_{2}^{bc^{y}d^{y}},\{c^{\rho}d^{-2y-\rho},bc^{y+\rho}d^{-y-\rho}\})$, and the three lepton mixing angles are required to be compatible with the experimental data at $3\sigma$ level~\cite{Esteban:2016qun}.
}
\end{figure}

\begin{figure}[t!]
\centering
\includegraphics[width=0.98\textwidth]{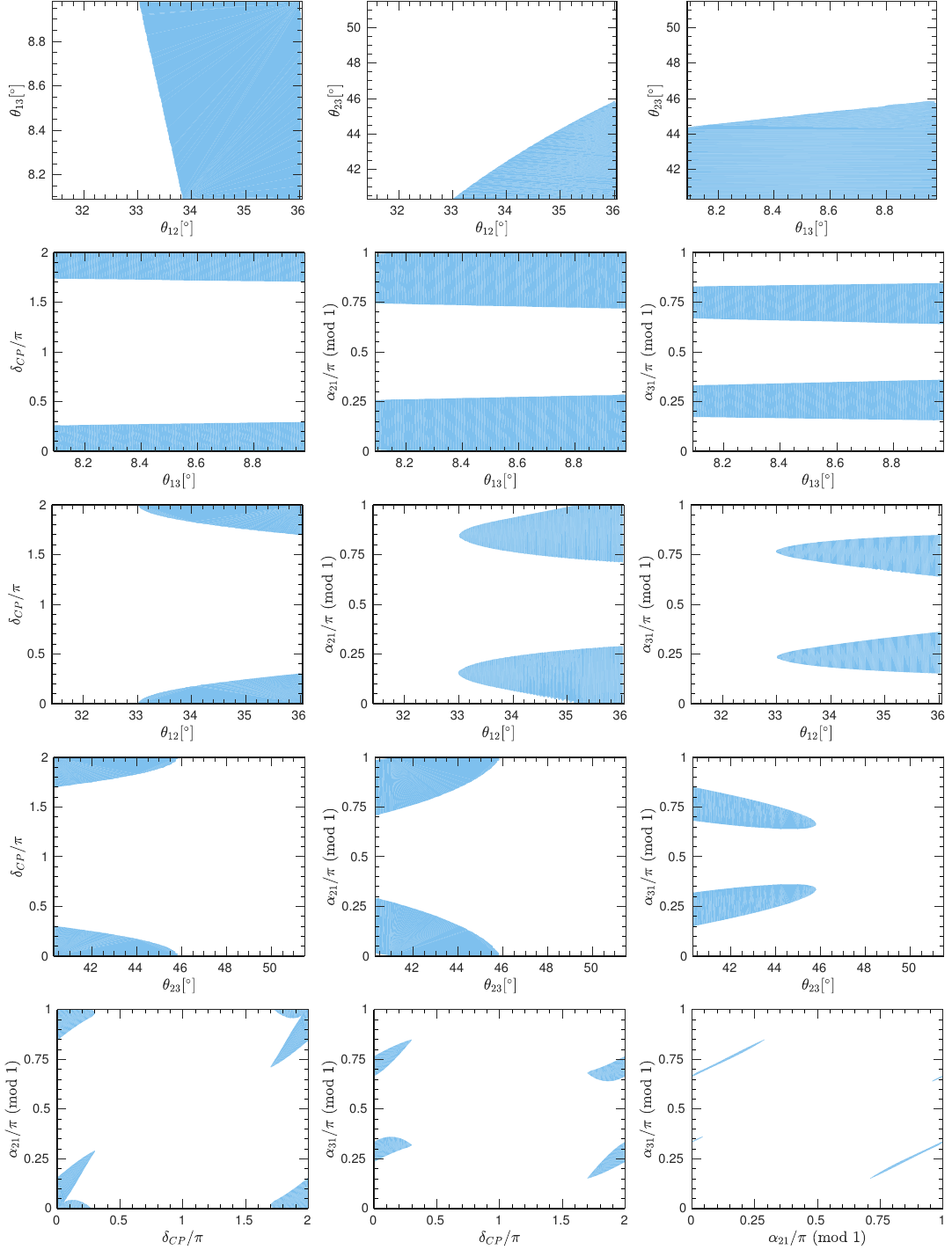}
\caption{\label{fig:U_II_1_correlations}
Correlations between different mixing parameters for the mixing pattern $U_{II, 1}$ with $\varphi_{4}=\pi/3$, where the residual symmetry is  $(G_{l},G_{\nu},X_{\nu})=(Z_{2}^{bc^{x}d^{x}},Z_{2}^{abc^{y}}, \{c^{\rho}d^{2y+2\rho},abc^{y+\rho}d^{2y+2\rho}\})$, and the three lepton mixing angles are required to be compatible with the experimental data at $3\sigma$ level~\cite{Esteban:2016qun}.
}
\end{figure}

\begin{figure}[t!]
\centering
\includegraphics[width=0.98\textwidth]{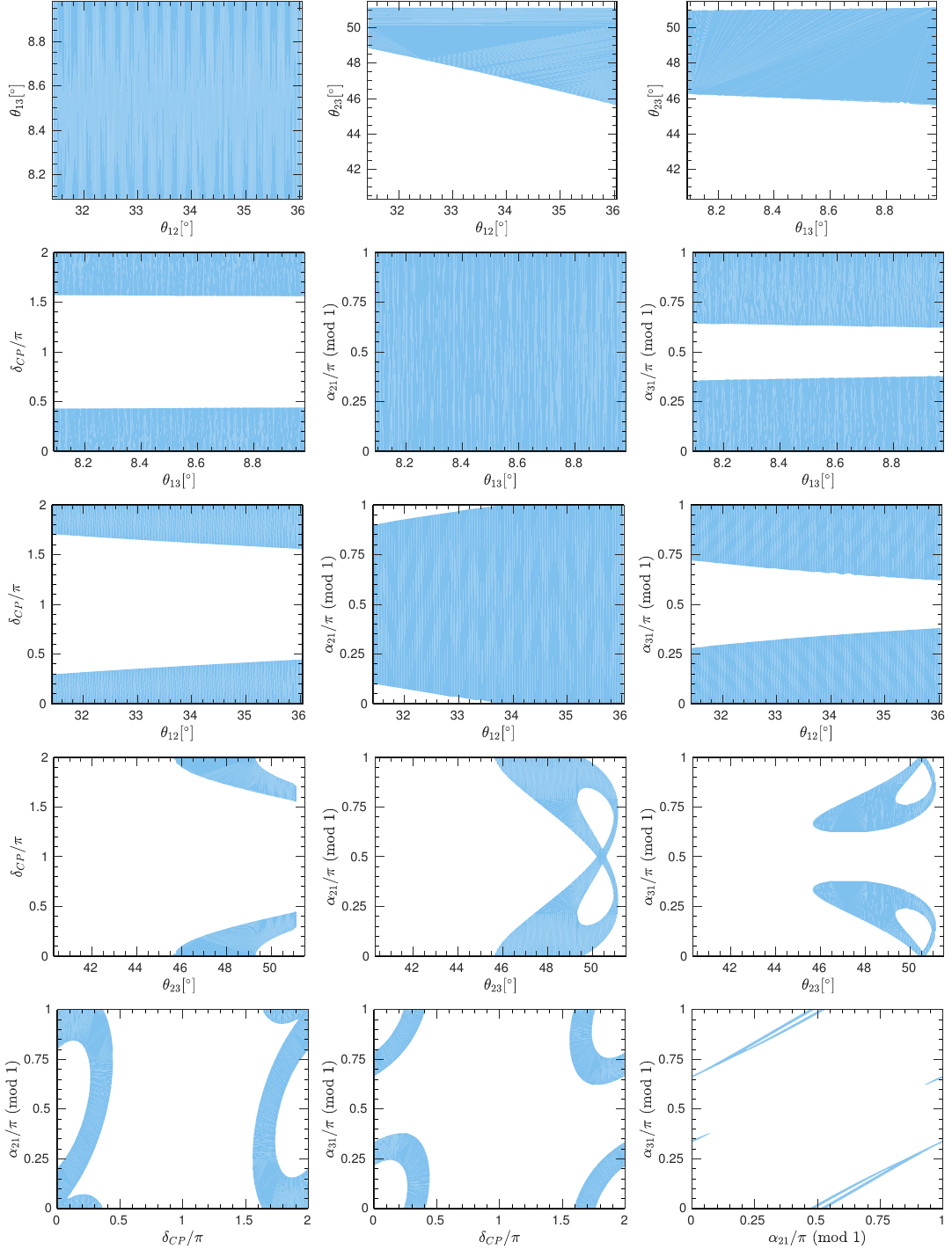}
\caption{\label{fig:U_II_2_correlations}
Correlations between different mixing parameters for the mixing pattern $U_{II, 2}$ with $\varphi_{4}=\pi/3$, where the residual symmetry is  $(G_{l}, G_{\nu}, X_{\nu})=(Z_{2}^{bc^{x}d^{x}},Z_{2}^{abc^{y}},\{c^{\rho}d^{2y+2\rho},abc^{y+\rho}d^{2y+2\rho}\})$ , and the three lepton mixing angles are required to be compatible with the experimental data at $3\sigma$ level~\cite{Esteban:2016qun}.
}
\end{figure}

\begin{figure}[t!]
\centering
\includegraphics[width=0.98\textwidth]{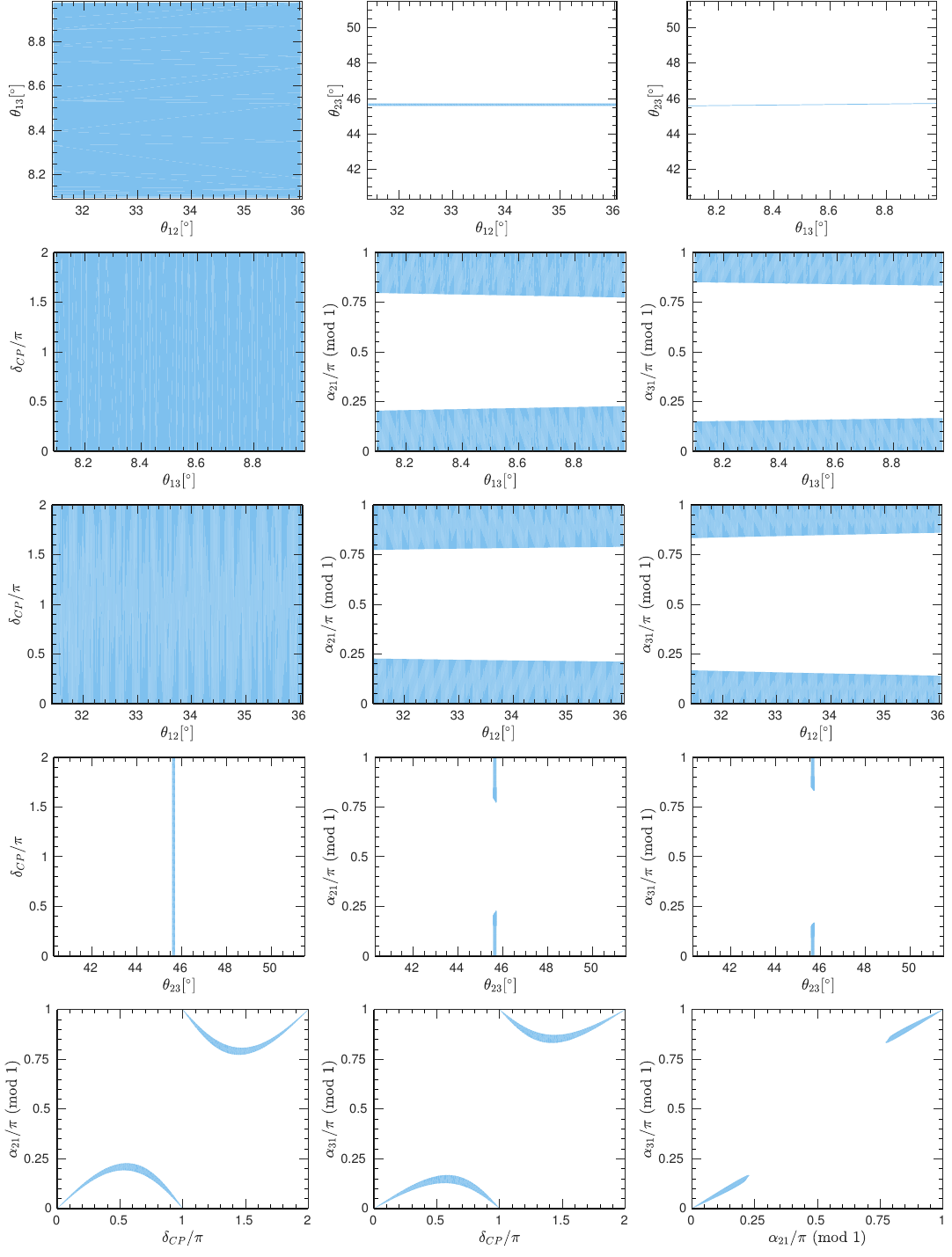}
\caption{\label{fig:U_III_2_correlations}
Correlations between different mixing parameters for the mixing pattern $U_{III, 2}$ with $\varphi_6=0$, where the residual symmetry is  $\{G_{l},G_{\nu},X_{\nu}\}=\{Z_{2}^{bc^{x}d^{x}},Z_{2}^{c^{n/2}},c^{\gamma}d^{\rho}\}$, and the three lepton mixing angles are required to be compatible with the experimental data at $3\sigma$ level~\cite{Esteban:2016qun}. }
\end{figure}

\begin{figure}[t!]
\centering
\includegraphics[width=0.98\textwidth]{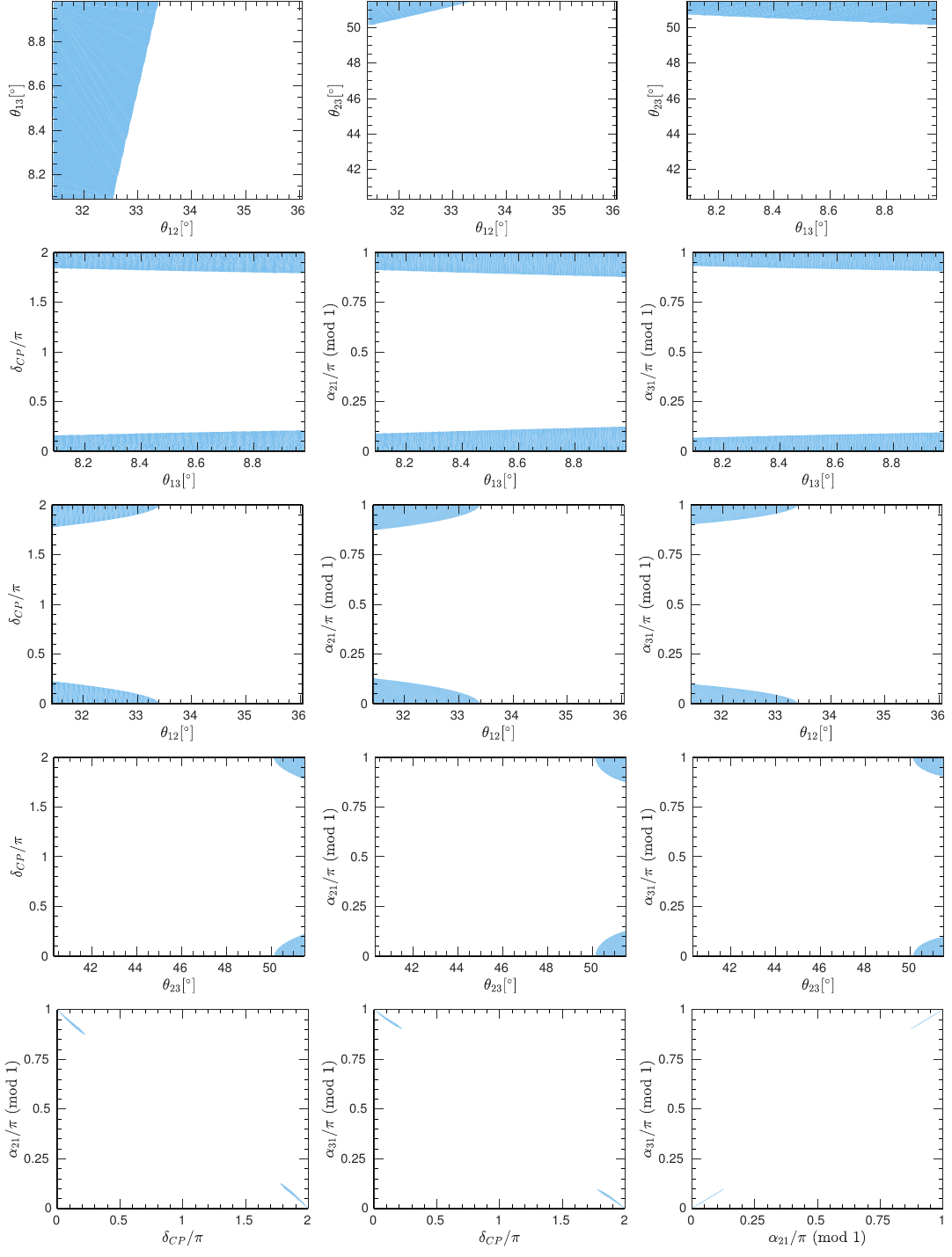}
\caption{\label{fig:U_III_1_correlations}
Correlations between different mixing parameters for the mixing pattern $U_{III, 3}$ with $\varphi_6=0$, where the residual symmetry is  $\{G_{l},G_{\nu},X_{\nu}\}=\{Z_{2}^{bc^{x}d^{x}},Z_{2}^{c^{n/2}},c^{\gamma}d^{\rho}\}$, and the three lepton mixing angles are required to be compatible with the experimental data at $3\sigma$ level~\cite{Esteban:2016qun}.
}
\end{figure}

\end{document}